\documentclass{jfm}

\usepackage{graphicx}
\usepackage{natbib}
\usepackage{amsmath}
\usepackage{psfrag}
\usepackage{psfragx}
\usepackage{epsfig}
\usepackage{xcolor,import}
\usepackage{pstool}
\usepackage{placeins}
\usepackage{amssymb}
\usepackage{mathabx}
\usepackage{graphicx,xcolor}        
\usepackage{subfig}                             
\usepackage{wasysym}        
\usepackage{floatrow}
\usepackage{dashrule}
\usepackage{tikz}
\usepackage{booktabs}

\usetikzlibrary{arrows,decorations.markings}
\usetikzlibrary{calc}
\ifCUPmtlplainloaded \else
  \checkfont{eurm10}
  \iffontfound
    \IfFileExists{upmath.sty}
      {\typeout{^^JFound AMS Euler Roman fonts on the system,
                   using the 'upmath' package.^^J}%
       \usepackage{upmath}}
      {\typeout{^^JFound AMS Euler Roman fonts on the system, but you
                   dont seem to have the}%
       \typeout{'upmath' package installed. JFM.cls can take advantage
                 of these fonts,^^Jif you use 'upmath' package.^^J}%
      }
  \else
  \fi
\fi

\ifCUPmtlplainloaded \else
  \checkfont{msam10}
  \iffontfound
    \IfFileExists{amssymb.sty}
      {\typeout{^^JFound AMS Symbol fonts on the system, using the
                'amssymb' package.^^J}%
       \usepackage{amssymb}%
         
         \let\geq=\geqslant
      }{}
  \fi
\fi

\ifCUPmtlplainloaded \else
  \IfFileExists{amsbsy.sty}
    {\typeout{^^JFound the 'amsbsy' package on the system, using it.^^J}%
     \usepackage{amsbsy}}
    {}
\fi

\newcommand\Rey{\mbox{\textit{Re}}}  

\newsavebox{\astrutbox}
\sbox{\astrutbox}{\rule[-5pt]{0pt}{20pt}}

\tikzset{%
  >=latex,
  inner sep=0pt,%
  outer sep=2pt,%
  mark coordinate/.style={inner sep=0pt,outer sep=0pt,minimum size=3pt,
    fill=black,circle}%
}
\usepackage{amsmath}
\usetikzlibrary{arrows}
\usepackage{pgfplots}
\usetikzlibrary{calc,fadings,decorations.pathreplacing}
\usepackage{commath}

\title[]{Lumley Decomposition of the Turbulent Round Jet Far-field. Part 2 - Dynamics}

\author[Azur Hod\v zi\' c, Knud Erik Meyer, William K. George, and Clara M. Velte]%
{Azur Hod\v zi\' c$^1$
  \thanks{Email address for correspondence: azuhod@mek.dtu.dk}, Knud Erik Meyer$^1$,\\
 William K. George$^2$, and Clara M. Velte$^1$}

\shortauthor{A. Hod\v zi\' c, K. E. Meyer, W. K. George and C. M. Velte}

\affiliation{$^1$Department of Mechanical Engineering, Technical University of Denmark,
2800, Kgs. Lyngby, Denmark\\[\affilskip]
$^2$Department of Aeronautics, Imperial College London, South Kensington Campus, London SW7 2AZ, UK}

\pubyear{2010}
\volume{650}
\pagerange{119--126}
\date{?; revised ?; accepted ?. - To be entered by editorial office}
\begin{document}

\maketitle

\begin{abstract}
In the current work the reconstruction of the far-field region of the turbulent axi-symmetric jet is performed in order to investigate the modal turbulence kinetic energy production contributions. The reconstruction of the field statistics is based on a semi-analytical Lumley Decomposition (LD) of the PIV sampled field using stretched amplitude decaying Fourier modes (SADFM), derived in \cite{Hodzic2019_part1}, along the streamwise coordinate. It is shown that, a wide range of modes obtain a significant amount of energy directly from the mean flow, and are therefore not exclusively dependent on a Richardson-like energy cascade even in the $\kappa$-range in which the energy spectra exhibit the $-5/3$-slope. It is observed that the $-7/3$-range in the cross-spectra is fully reconstructed using a single mode in regions of high mean shear, and that shear-stresses are nearly fully reconstructed using the first two modes. These results indicate that most of the energy production related to shear-stresses is related to the first LD mode. 
\end{abstract}
\section{Introduction}
Due to its symmetry and absence of solid boundaries, the turbulent round jet is in many ways an ideal flow for studying turbulence. Turbulent jets have been studied for almost half a century. The first detailed turbulence measurements (using hot-wires) in jet mixing layers were initially investigated by \cite{Corrsin1955} and \cite{Townsend1956}. They reported an intermittency surface dividing the turbulent/non-turbulent interface at the periphery of the jet. \cite{Wygnanski1969a} explored the similarity region of the jet measuring first- and second order moments.
Subsequent experiment using flying hot-wire to reduce cross-flow errors by \cite{Panchapakesan1993a} and \cite{Hussein1994} gave very different results, in large part because of the size of the enclosure relative to the jet. \cite{Hussein1994} performed extensive measurements and analysis of the jet flow, in particular, and demonstrated the impact of back-flow for confined jets, and its impact on statistics and self-similarity. The latter also demonstrated the non-isotropy of the velocity derivatives and dissipation.

\cite{george201750} (and especially Appendix II) provides an extensive review of the application of POD since its introduction by Lumley in 1967, so we provide only a brief summary here. Past investigations conducted in the vicinity of the round jet potential core utilizing the POD were conducted by \cite{leib1984application}, \cite{Glauser1987}, \cite{Citriniti2000}, \cite{Gamard2002}, \cite{Arndt1997}. The measurements were performed with hot-wire rakes allowing a spatio-temporal decomposition of the flow field cross-plane. The decomposition was applied in order to identify turbulent structures and their relation to radiated noise. These works demonstrated conclusively that energy was concentrated in a small number of POD modes, but many Fourier modes in both time and in the azimuthal direction. The results of \cite{Citriniti2000} identified vortex rings participating in rapid ejection events, confirming the conjecture of \cite{Glauser1987}. Furthermore, the reconstructed fields indicated the presence of stream-wise vortex filaments in relation to the ejection events. \cite{Jung2004} and \cite{Gamard2004} expanded these studies to cover the $2-69\,\mathrm{D}$ range from the nozzle, for three different Reynolds numbers, and identified the modal evolution of the dynamics of the flow. Their decomposition reached an asymptotic state faster than the single-point statistics.


\cite{Wanstrom2009} took advantage of the self-similar nature of the far-field turbulence identified by \cite{Ewing2007}, and performed a Fourier decompositions of the flow in the similarity scaled streamwise and azimuthal directions. Her work (see also \cite{Wanstrom2006}, \cite{Wanstrom2007}, \cite{Wanstrom2012}) revealed that, when scaled by the centerline velocity and interpolated onto the similarity coordinate system, the resulting flow field becomes homogeneous. This confirmed the results of \cite{Ewing2007} who showed that the correlation function of the scaled velocities is independent of the streamwise similarity coordinate. As originally suggested in the work of \cite{Ewing1995}, \cite{Wanstrom2009} concluded that the modes in the streamwise direction of the scaled field were Fourier modes. The main contribution of the work of \cite{Wanstrom2009} was in the decomposition of the turbulent flow field in the streamwise and cross-plane direction. \cite{Wanstrom2009} further hypothesized that multiple modes could tap into the energy extraction process from the mean flow directly. 

Most recently the studies of \cite{Towne2018} and \cite{Schmidt2017} performed space-time decompositions of the turbulent jet from LES data in order to demonstrate the effects of missing dimensions of the decompositions as well as to characterize the modal decomposition of the flow in the near- and intermediate region. The work of \cite{Mullyadzhanov2018} demonstrated propagating helical waves from a $\Rey=5940$ pipe jet until $x/D=40$.

The current study of the turbulence in the far-field region of the jet is a continuation of the work in \cite{Hodzic2019_part1} where the Lumley Decomposition (LD) was applied in order to decompose the flow into stretched amplitude decaying Fourier modes (SADFM) along the streamwise coordinate and into numerical modes in the transverse direction of the flow. The current work introduces the Galerkin projection of turbulence kinetic energy transport equation in general coordinates by applying a tensor notation to a basis in $L^2\left(\Omega,\mathbb{C}^3\right)$. The production term is analyzed and expanded by the eigenfunctions from experimental data, and used to show the modal contribution to the energy production term as well as allowing an in-depth study if the modal building blocks of the component spectra. The production term and the moments of the flow are analyzed in order to investigate the TKE transport from the mean flow to the eigenfunctions. 

In the current work the the experimental setup from \cite{Hodzic2019_part1} is summarized briefly, whereafter the Galerkin projection of the turbulence kinetic energy equation is introduced in curvilinear coordinates. The reconstruction of the SADFM spectra are then performed using modal building blocks from the LD. The reconstruction of the $-5/3$- and $-7/3$-ranges of the component energy- and cross spectra are then analyzed. This is followed by an energy production analysis which estimates the degree to which individual modes are able to obtain energy directly from the mean flow. Then a non-linear transport analysis is performed based on a classical model of the one-dimensional spectrum together with the reconstruction of the single-point statistics.
\section{Experimental procedure and data processing\label{sec:experimental_procedure}}\noindent
The experimental setup for the acquisition of the current data was described in \cite{Hodzic2019_part1}, and will only briefly be summarized in the current section. It used the same jet reported in earlier studies by \cite{Gamard2004} and \cite{Wanstrom2009}. The dataset $E_1$ from \cite{Hodzic2019_part1} is used for the analysis of the dynamics in the current work. The jet was driven by a fan supplying air into the jet box with inner dimensions $58.5\times58.5\times59\,\mathrm{cm^3}$. The jet-nozzle diameter was $D=1\mathrm{cm}$ creating a $\Rey=20\,000$ jet based on the nozzle exit velocity and nozzle diameter. 

DynamicSudio v4.0 was applied for data processing using adaptive correlation, resulting in a final grid of ${32\times 32\,\mathrm{pix}}$ with interrogation areas with $50\%$ overlap. The dimensions of the interrogation area for $E_1$ were ${\Delta^2 = \left(2.5\,\mathrm{mm}\right)^2}$ and ${\left(3.0\,\mathrm{mm}\right)^2}$ for cameras 1 and 2, respectively. Window shifting with moving averages was applied.  
%
\section{Galerkin projection of the velocity field}
In order to expand the governing equations in general coordinates in terms of an orthogonal set of vector-valued basis functions, ${L^2_w\left(\Omega,\mathbb{C}^3\right):=\left\lbrace \overline{\Phi}:\Omega\rightarrow\mathbb{C}^3 \rvert \int_\Omega \lVert \overline{\Phi} \rVert^2wd\mu <\infty\right\rbrace}$ (see \cite{Hodzic2019_part1}), the first part of this section will, for the sake of completeness, include the necessary aspects of the tensor formulation of the Lumley Decomposition, (for the full description see \cite{Hodzic2019_part1}). The tensor form of the Lumley Decomposition is given by \cite{Hodzic2019_part1}
\begin{eqnarray}
\int_\Omega R^i_{\cdot\hat{i}}\varphi^{\hat{i}}\hat{w}\sqrt{\widehat{Z}}d\mu^{\widehat{4}}=\lambda \varphi^i,
\label{eq:LD_arithmetic}
\end{eqnarray}
where $\sqrt{\widehat{Z}}$ is the volume element, $R^i_{\cdot\hat{i}}=\left\langle v^iv_{\hat{i}}\right\rangle$ is the two-point, two-time mixed correlation tensor and the hat, $(\hat{\cdot})$, indicates the coordinate of integration. The upper- and lower indices denote covariant and contravariant components, respectively, and repeated indices in a term invoke the Einstein summation convention. The orthogonality of the modes with respect to the $L^2\left(\Omega,\mathbb{C}^3\right)$-inner product can then be expressed in terms of the Kronecker delta (mixed and lowered indices)
\begin{equation}
\delta^\alpha_\beta=\left(\overline{\Phi}^\alpha,\overline{\Phi}_\beta\right)_w\, ,\,\delta_{\alpha\beta}=\left(\overline{\Phi}_\alpha,\overline{\Phi}_\beta\right)_w.\label{eq:orthogonality_eigenfunctions}
\end{equation}
The fluctuating part of the velocity vector, $\overline{v}=v^i\overline{z}_i$, can then be decomposed in terms of the eigenfunctions
\begin{equation}
\overline{v} = v^\alpha\overline{\Phi}_\alpha,\label{eq:decomposition_basis}
\end{equation}
and the ensemble averaged kinetic energy of the field is given by a summation over $\alpha$
\begin{equation}
\sum_{\alpha=1}^n\lambda^\alpha=\sum_{\alpha=1}^n\left\langle \left|v^\alpha\right|^2\right\rangle=\left\langle v^\alpha v_\alpha^*\right\rangle,\label{eq:eigenvalues_modal_coefficients}
\end{equation}
where $n$ is the dimensionality of the $L^2_w\left(\Omega,\mathbb{C}^3\right)$-space. As in \cite{Hodzic2019_part1} the data is analyzed in stretched spherical coordinates (SSC), $z^{i}$, which are related to the Cartesian coordinates, $z^{i'}$, by
\begin{eqnarray}
x(\xi,\theta,\phi) &=& Ce^{\xi}\cos\theta+x_0,\label{eq:x_SSC}\\
y(\xi,\theta,\phi) &=& Ce^{\xi}\sin\theta\cos\phi,\label{eq:y_SSC}\\
z(\xi,\theta,\phi) &=& Ce^{\xi}\sin\theta\sin\phi\label{eq:z_SSC}.
\end{eqnarray}
The covariant- and contravariant metric tensors are defined as
\begin{equation}
z_{ij} = \begin{Bmatrix}
\left(Ce^{\xi}\right)^{2} & 0 & 0\\
0 & \left(Ce^{\xi}\right)^{2} & 0\\
0 & 0 & \left(Ce^{\xi}\sin\theta\right)^{2}
\end{Bmatrix},\label{eq:covariant_metric_tensor_SSC}
\end{equation}
and
\begin{equation}
z^{ij} = \begin{Bmatrix}
\left(Ce^{\xi}\right)^{-2} & 0 & 0\\
0 & \left(Ce^{\xi}\right)^{-2} & 0\\
0 & 0 & \left(Ce^{\xi}\sin\theta\right)^{-2}
\end{Bmatrix},\label{eq:contravariant_metric_tensor_SSC}
\end{equation}
where the volume element is evaluated to be $\sqrt{Z}=\left(Ce^\xi\right)^3\sin\theta$.

In order to investigate how modes obtain their energy, and if this process can be described by the Richardson cascade model, the governing equations are expanded by the eigenfunctions. For an constant density flow field the continuity equation is defined as
\begin{equation}
\nabla_i V^i = 0,
\end{equation}
and the Navier-Stokes equations for a fluid with constant material properties in curvilinear coordinates are
\begin{equation}
\frac{\partial V^i}{\partial t}+V^j\nabla_jV^i=-\frac{1}{\rho}\nabla^ip+\nu\nabla^j\nabla_jV^i.
\end{equation}
The ensemble averaged turbulence kinetic energy equation is then defined as, \cite{Hodzic2019_part1}
\begin{eqnarray}
\underbrace{\frac{D K_t}{D t}}_{I}+\underbrace{\left\langle v_iv^j\right\rangle\nabla_j\left\langle V^i\right\rangle}_{II}+\underbrace{\frac{1}{2}\nabla_j\left\langle v_iv^iv^j\right\rangle}_{III} &=& -\underbrace{\frac{1}{\rho}\nabla_i\left\langle v^ip\right\rangle}_{IV}+\label{eq:turb_energy_terms}\\
&+&\underbrace{\nabla_j\left\langle v_i\tau^{ij}\right\rangle}_{V}-\underbrace{\left\langle \tau^{ij}\nabla_jv_i\right\rangle}_{VI},\nonumber
\end{eqnarray}
where $D/Dt$ and $\tau^{ij}$ are the material derivative and the second-order contravariant stress-tensor. 

With the aim of gaining a deeper understanding of the energy transport in the turbulent jet far-field, the velocity field is spanned by the modes themselves. In the current work, the procedure is to then express \eqref{eq:turb_energy_terms} for the jet far-field using the LD eigenfunctions, in general coordinates and not least to analyze the production term ($II$ in \eqref{eq:turb_energy_terms}) from the perspective of its modal representation. From this form, it is possible to reconstruct the terms of the energy equation, \eqref{eq:turb_energy_terms}, and analyze both the individual contributions of the modes as well as the cumulative sums of the modes. 

With this purpose in mind, the set of vector valued basis functions is restricted to the weighted space $L_w^2\left(\Omega,\mathbb{C}^3\right)$. From the invariance of the fluctuating part of the velocity we can decompose $\overline{v}$ as
\begin{equation}
\overline{v} = v^j\overline{z}_j=v^\alpha\overline{\Phi}_\alpha,\label{eq:velocity_components_relation}
\end{equation}
where $\overline{\Phi}_\alpha=\varphi^j_\alpha\overline{z}_j$. By suppressing the covariant basis vectors, the relation \eqref{eq:velocity_components_relation} yields
\begin{equation}
v^j=v^\alpha\varphi^j_\alpha.\label{eq:shift_tensor}
\end{equation}
This means that $\varphi^j_\alpha$ plays the role of a shift tensor (see \cite{Grinfeld2013}) and is thus an operator that projects a tensor onto the $n$-dimensional manifold spanned by the eigenfunctions. The term manifold, in this particular case, refers to one of zero curvature everywhere - which is a direct consequence of the linearity of the LD operator. 
\subsection{Eigenfunctions expressed in terms of SADFM}
The contravariant components of the eigenfunctions in terms of the SADFM are given by \cite{Hodzic2019_part1} 
\begin{equation}
\varphi^{j}_\alpha = \psi^j_\alpha e^{i\left(t\omega+\kappa\xi+m\phi-2\xi\right)} = \frac{\widetilde{\psi}^{j}_\alpha e^{i\left(t\omega+\kappa\xi+m\phi\right)-2\xi}}{\sqrt{C^5\Vol\sin\theta}},\label{eq:eigenfunctions}
\end{equation}
where $i=\sqrt{-1}$ and $\Vol = 2\pi TL_\xi$. Note that for a Fourier-based formulation of eigenfunctions, \eqref{eq:eigenfunctions}, the forms \eqref{eq:velocity_components_relation} and \eqref{eq:shift_tensor} should be understood in terms of an integration/sum over the Fourier domain unless a spectral representation of the reconstructed field is sought. The eigenfunctions in \eqref{eq:eigenfunctions} are thus a product of the numerical $\theta$-dependent term, say $\widetilde{\psi}^{j\alpha}/\sqrt{C^5\Vol\sin\theta}$, and the SADFM which are represented by the $\xi$-dependent part in \eqref{eq:eigenfunctions}. 
\subsection{Modal expansions of energy spectra and Reynolds stresses}
The local Reynolds stresses can be expressed by expanding the density normalized Reynolds stress-tensor $\Phi^i_{\cdot j}=\left\langle v^iv_j\right\rangle$ in terms of the eigenfunctions, \eqref{eq:shift_tensor}
\begin{equation}
\Phi^i_{\cdot j} = \lambda\varphi^{i\alpha}\varphi^*_{j\alpha}.\label{eq:shear_stress_galerkin}
\end{equation}
Contracting \eqref{eq:shear_stress_galerkin} yields the measure of TKE at any given point in the domain
\begin{equation}
\Phi^i_{\cdot i}=\lambda\varphi^{i\alpha}\varphi^*_{i\alpha} = \underbrace{\lambda\varphi^{1\alpha}\varphi^*_{1\alpha}}_{\Phi^1_{\cdot 1}}+\underbrace{\lambda\varphi^{2\alpha}\varphi^*_{2\alpha}}_{\Phi^2_{\cdot 2}}+\underbrace{\lambda\varphi^{3\alpha}\varphi^*_{3\alpha}}_{\Phi^3_{\cdot 3}}\, ,\label{eq:TKE_galerkin}
\end{equation}
such that the normal stresses are given by evaluating \eqref{eq:shear_stress_galerkin} for a specific choice of $i=j$, i.e. $\Phi^1_{\cdot 1}$, $\Phi^2_{\cdot 2}$, and $\Phi^3_{\cdot 3}$, for the three coordinate directions. The shear-stresses are obtained from \eqref{eq:shear_stress_galerkin} by setting $i\neq j$. 

For the jet far-field where a Fourier-based decomposition is applied, both \eqref{eq:shear_stress_galerkin} and \eqref{eq:TKE_galerkin} are a function of the streamwise coordinate, $\xi$, $\theta$ as well as the azimuthal coordinate $\phi$. Note that the expressions are also a function of $\omega$, $\kappa$, and $m$ (see \eqref{eq:eigenfunctions}). Therefore \eqref{eq:shear_stress_galerkin} and \eqref{eq:TKE_galerkin} are in fact spectra over  $\omega$, $\kappa$, and $m$ through $\widetilde{\psi}^{i\alpha}$ and $\widetilde{\psi}_{j\alpha}^*$. Substituting \eqref{eq:eigenfunctions} into \eqref{eq:shear_stress_galerkin} yields
\begin{equation}
\Phi^i_{\cdot j} =  \frac{\lambda\widetilde{\psi}^{i\alpha}\widetilde{\psi}^*_{j\alpha}}{C^5\Vol e^{2\xi}\sin\theta},\label{eq:shear_stress_galerkin_fourier}
\end{equation}
where the complex exponentials equal out due to the product of complex conjugate pairs. Integrating out the $\omega$- and $\kappa$-dependencies and summing over $m$ yields the local Reynolds stresses in the flow. Note that \eqref{eq:shear_stress_galerkin_fourier} evolves downstream proportionally to $e^{-2\xi}$, as seen from the substitution of \eqref{eq:eigenfunctions}. The non-dimensionalized version of \eqref{eq:shear_stress_galerkin_fourier} is then, \cite{Hodzic2019_part1}
\begin{equation}
\widetilde{\Phi}^i_{\cdot j} = \widetilde{\lambda}\widetilde{\psi}^{i\alpha}\widetilde{\psi}^*_{j\alpha},\label{eq:shear_stress_galerkin_fourier_non_dim}
\end{equation}
where $\widetilde{\lambda} =\lambda/(CB^2M_0)$, where $B=2B_u/\sqrt{\pi}=6.5$ is the velocity decay rate, \cite{Hussein1994}, where $B_u=5.76$, \cite{Hodzic2019_part1} and $M_0$ is the momentum flux at the nozzle exit. In the current work a number in parenthesis, $(\alpha)$, placed above an object will be used to designate the $\alpha$-th component of the statistics, e.g. the component of \eqref{eq:shear_stress_galerkin_fourier_non_dim} related to the $\alpha$-th eigenvalue is designated by $\widetilde{\Phi}^{i(\alpha)}_{\cdot j}$.
\FloatBarrier
\subsection{Turbulence kinetic energy transport equation\label{sec:TKE_equation}}
Expanding the energy equation in terms of the eigenfunctions by combining \eqref{eq:turb_energy_terms} and \eqref{eq:shift_tensor}, yields the following expressions for the terms in \eqref{eq:turb_energy_terms} 
\begin{subequations}
\begin{align}
I &: \frac{D}{Dt}\frac{1}{2} \left\langle v_\alpha v^{\beta*} \varphi^\alpha_i\varphi^{i*}_\beta\right\rangle,\\
II &:\left\langle v_{\alpha}  v^{\beta*}\varphi^\alpha_i\varphi^{j*}_\beta\right\rangle\nabla_j \left\langle V^i\right\rangle, \\
III &: \frac{1}{2}\nabla_j\left\langle v_\alpha v^{\beta *} v^\gamma\varphi^\alpha_i\varphi^{i*}_\beta\varphi^j_\gamma\right\rangle,\\
IV &: \frac{1}{\rho}\nabla_i\left\langle v^\alpha\varphi^i_\alpha p\right\rangle,\\
V &:\nu\left( \nabla_j\left\langle v^\alpha\varphi_{i\alpha}\nabla^jv^{\beta *}\varphi^{i*}_\beta \right\rangle +\nabla_j\left\langle v^\alpha\varphi_{i\alpha}\nabla^i v^{\beta *}\varphi^{j*}_\beta\right\rangle\right),\\
VI &:\nu\left\langle\nabla^j v^\alpha\varphi^i_\alpha\nabla_j v^{\beta *}\varphi^*_{i\beta}+\nabla^i v^\alpha\varphi^j_\alpha\nabla_j v^{\beta*}\varphi^*_{i\beta}\right\rangle.
\end{align}
\end{subequations}
where $( ^*)$ designates complex conjugation and the terms in \eqref{eq:turb_energy_terms}. Since the eigenfunctions are deterministic and the coefficients are coordinate independent, we can rewrite $I-VI$ in order to obtain the following forms of the ensemble averaged energy equation
\begin{subequations}
\begin{align}
I &: \frac{1}{2}\frac{D}{Dt} \varphi^\alpha_i\varphi^{i*}_\alpha,\\
II &:\lambda \varphi^\alpha_i\varphi^{j*}_\alpha\nabla_j \left\langle V^i\right\rangle\label{eq:galerkin_production_term}, \\
III &: \frac{1}{2}\left\langle v_\alpha v^{\beta *}\label{eq:galerkin_transport_term} v^\gamma\right\rangle\varphi^j_\gamma\nabla_j\varphi^\alpha_i\varphi^{i*}_\beta,\\
IV &: \frac{1}{\rho}\varphi^i_\alpha\left\langle v^\alpha\nabla_i p\right\rangle,\\
V &: \nu\lambda\nabla_j\left(\varphi^\alpha_i\nabla^j\varphi_\alpha^{i*}+ \varphi^\alpha_i\nabla^i\varphi^{j*}_\alpha\right),\\
VI &: \nu\lambda \left( \nabla^j\varphi^{i\alpha}\nabla_j\varphi^{*}_{i\alpha}+\nabla^i\varphi^{j\alpha}\nabla_j\varphi^{*}_{i\alpha}\right)\label{eq:galerkin_dissipation_term}.
\end{align}
\label{eq:galerkin_energy_terms}
\end{subequations}
\noindent
In \eqref{eq:galerkin_energy_terms} a summation over repeated indices is implied, where the eigenvalue, $\lambda$, is to be interpreted as $\lambda^\alpha$ and should therefore be included in the summation (the $\alpha$ was suppressed in order to accommodate the Einstein summation convention, which does not allow three repeated indices).

This form of the TKE transport, \eqref{eq:galerkin_energy_terms}, for a fluid with constant material properties holds in any well-defined coordinate system. It can therefore be applied to any flow field, for which the eigenfunctions belong to the space of square integrable vector-valued functions, $\overline{\Phi}^\alpha \in L^2_w(\Omega,\mathbb{C}^3)$. Note how the tensor notation provides an overview of the component parts (both velocity- and modal components) of the various terms of \eqref{eq:galerkin_energy_terms} without the distractions of introducing a specific coordinate system. The interpretation of the terms is, however, subject to the definition of the $L_w^2(\Omega,\mathbb{C}^3)$-inner product, which is determined by the choice of weight function. For a unit weight the $\lambda^\alpha$ represent the total turbulence kinetic energy of the field related to mode $\alpha$, as opposed to the case of a non-unit weight where \eqref{eq:galerkin_energy_terms} should be viewed as a weighted energy equation. This is analogous to applying a window function in order to reduce spectral leakage when producing energy spectra of a turbulent velocity field in which case the window alters the energy of the field unless the spectra are scaled in order to compensate for this effect.

For a self-adjoint LD operator, the eigenvalues are guaranteed to be non-negative. In this case $\overline{\Phi}^\alpha$ cannot be zero everywhere in the domain for any $\alpha$. For turbulent flows characterized by a self-adjoint LD operator this directly implies either 1) \textit{all modes} away from nodes must exchange energy with the mean flow in the presence of non-zero mean gradients or 2) the energy production is exactly zero for a given $\alpha$ due to the positive and negative contributions of the four terms which $II$ consists of in the jet far-field 
\begin{eqnarray}
\mathcal{P}&=& \lambda\Big(\varphi_1^\alpha\varphi^{1*}_\alpha\nabla_1\left\langle V^1\right\rangle + \varphi_2^\alpha\varphi^{2*}_\alpha\nabla_2\left\langle V^2\right\rangle\nonumber\\
&+& \varphi_1^\alpha\varphi^{2*}_\alpha\nabla_2\left\langle V^1\right\rangle + \varphi_2^\alpha\varphi^{1*}_\alpha\nabla_1\left\langle V^2\right\rangle\Big). \label{eq:PROD_streamwise}
\end{eqnarray}
The degree to which this occurs in different areas of the domain is determined by the corresponding eigenvalue-eigenfunction combination. However, in flows with non-zero mean gradients - and as the results of this work suggest - the direct energy transport from the mean field to the modes is closer to being the norm rather than the exception. This supports the hypothesis in \cite{Wanstrom2009} and \cite{Wanstrom2006} stating that multiple modes are able to obtain their energy directly from the mean flow. 
\\
\\
Term $III$ governs the non-linear energy transport caused by velocity fluctuations. We see that the term regulates the energy fluxes within- and across modes and this energy exchange is not directly dependent on the mean TKE, $\lambda$, but is instead dependent on the instantaneous TKE with a $v^\gamma$ modulation, where the latter works as a Lagrangian multiplier together with $\varphi^j_\gamma$. The process is governed by the modal interactions between modes $\alpha$, $\beta$, and $\gamma$, which in the case of Fourier modes is traditionally coined as \textit{triadic interactions}. Since all modes must have non-zero coefficients (as they have non-zero eigenvalues), it means that the coefficients in \eqref{eq:galerkin_production_term} determine the intensity of the modal energy transport across the jet. The key, however, to understanding the spatial distribution of the energy transport can be found in the eigenfunctions. We can immediately infer that the energy transfer is zero at all nodes of $\varphi^j_\gamma$. We see that there is no energy transfer when the spatial derivatives of $\varphi^\alpha_i\varphi^{i*}_\beta$ are zero. We also observe that the energy transfer is maximized in flow regions where the local extrema of $\varphi^j_\gamma$ coincide with the maximum gradient of $\varphi^\alpha_i\varphi^{i*}_\beta$. For $\alpha=\beta=\gamma$ expression \eqref{eq:galerkin_production_term} defines the non-linear energy transfer within a given mode - or in other words the extent to which the individual modes redistribute their energy over the domain (space and time). 

If $\alpha$, $\beta$, and $\gamma$ in \eqref{eq:galerkin_transport_term} correspond to the most energetic modes, we would expect the eigenfunctions to exhibit negligible spatial gradients. This is because the distance between the local extrema would be relatively large i.e. the number of local extrema would be relatively small over the span of the flow, which would infer smaller gradients. For low-energy eigenfunctions, where the number of local extrema is generally high over the span of the flow we see from \eqref{eq:galerkin_transport_term} that the non-linear energy transfer would be significant. On average, energy transfer would be greatest between modes with similar energy-levels, a phenomenon that is most commmonly seen at high mode numbers (corresponding to low energies). Since $\varphi^\alpha_i$ are continuously varying over the domain, it shows that the process of energy transport is very dynamic indeed, and for most turbulent flows it is continuously varying. It is especially interesting to classify the extent of energy exchanged between adjacent modes - especially those which are closely spaced in terms of energy in $\kappa$-regions where the energy spectra exhibit the $-5/3$-slope which traditionally is related to a so-called inertial subrange.
\\
\\
Terms $V$ and $VI$ are the modal representations of the viscous transport and dissipation. Both terms are dependent on nonzero modal gradients in order to be present in the flow. Furthermore, since both terms are proportional to $\lambda$ and are finite we note that the suprema of $V$ and $VI$ must necessarily be bounded by the opposing characteristics of the eigenvalues (representing the turbulence kinetic energy) and modal gradients; large eigenvalues must be balanced by small modal gradients in order to ensure a balanced energy dissipation. We also note that since the eigenvalues are non-zero dissipation also remains non-zero for all modes, unless $ \nabla^j\varphi^{i\alpha}\nabla_j\varphi^{*}_{i\alpha}=-\nabla^i\varphi^{j\alpha}\nabla_j\varphi^{*}_{i\alpha}$ everywhere in the domain for a given $\alpha$. 

Since the dissipation takes place predominantly at high mode numbers then unlike the energy production, it will necessarily be distributed more or less homogeneously across the entire domain, due to the increasing number of local extrema related to high mode numbers. In fact, high mode number eigenfunctions have non-negligible complex parts, \cite{Hodzic2019_part1}. The dissipation of a given mode may therefore be related to the imaginary part of the modes, and this characteristic may potentially be used as an indicator for the level of dissipation a given mode produces. If there is a direct relation between the magnitude of the imaginary part and the level of dissipation a given mode produces, it may have a profound effect on efficient modeling of dissipation in general turbulence flows. Moreover, it could be used to determine how forcing - such as accelerations - affect the dissipation in turbulent flows and which modes are responsible for these effects.
\FloatBarrier
\subsection{Reconstruction of the Velocity field\label{sec:reconstruction}}
In \cite{Wanstrom2009} it was suggested that all modes of the jet are able to obtain a significant part of their energy directly from the mean flow, thereby circumventing a Richardson-like energy cascade from low- to high mode numbers. 
\begin{figure}[h]
    \centering      
    \subfloat[]{\includegraphics[width=0.30\linewidth]{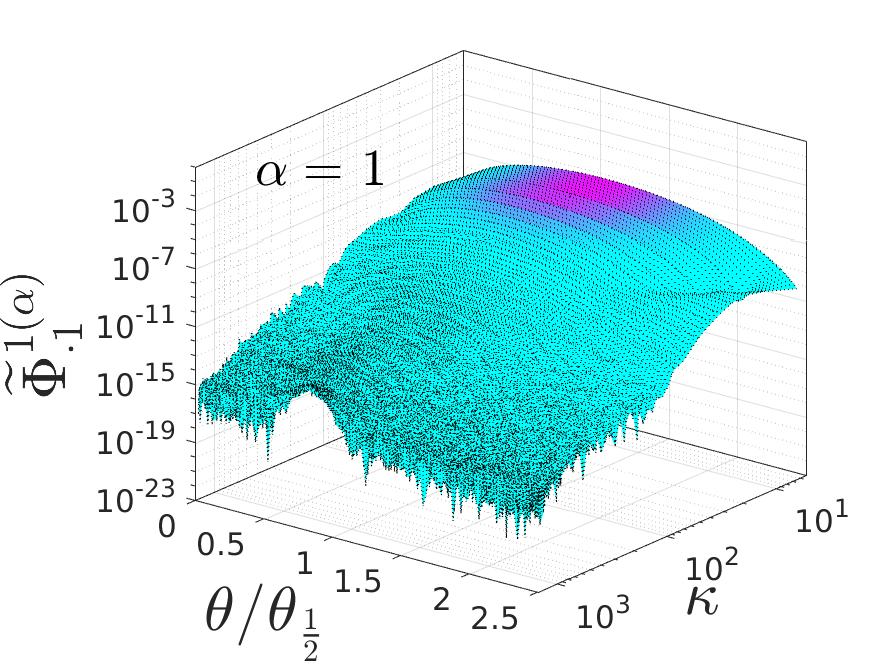}\label{fig:surf_single_spectra_uu_1}}
     \subfloat[]{\includegraphics[width=0.30\linewidth]{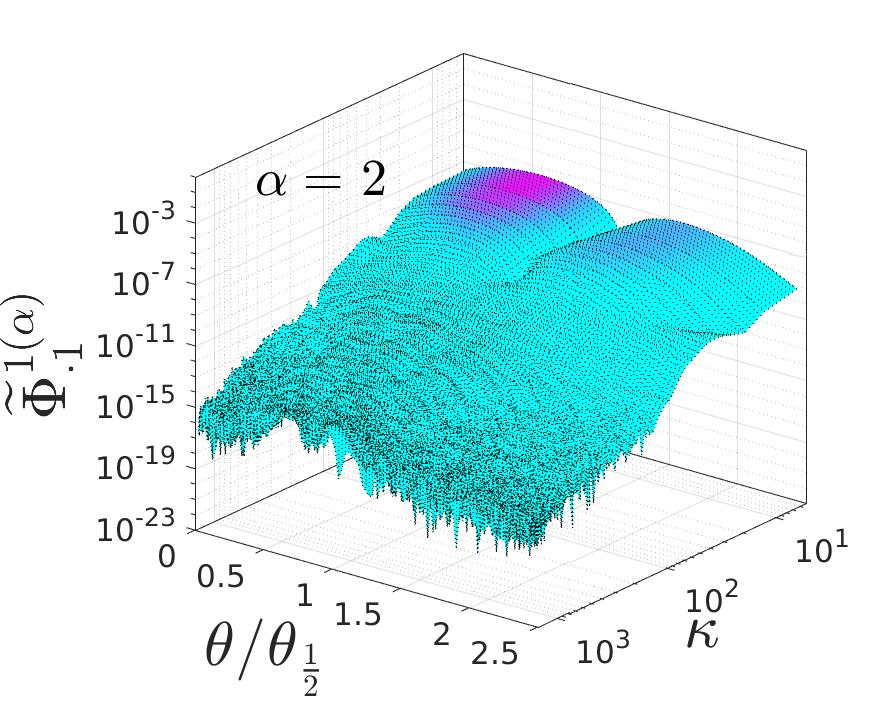}\label{fig:surf_single_spectra_uu_2}}
         \subfloat[]{\includegraphics[width=0.30\linewidth]{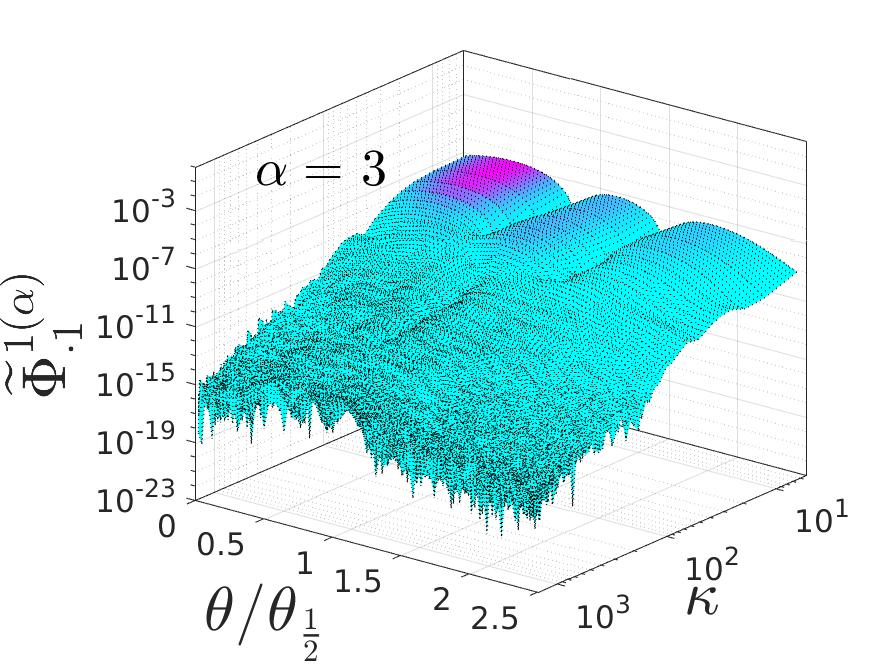}\label{fig:surf_single_spectra_uu_3}}\\
    \subfloat[]{\includegraphics[width=0.30\linewidth]{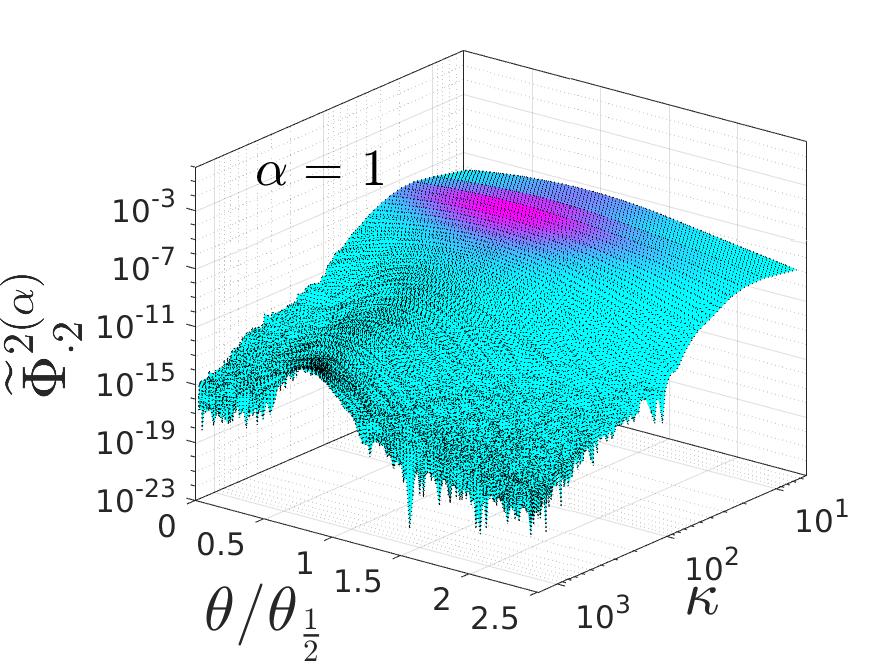}\label{fig:surf_single_spectra_vv_1}}
     \subfloat[]{\includegraphics[width=0.30\linewidth]{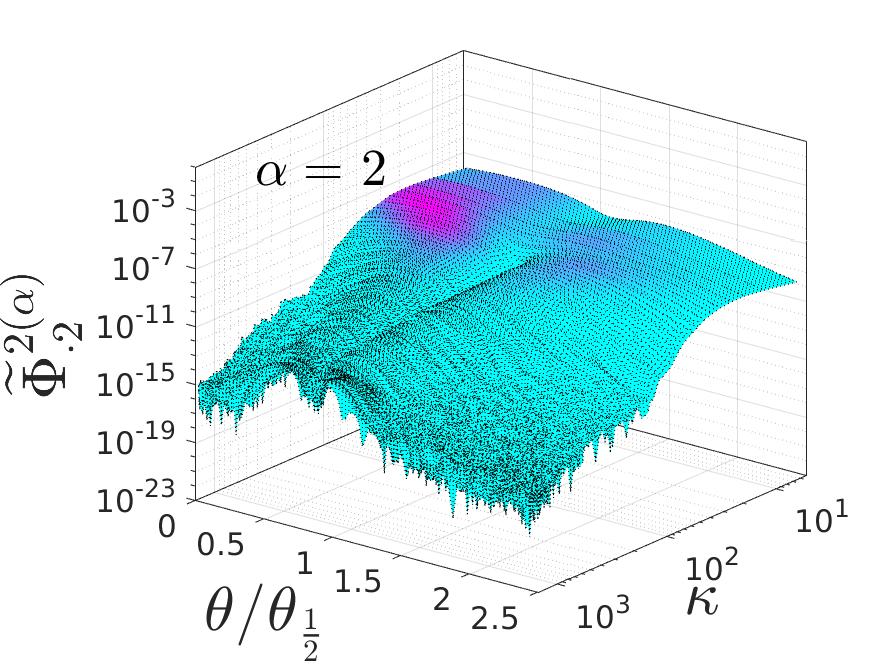}\label{fig:surf_single_spectra_vv_2}}
         \subfloat[]{\includegraphics[width=0.30\linewidth]{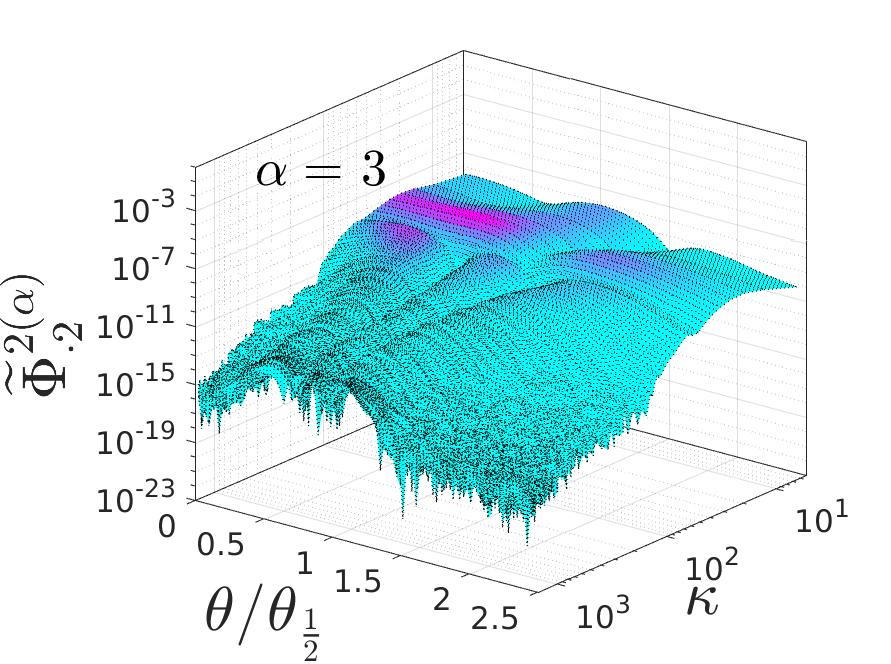}\label{fig:surf_single_spectra_vv_3}}\\
    \subfloat[]{\includegraphics[width=0.30\linewidth]{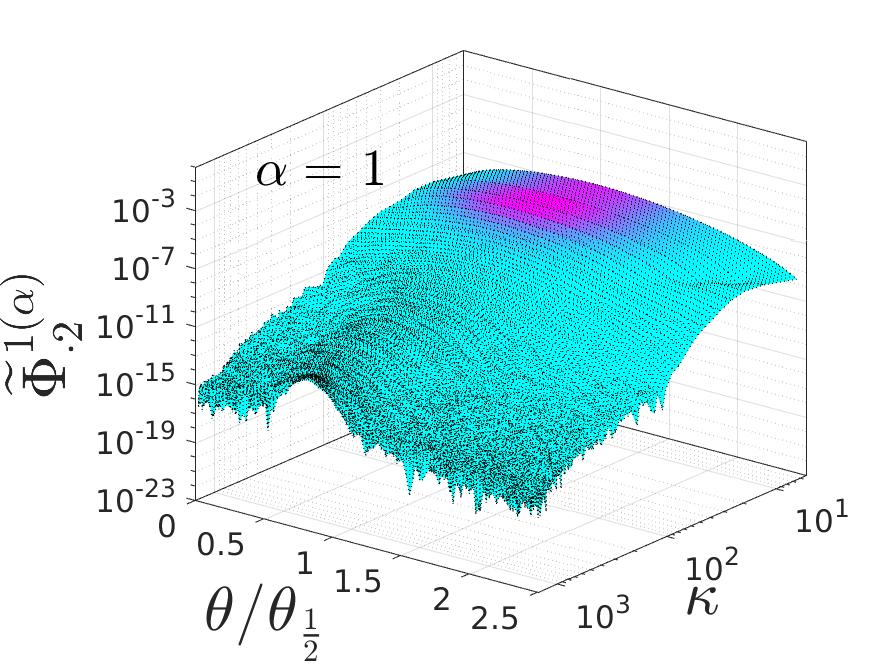}\label{fig:surf_single_spectra_uv_1}}
     \subfloat[]{\includegraphics[width=0.30\linewidth]{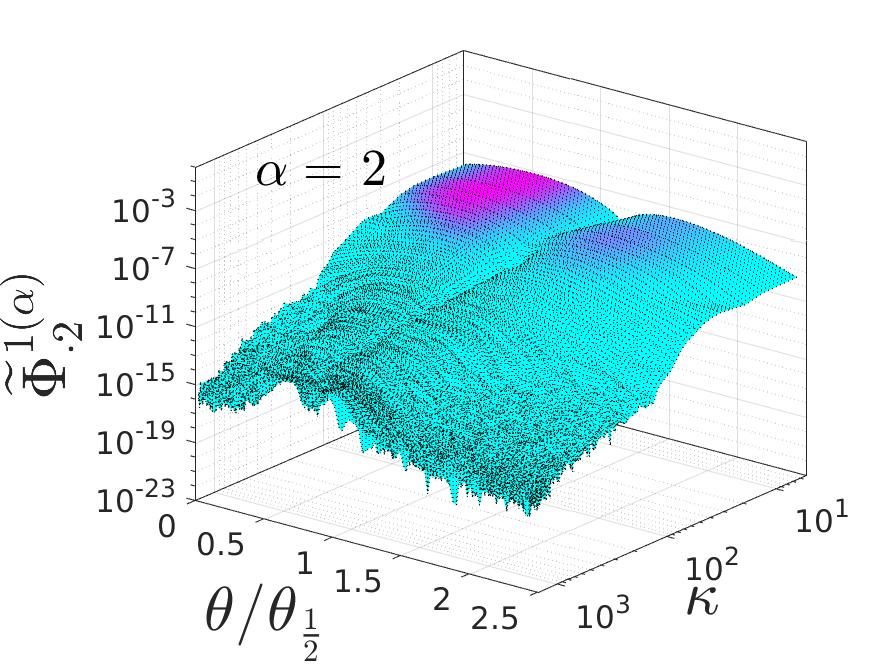}\label{fig:surf_single_spectra_uv_2}}
         \subfloat[]{\includegraphics[width=0.30\linewidth]{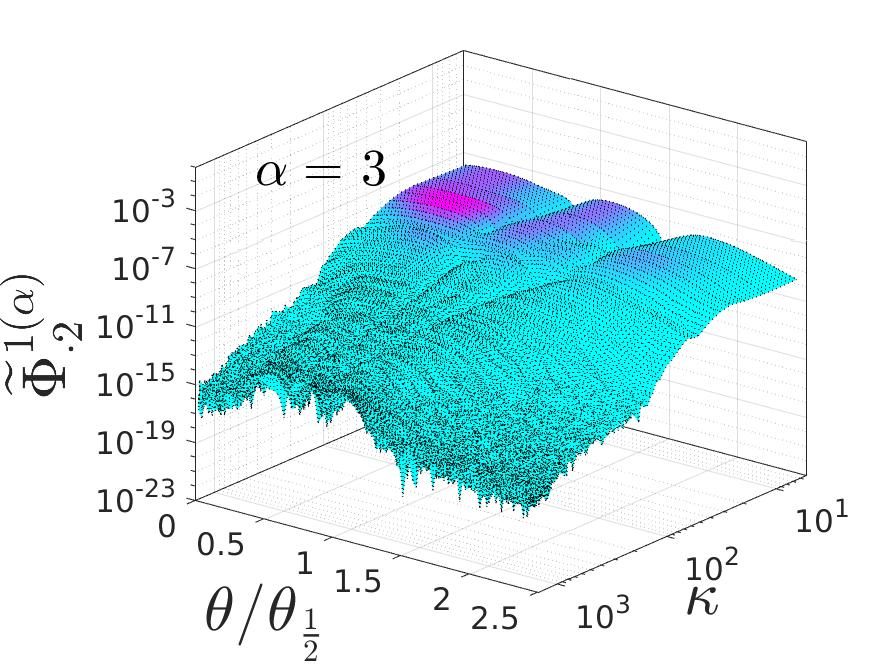}\label{fig:surf_single_spectra_uv_3}}                     
\caption{Modal contributions to the reconstruction of the the normalized spectra, $\widetilde{\Phi}^{i(\alpha)}_{\cdot,j}$, of modes $n=[1:3]$. \label{fig:spectra_uu_reconstructed_SSC}}
\end{figure}
%
\noindent
From the current measurements we can examine directly the modal contributions to the reconstruction of the energy production term, \eqref{eq:galerkin_production_term}, as well as the variations of these contributions across the jet. 

Note that while the mean turbulence kinetic energy is dominated by the streamwise Reynolds-stresses the energy production term is dominated by the shear-stresses. This characterization of the principal modal contributions to the energy production will provide information regarding the effect of the mean shear gradient - which varies greatly across the jet. 

Also, we can examine the modal contributions to the component energy density spectra and cross-spectra. The building blocks required to reconstruct the characteristic $-5/3$-range across the span of the flow will be provided in order to investigate which modes are dominant participants in the energy transfer between scales. Finally, in order to gain a broader understanding of the self-similarity hypothesis posed by \cite{Lumley1967} in terms of its modal building blocks, the $-7/3$-range of the cross-spectrum will be reconstructed. This discussion will lead to the notion of modal self-similarity.

The individual contributions of modes $1-3$ to the spectra $\widetilde{\Phi}^{1}_{\cdot,1}$, $\widetilde{\Phi}^{2}_{\cdot,2}$, and $\widetilde{\Phi}^{1}_{\cdot,2}$ are shown in figure \ref{fig:spectra_uu_reconstructed_SSC}. The subfigures herein give an overview of the individual contributions of the first three modes for all wavenumbers and illustrate the contribution of the first mode to the spectra. The role of the $\theta$-component of the modes as well as the contribution of this component to the energy production will be demonstrated more clearly in the paragraphs to follow, but it is clear from the contour lines in figures \ref{fig:surf_single_spectra_uu_1} and \ref{fig:surf_single_spectra_vv_1}, that unlike $\widetilde{\Phi}^{1(1)}_{\cdot,1}$ the peak of $\widetilde{\Phi}^{2(1)}_{\cdot,2}$ occurs at higher wavenumbers in the production range. As noted earlier in \cite{Wanstrom2009}, this is consistent with what is observed in homogeneous isotropic turbulence, \cite{Tennekes1972}, where it is a direct consequence of incompressibility. It is probably significant that it is observed in this turbulence as well. 
%
%
%
%
\begin{figure}[h]
    \centering      
\subfloat[]{\includegraphics[width=0.30\linewidth]{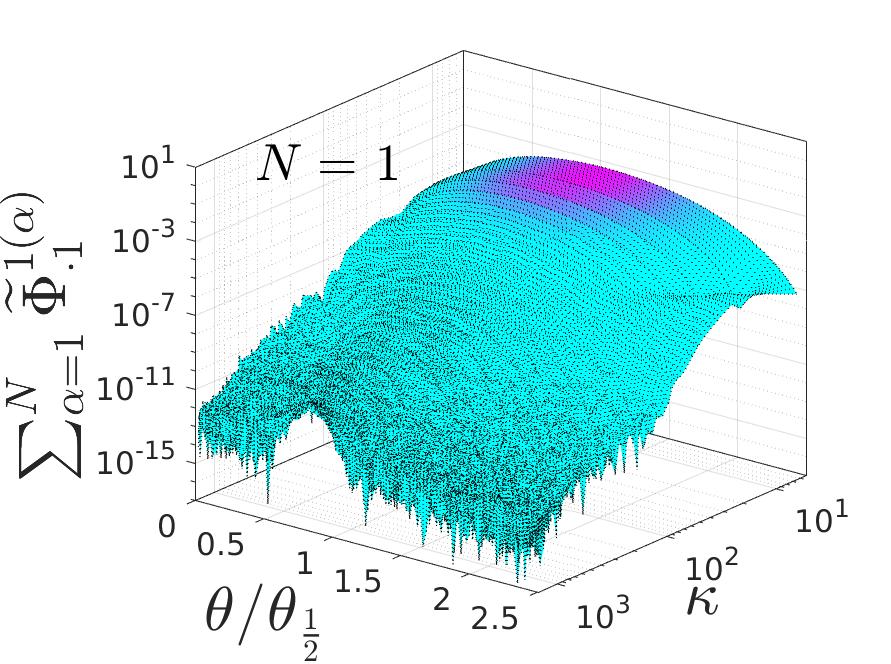}\label{fig:surf_cumsum_spectra_uu_1}}
\subfloat[]{\includegraphics[width=0.30\linewidth]{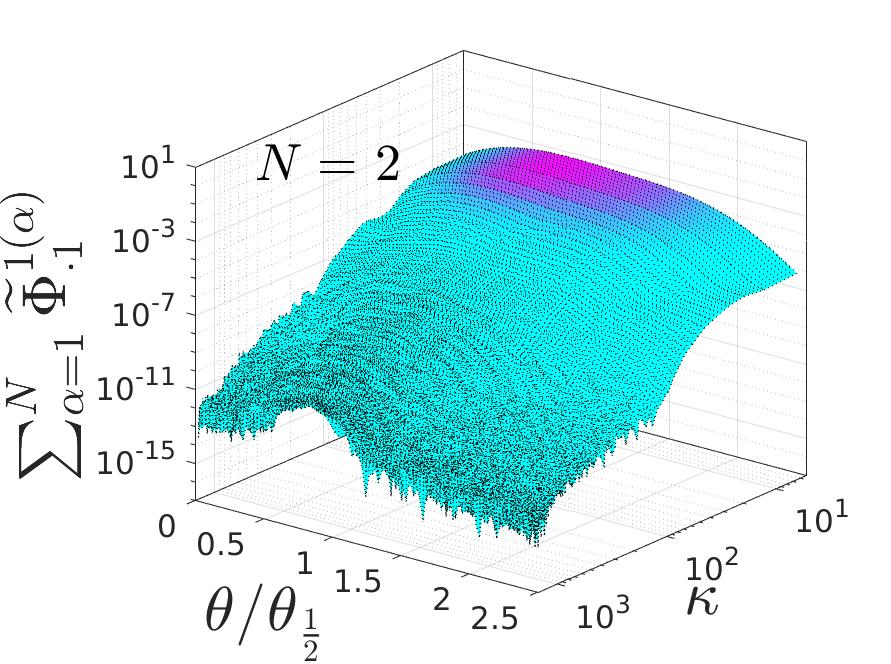}\label{fig:surf_cumsum_spectra_uu_2}}
\subfloat[]{\includegraphics[width=0.30\linewidth]{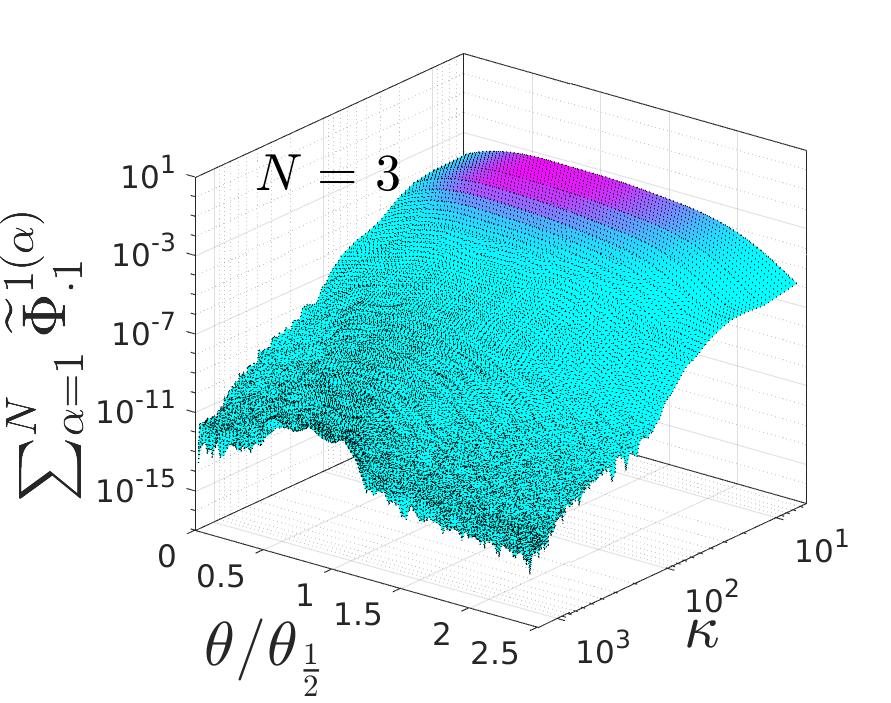}\label{fig:surf_cumsum_spectra_uu_3}}\\
\subfloat[]{\includegraphics[width=0.30\linewidth]{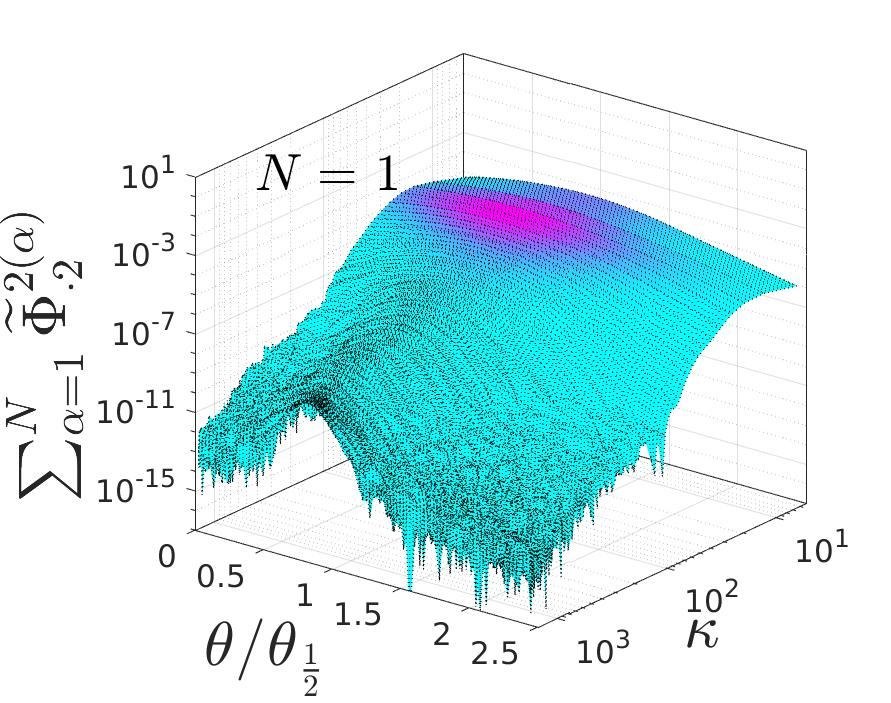}\label{fig:surf_cumsum_spectra_vv_1}}
\subfloat[]{\includegraphics[width=0.30\linewidth]{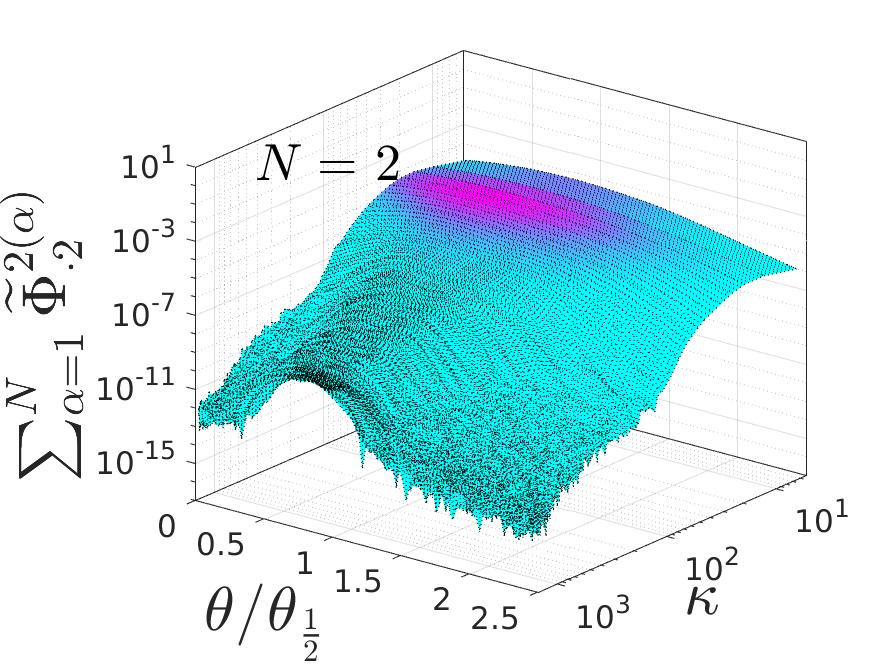}\label{fig:surf_cumsum_spectra_vv_2}}
\subfloat[]{\includegraphics[width=0.30\linewidth]{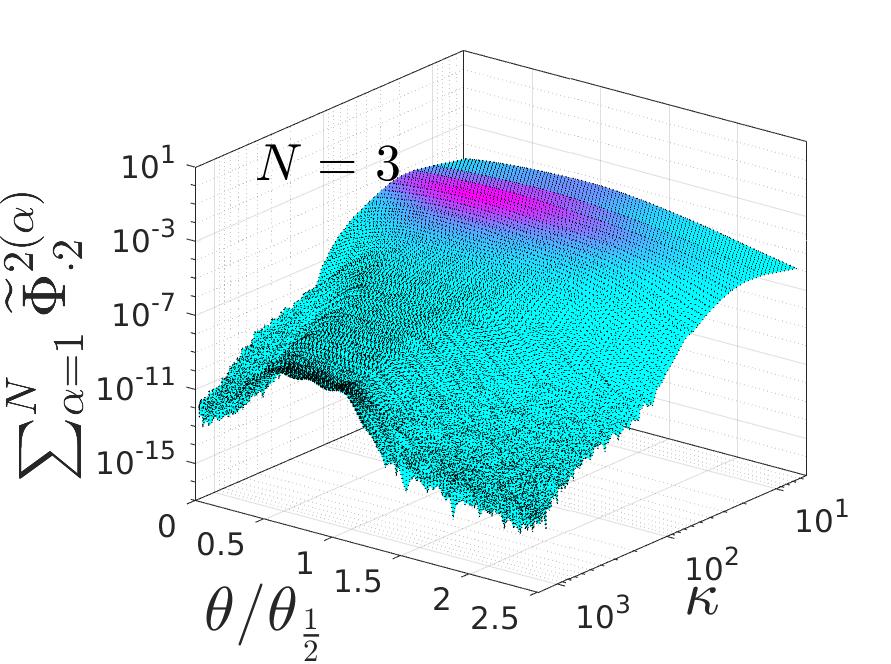}\label{fig:surf_cumsum_spectra_vv_3}}\\
\subfloat[]{\includegraphics[width=0.30\linewidth]{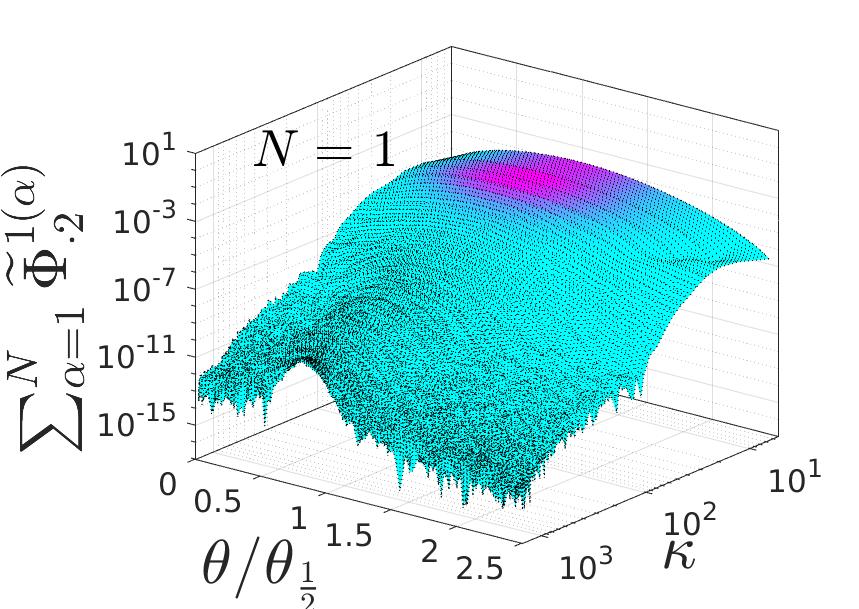}\label{fig:surf_cumsum_spectra_uv_1}}
\subfloat[]{\includegraphics[width=0.30\linewidth]{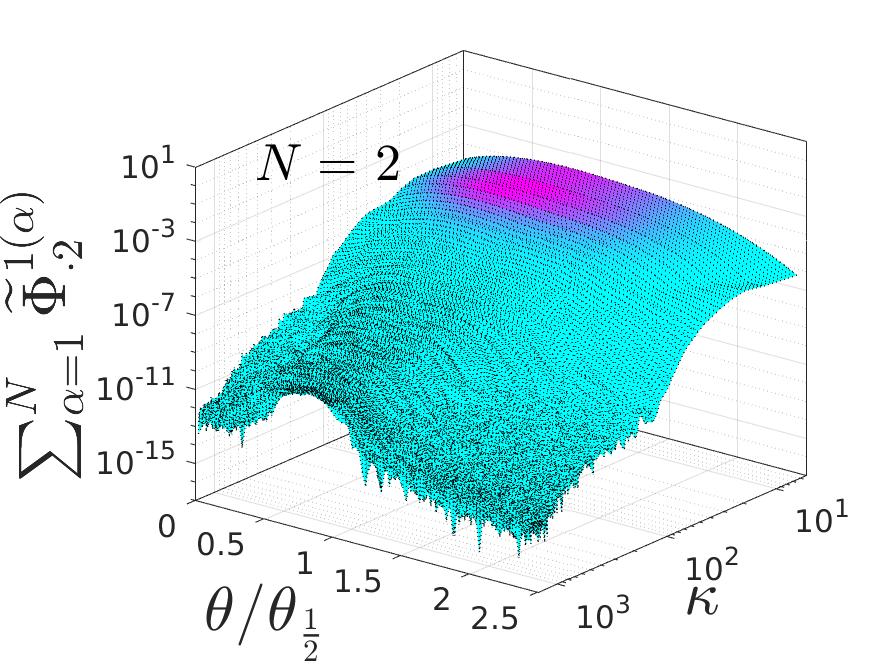}\label{fig:surf_cumsum_spectra_uv_2}}
\subfloat[]{\includegraphics[width=0.30\linewidth]{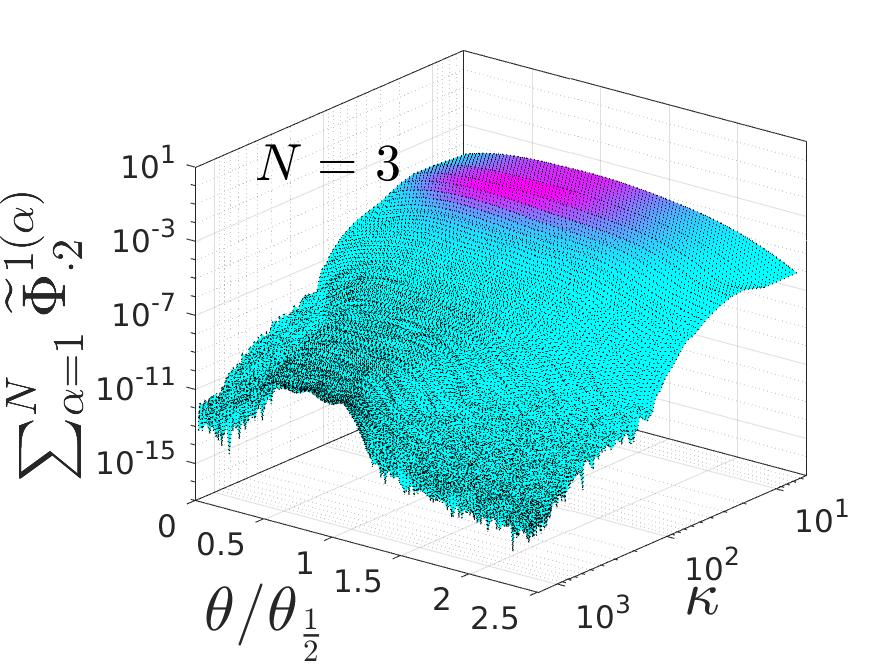}\label{fig:surf_cumsum_spectra_uv_3}}
\caption{Modal cumulative contributions to the reconstruction of the normalized cumulative sums, $\sum_{\alpha=1}^{N}\widetilde{\Phi}^{i(\alpha)}_{\cdot,j}$, of the cumulative sums $N=[1,2,3]$. \label{fig:spectra_cumulative_uu_reconstructed_SSC}}
\end{figure}
%

Comparing the spectrum peaks of figures \ref{fig:surf_single_spectra_uu_1} and \ref{fig:surf_single_spectra_vv_1} indicates that significant energy production through mean shear can only occur from the wavenumber range where significant contributions from both $\widetilde{\Phi}^{1(1)}_{\cdot,1}$ and $\widetilde{\Phi}^{2(1)}_{\cdot,2}$ overlap. Therefore, the contribution of the first mode to the reconstruction of the cross-spectrum, $\widetilde{\Phi}^{1(1)}_{\cdot,2}$, in figure \ref{fig:surf_single_spectra_uv_1} shows that the maximum energy production is related to slightly higher wavenumbers than for the peak in $\widetilde{\Phi}^{1(1)}_{\cdot,1}$. This means that the most significant energy production from normal- and shear stresses occurs at different wavenumbers. However, aliasing of high wavenumber energy into lower wavenumbers of one-dimensional spectra introduces uncertainties in this regard. 

Upon closer inspection - and in stark contrast to the low-wavenumber region - it is seen in figure \ref{fig:spectra_uu_reconstructed_SSC} that each modal contribution to the reconstruction of the spectra retains its topology in the inertial subrange. This should be interpreted as \textit{modal self-similarity} across wavenumbers, and can be considered as statistical support of the notion of a time averaged scale invariance across the jet-width. It can be considered an extension of the idea proposed by \cite{Lumley1967} for constant shear flows when modeling the cross-spectral similarity in the inertial subrange. The current results show that the building blocks of these spectra admit to this similarity as well. The modal self-similarity of higher modes in fact seems independent of the large velocity gradient variations over the span of the flow. Symmetry of the turbulence is suggested by the similarity between the respective modal contributions to $\widetilde{\Phi}^{1(\alpha)}_{\cdot,1}$ and $\widetilde{\Phi}^{2(\alpha)}_{\cdot,2}$ in figure \ref{fig:spectra_uu_reconstructed_SSC} in the inertial subrange. 

The cumulative sums of the modal contributions to the reconstruction of energy density- and cross-spectra are shown in figure \ref{fig:spectra_cumulative_uu_reconstructed_SSC}. It is characteristic that various modes are rebuilding the spectra initially from $\theta/\theta_0=1$. It is seen that these modes are very efficiently constructing a \textit{flat} topology around the region of one half-width, which is related to it being the most energetic region in the flow. It is also evident that the high wavenumber regions very close to the centerline as well as at the edge of the jet are very slowly reconstructed - which will become more evident later, when the Reynolds stresses are reconstructed.

\begin{figure}[h]
    \centering      
    \subfloat[]{\includegraphics[width=0.40\linewidth]{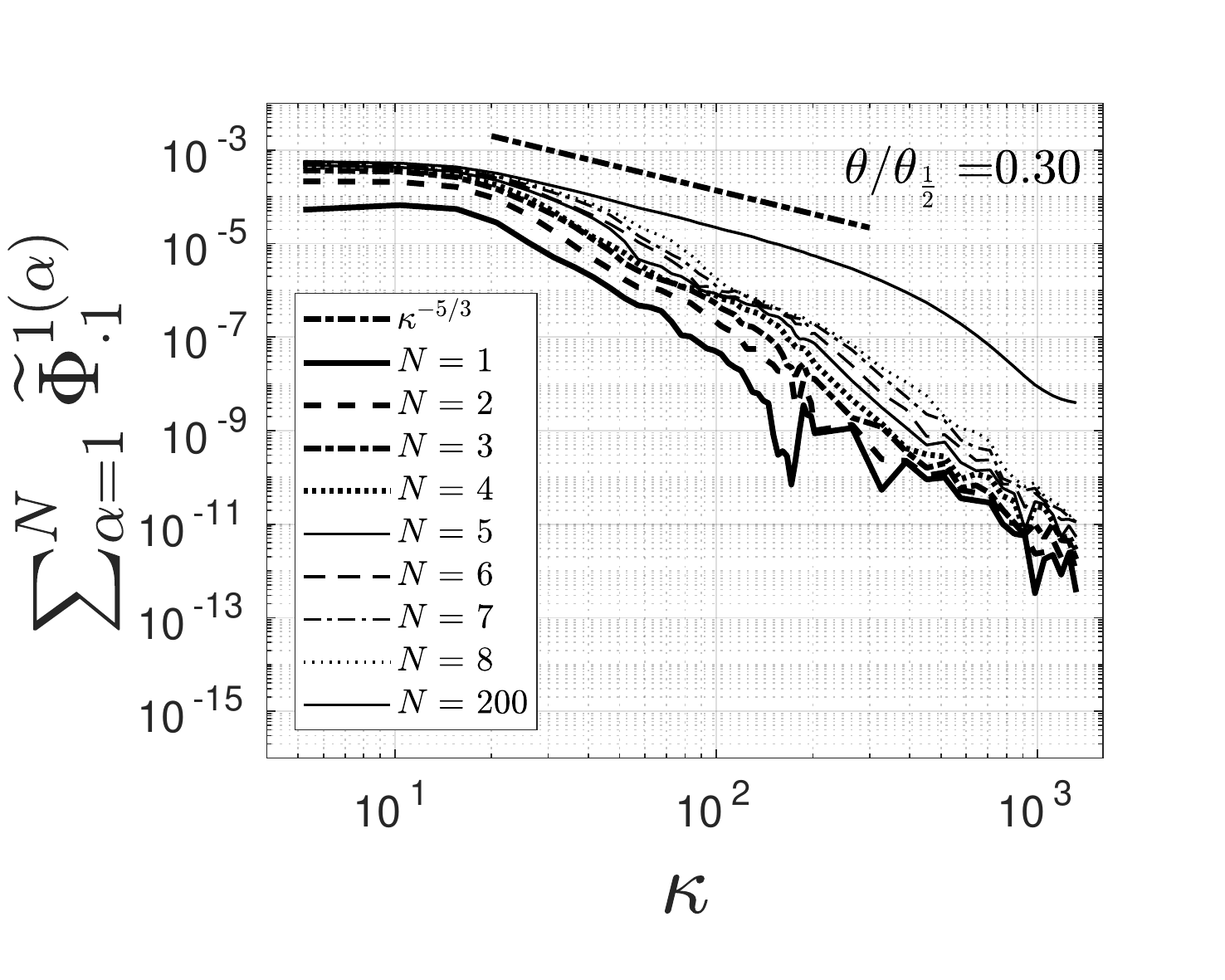}\label{fig:spectra_uu_30}}
    \subfloat[]{\includegraphics[width=0.40\linewidth]{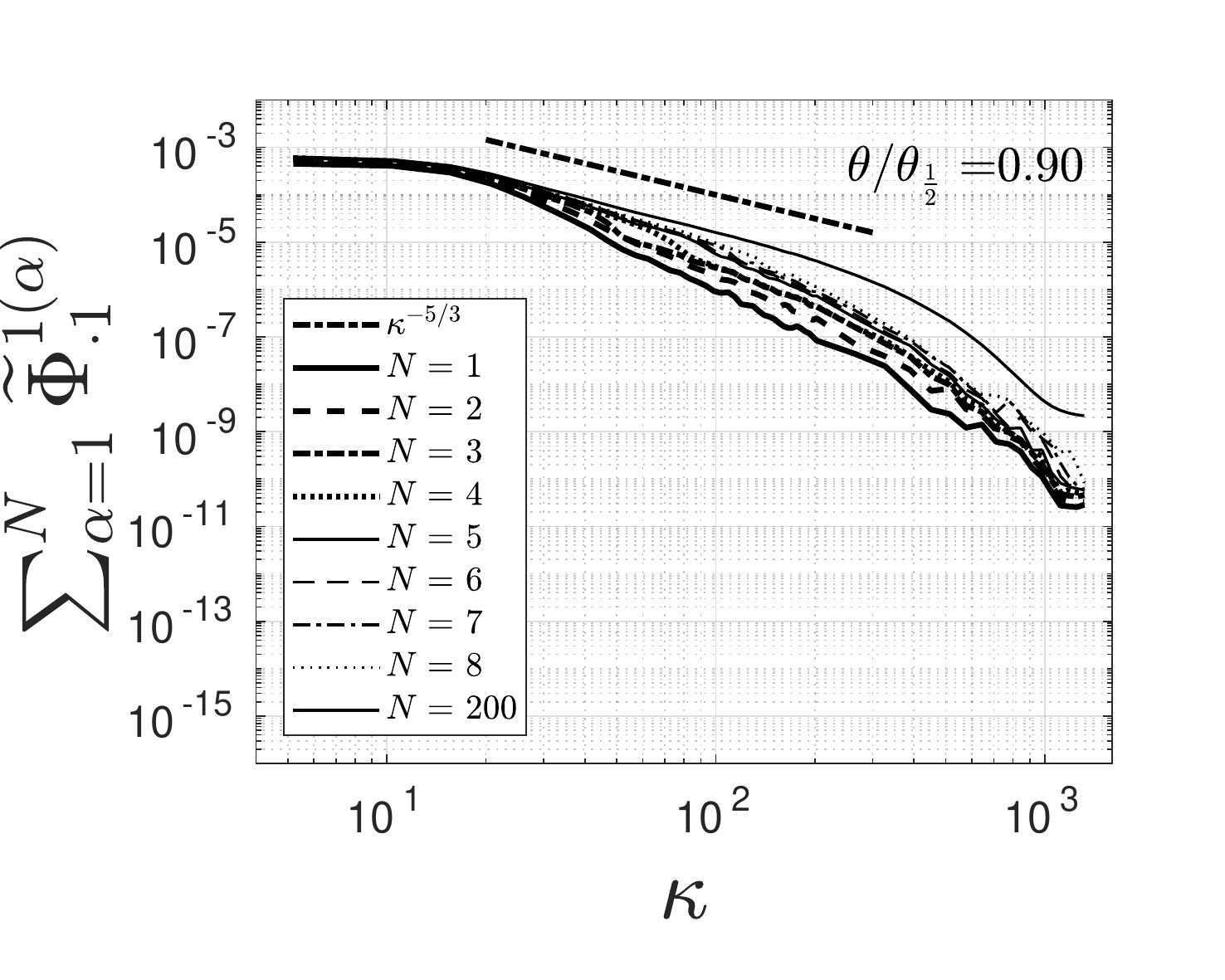}\label{fig:spectra_uu_90}}\\
    \subfloat[]{\includegraphics[width=0.40\linewidth]{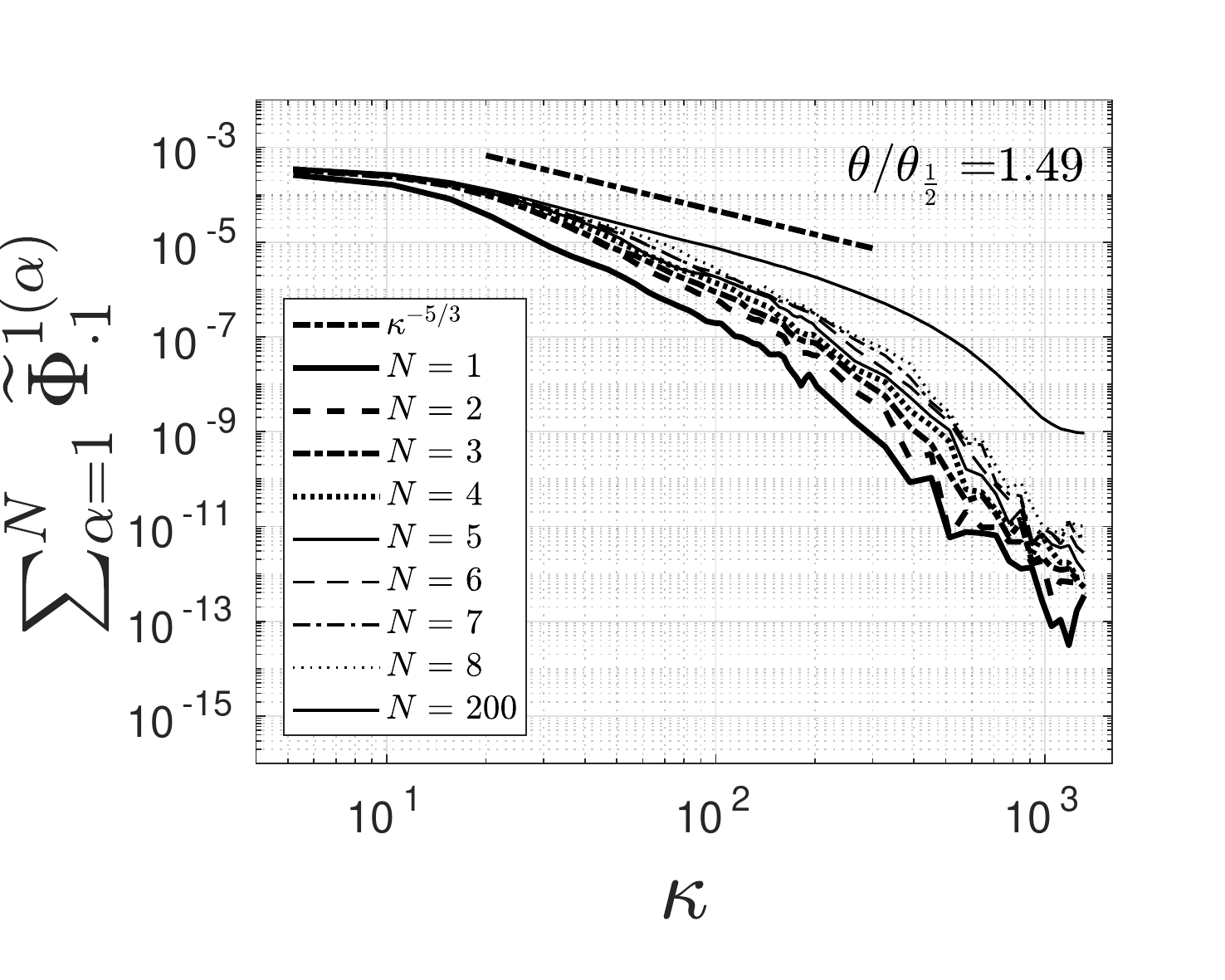}\label{fig:spectra_uu_149}}
    \subfloat[]{\includegraphics[width=0.40\linewidth]{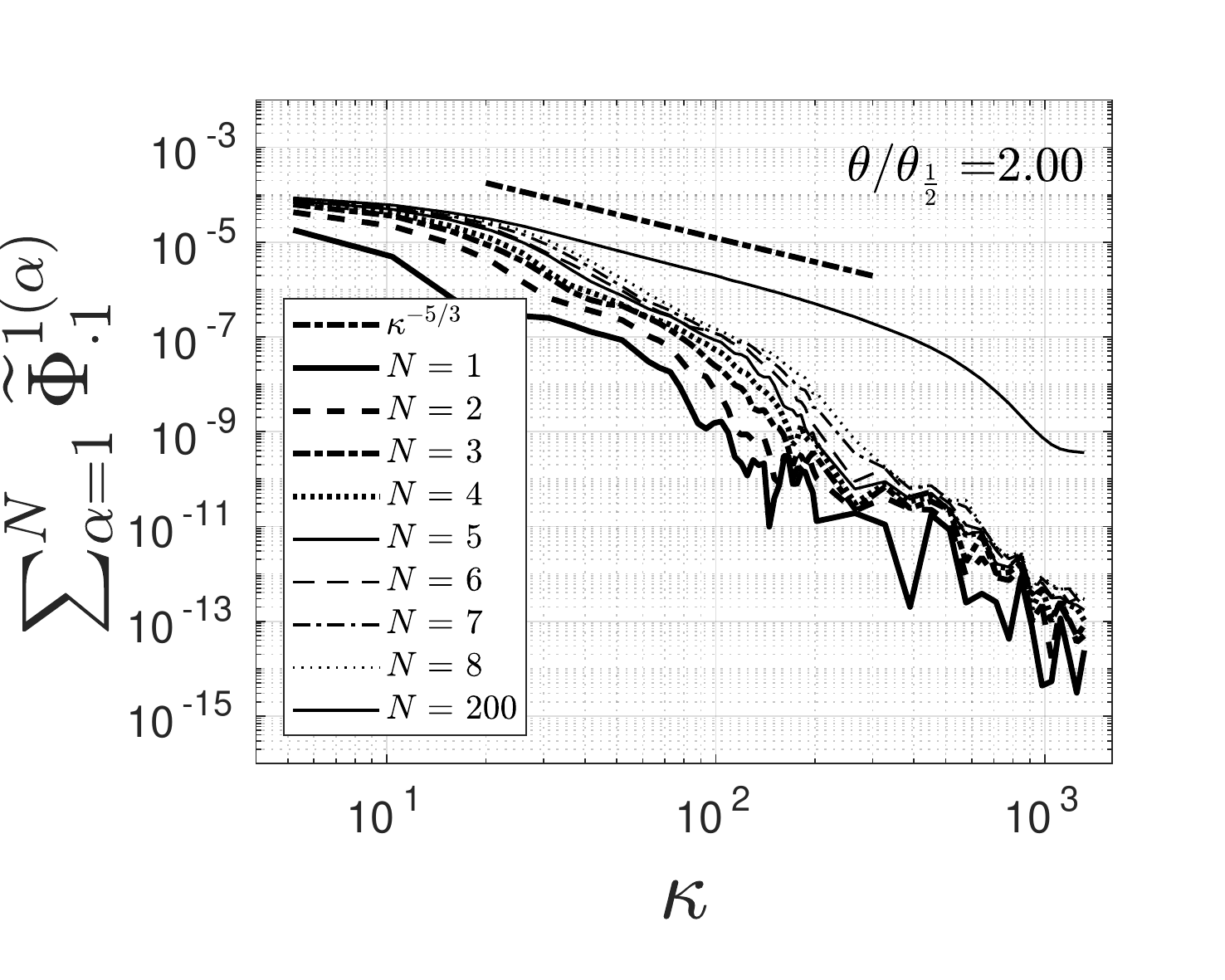}\label{fig:spectra_uu_200}}
\caption{Cumulative modal components of single-point spatial spectra, $\sum_{\alpha=1}^N\widetilde{\Phi}^{1(\alpha)}_{\cdot,1}$, at various spanwise coordinates, $\theta/\theta_{\frac{1}{2}}=[0.30,0.90,1.49,2.00]$. \label{fig:cumulative_spectra_uu_reconstructed_SSC}}
\end{figure}

\begin{figure}[h]
    \centering      
    \subfloat[]{\includegraphics[width=0.40\linewidth]{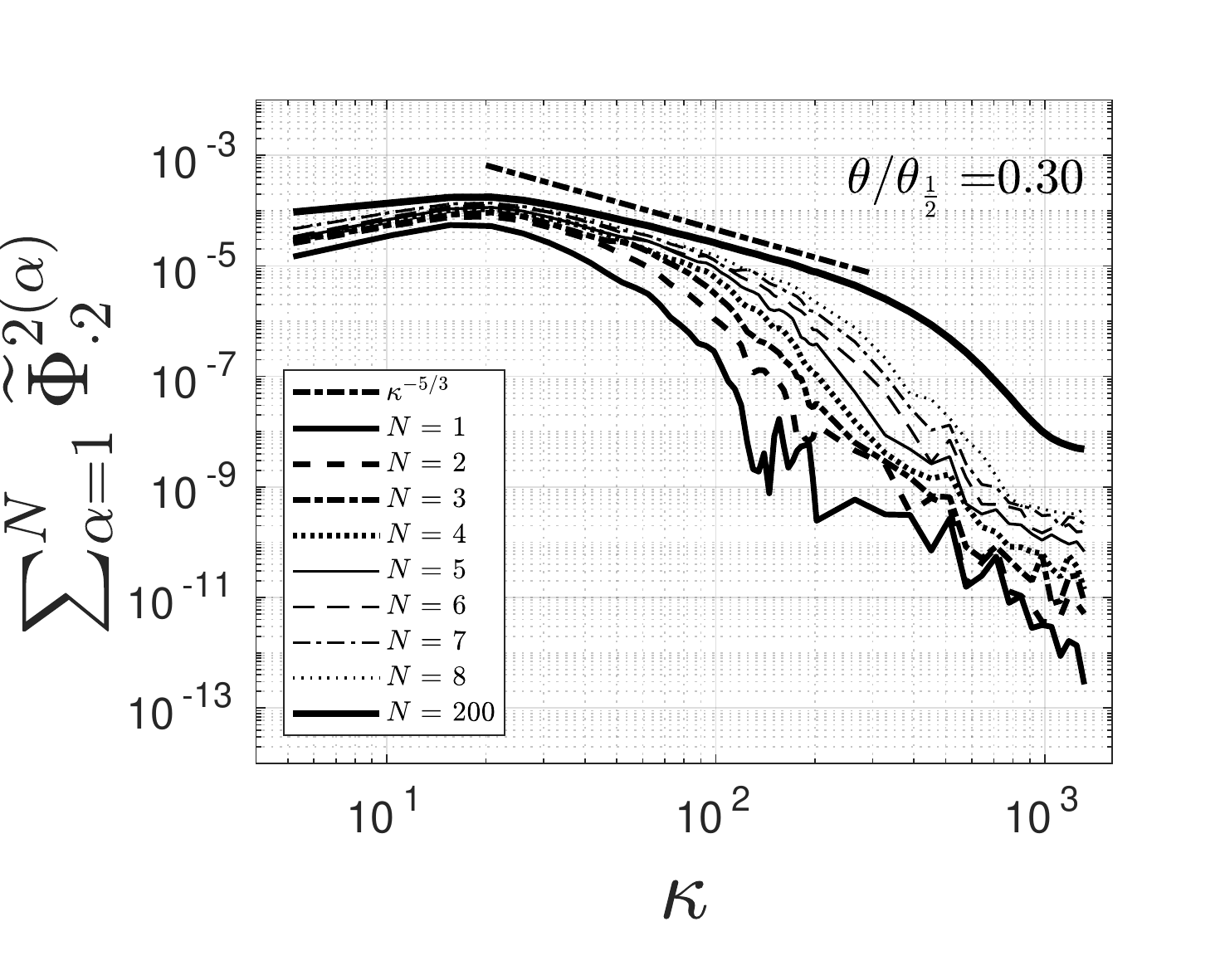}\label{fig:spectra_vv_30}}
    \subfloat[]{\includegraphics[width=0.40\linewidth]{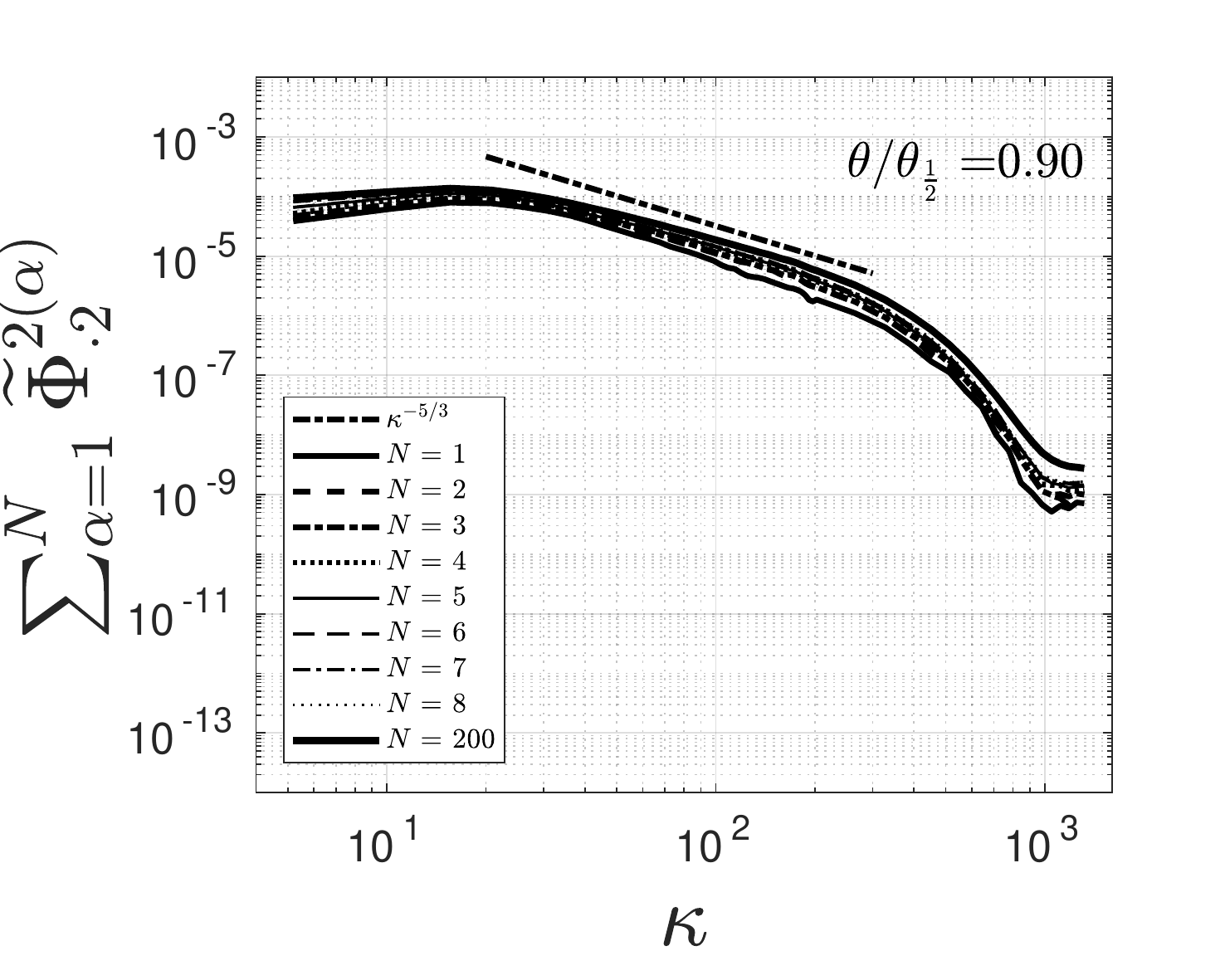}\label{fig:spectra_vv_90}}\\
    \subfloat[]{\includegraphics[width=0.40\linewidth]{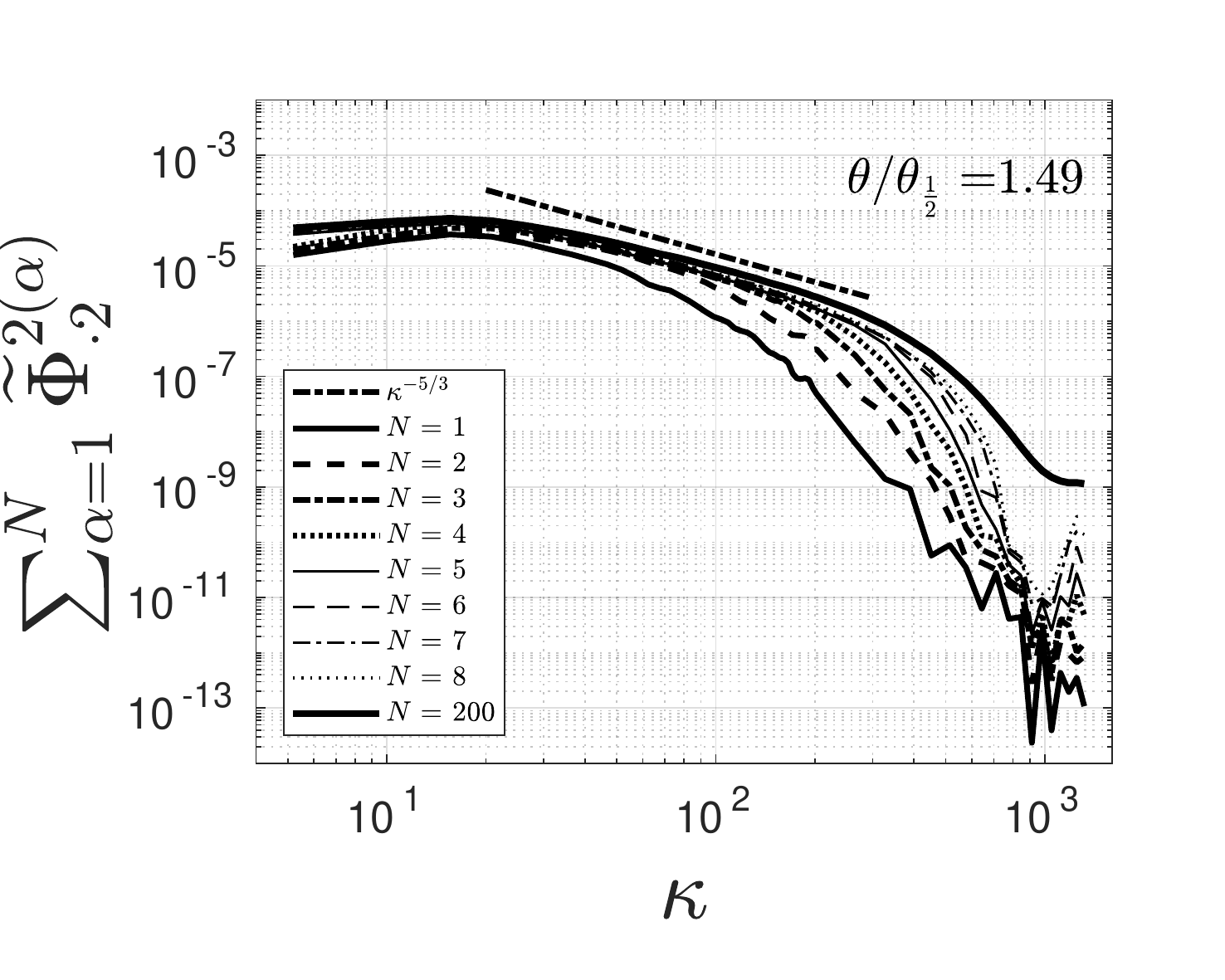}\label{fig:spectra_vv_149}}
    \subfloat[]{\includegraphics[width=0.40\linewidth]{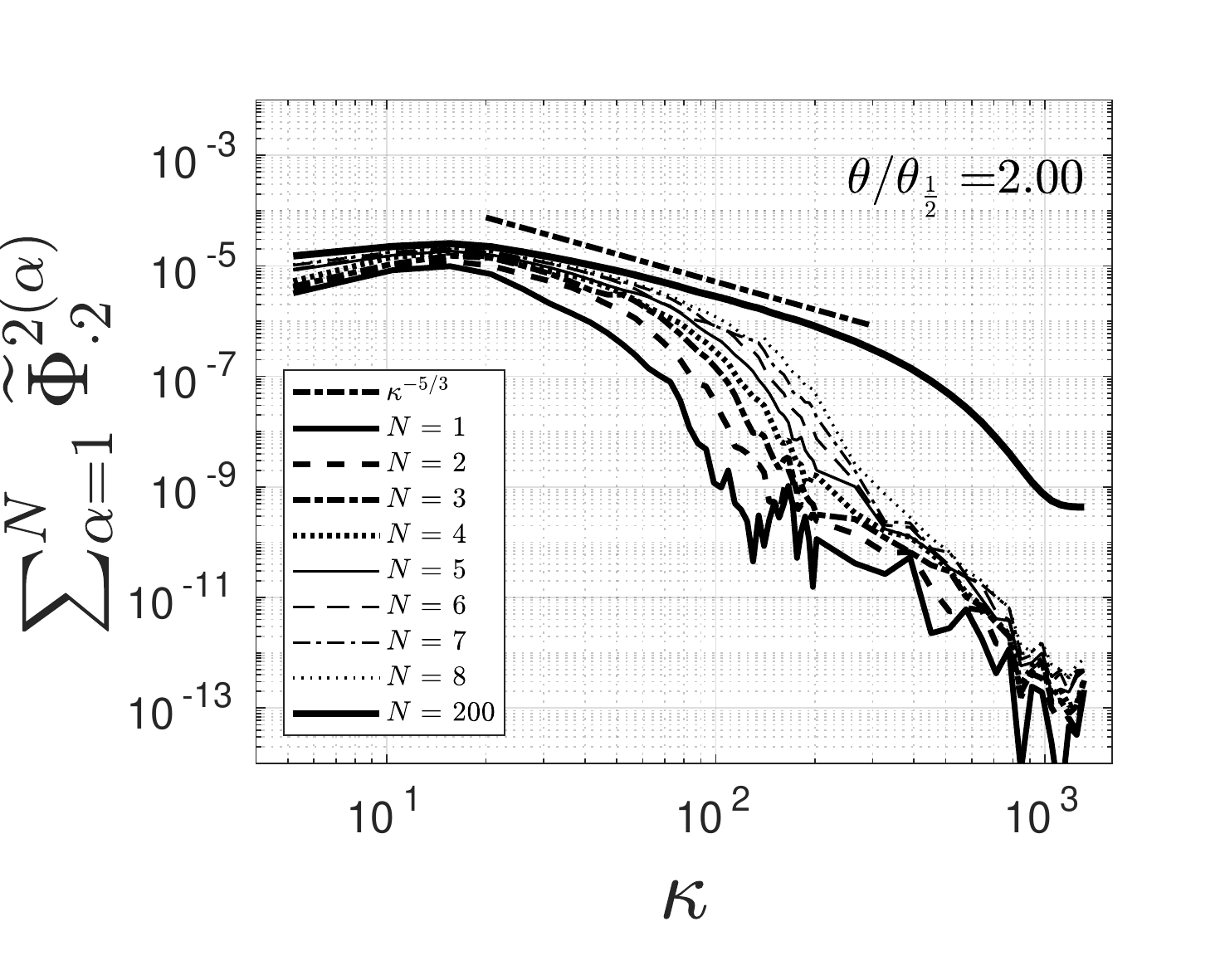}\label{fig:spectra_vv_200}}
\caption{Cumulative modal components of single-point spatial spectra, $\sum_{\alpha=1}^N\widetilde{\Phi}^{2(\alpha)}_{\cdot,2}$, at various spanwise coordinates, $\theta/\theta_{\frac{1}{2}}=[0.30,0.90,1.49,2.00]$. \label{fig:cumulative_spectra_vv_reconstructed_SSC}}
\end{figure}

\begin{figure}[h]
    \centering      
    \subfloat[]{\includegraphics[width=0.40\linewidth]{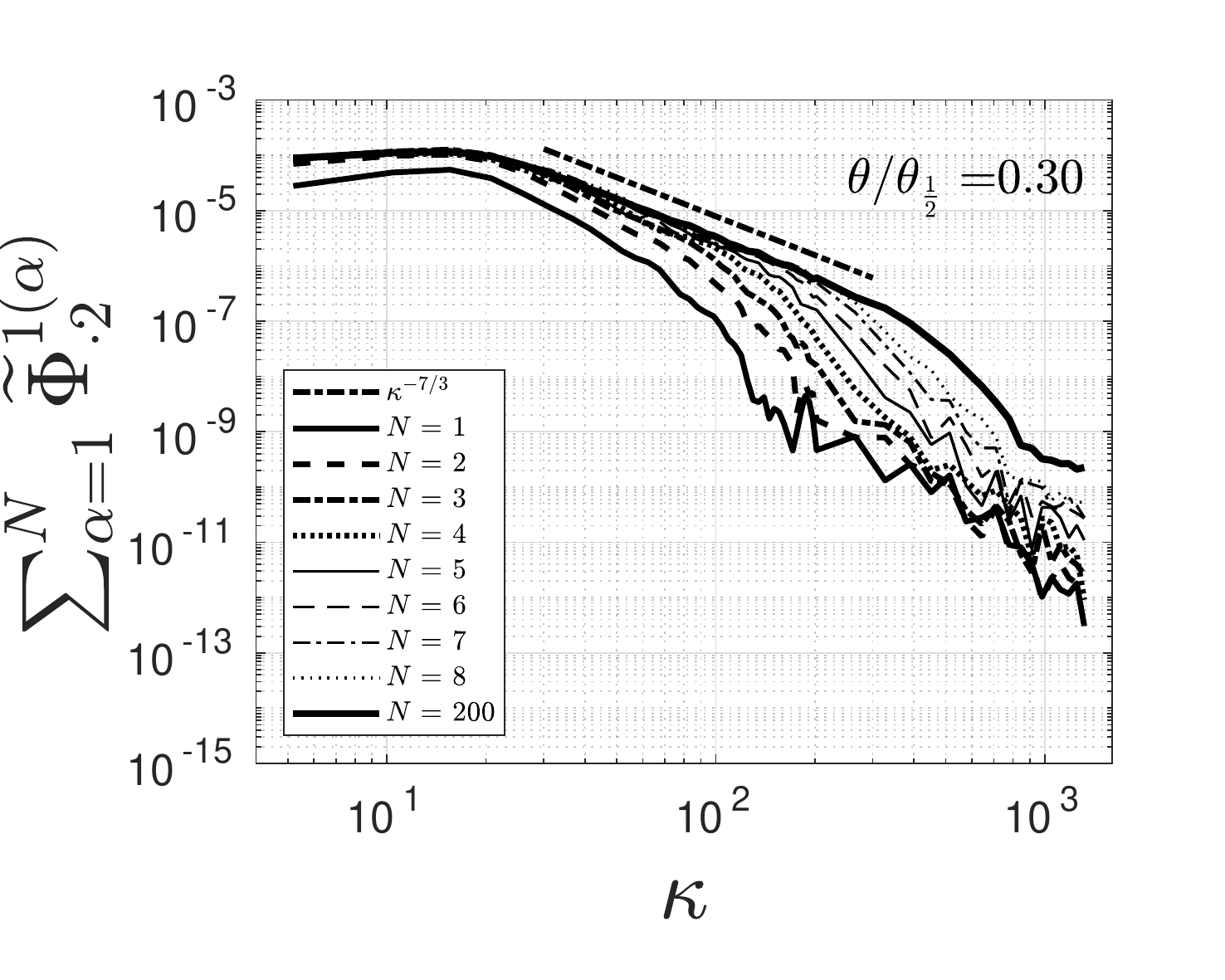}\label{fig:spectra_uv_30}}
    \subfloat[]{\includegraphics[width=0.40\linewidth]{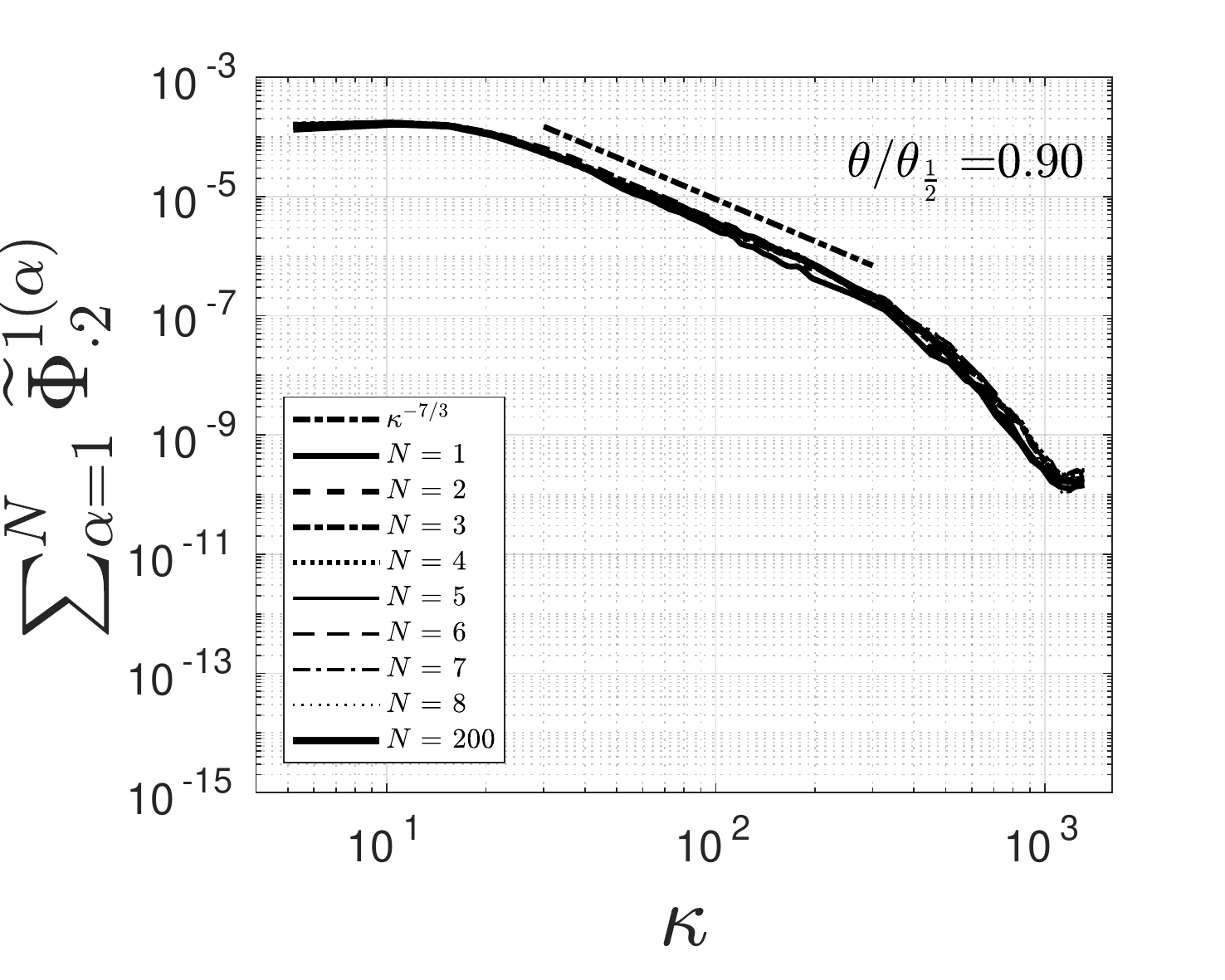}\label{fig:spectra_uv_90}}\\
    \subfloat[]{\includegraphics[width=0.40\linewidth]{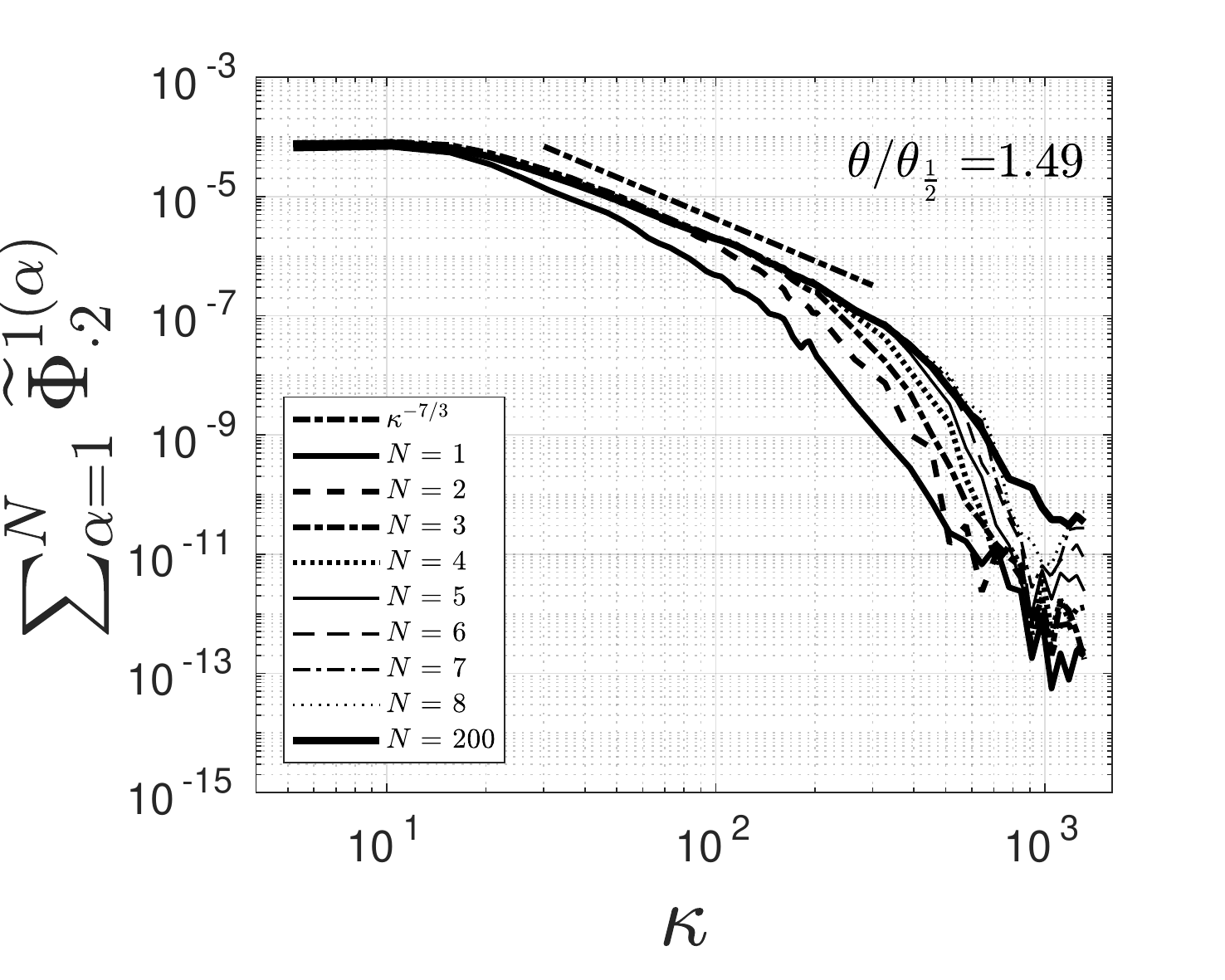}\label{fig:spectra_uv_149}}
    \subfloat[]{\includegraphics[width=0.40\linewidth]{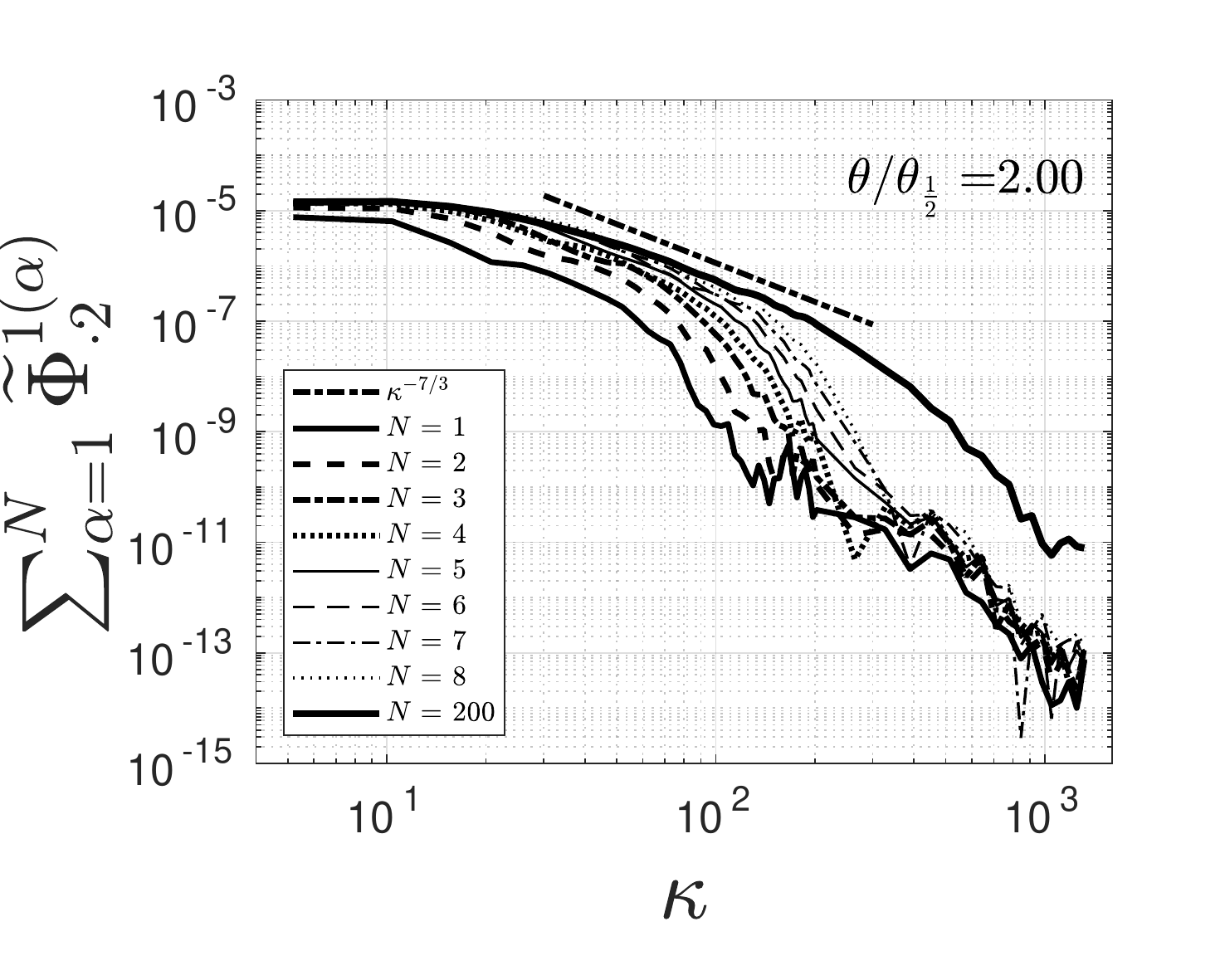}\label{fig:spectra_uv_200}}
\caption{Cumulative modal components of single-point spatial spectra, $\sum_{\alpha=1}^N\widetilde{\Phi}^{1(\alpha)}_{\cdot,2}$, at various spanwise coordinates, $\theta/\theta_{\frac{1}{2}}=[0.30,0.90,1.49,2.00]$. \label{fig:cumulative_spectra_uv_reconstructed_SSC}}
\end{figure}

Figures \ref{fig:cumulative_spectra_uu_reconstructed_SSC}-\ref{fig:cumulative_spectra_uv_reconstructed_SSC} show the cumulative modal contributions in the reconstruction of $\widetilde{\Phi}^{i}_{\cdot,j}$ for various $\theta$-position across the jet. These figures are relevant in order to analyze the reconstruction of the $-5/3$- and $-7/3$-slope ranges at various $\theta$-coordinates. Specifically it is interesting to investigate whether individual modal building blocks reconstruct these slopes, or whether it is a combination of modes that is required to achieve the reconstruction. From the fully reconstructed energy spectra in figures \ref{fig:cumulative_spectra_uu_reconstructed_SSC} and \ref{fig:cumulative_spectra_vv_reconstructed_SSC} it is seen that the $-5/3$-range is already manifested from $\theta/\theta_{\frac{1}{2}}=0.07$ and stretches across the entire width of the jet. The reconstruction at the centerline $\theta/\theta_{\frac{1}{2}}=0$ has not been included due to the singularity of the SSC at this location. However, the first eight individual modal contributions in figure \ref{fig:app_single_spectra_uu_reconstructed_SSC} show that none of these reflect the $-5/3$-slope. The cumulative sum of the of modal contributions to the reconstruction of the $\xi$-component spectrum is seen in figure \ref{fig:cumulative_spectra_uu_reconstructed_SSC} and shows the first eight modal contributions to $\widetilde{\Phi}^{1}_{\cdot,1}$, as well as the fully reconstructed spectrum. It is seen that the $-5/3$-slope is reconstructed only when contributions of higher modes are added. This means that the $-5/3$-slope in $\widetilde{\Phi}^{1}_{\cdot,1}$ is defined by low-energy modes, each of which are topologically different. 

%
\noindent
%
%

%
\noindent
Now the case is different for the reconstruction of the $\widetilde{\Phi}^{2}_{\cdot,2}$-spectrum. The cumulative modal contributions to $\widetilde{\Phi}^{2}_{\cdot,2}$ are shown in figure \ref{fig:cumulative_spectra_vv_reconstructed_SSC}. As one moves from the centerline towards the $\theta/\theta_{\frac{1}{2}}$-region, the number of modes contributing to the reconstruction of the $-5/3$-range is diminishing. Upon closer inspection it is seen that both modes $1$ and $2$ are required to reconstruct the $-5/3$-slope at $\theta/\theta_{\frac{1}{2}}=0.67$, whereas the $-5/3$-range is reconstructed entirely by the the first mode around $\theta/\theta_{\frac{1}{2}}=1$. 
%
%
\noindent
This is seen from the collapse  of the cumulative contributions to the reconstruction of $\widetilde{\Phi}^{2}_{\cdot,2}$ around $\theta/\theta_{\frac{1}{2}}=1$ which is shown in figure \ref{fig:spectra_vv_90}. The pattern is reversed as we move further towards the jet boundary, where multiple modes are again observed to contribute to the reconstruction of the $-5/3$-slope of the spectrum. From the role of the $\theta$-velocity component in the energy transport, it is clear that the $\theta$-components of the modes are optimized in terms of the energy transport. The reconstruction of the $-5/3$-region in the $\widetilde{\Phi}^{2}_{\cdot,2}$-spectrum using the first mode only is a direct consequence of this.  

Observing the reconstruction of the cross-spectra, $\widetilde{\Phi}^1_{\cdot,2}$, in figure \ref{fig:cumulative_spectra_uv_reconstructed_SSC}, similar traits to the reconstruction of $\widetilde{\Phi}^{2}_{\cdot,2}$ are seen. Note, however, that the modes reconstruct the $-7/3$-slope around $\theta/\theta_{\frac{1}{2}}=1$ with the first mode alone. This result is quite astonishing demonstrating that the shear-stress production is defined almost completely by the first LD mode. It will in fact later be shown that unlike the Reynolds normal stresses, the shear-stress profile will be reconstructed almost entirely with the first mode at all $\theta$-coordinates. This is also indicated by the cumulative reconstruction of cross-spectra, due to the collapse of the modal contributions over a very wide range of the jet-width. This indicates that each additional mode contributes very little to the total spectrum. These results not only confirm the spectral similarity from \cite{Lumley1967} which was somewhat indicated by the modal self-similarity discussed earlier, but also demonstrate that shear-stress producing structures in the inertial subrange are completely defined by the first LD mode in regions with high constant shear in the jet, namely around $\theta/\theta_{\frac{1}{2}}=1$. This correlates with the fact that high shear regions may produce significant TKE at a wide range of scales, as seen directly from \eqref{eq:galerkin_production_term} in case of large mean gradients.

\subsection{Quantification of modal self-similarity}
The self-similarity of eigenfunctions was originally observed in \cite{Gamard2004} and was noted in \cite{Wanstrom2009} and predicted by \cite{Ewing2007}. It has also been recently documented in \cite{Muralidhar2018} in their spatio-temporal proper orthogonal decomposition of a turbulent channel flow. Those results together with the ones presented in the current work indicate that modal self-similarity is a fundamental trait spanning across turbulent flows whenever Fourier-based decompositions are applied. 

In the following modal self-similarity is depicted from the perspective of parallelity of the numerical eigenfunctions with respect to an inner product over the $\theta$-coordinate. The inner product of the normalized numerical modes is written as
\begin{equation}
\left(\widetilde{\psi}^{i\alpha},\widetilde{\psi}_{i\beta}\right)_\theta = \int_0^{\theta_{\text{max}}}\widetilde{\psi}^{i\alpha}\widetilde{\psi}_{i\beta}^* d\theta,\label{eq:parallelity}
\end{equation}
\begin{figure}[htp]
\centering      
\subfloat[]{\includegraphics[width=0.40\linewidth]{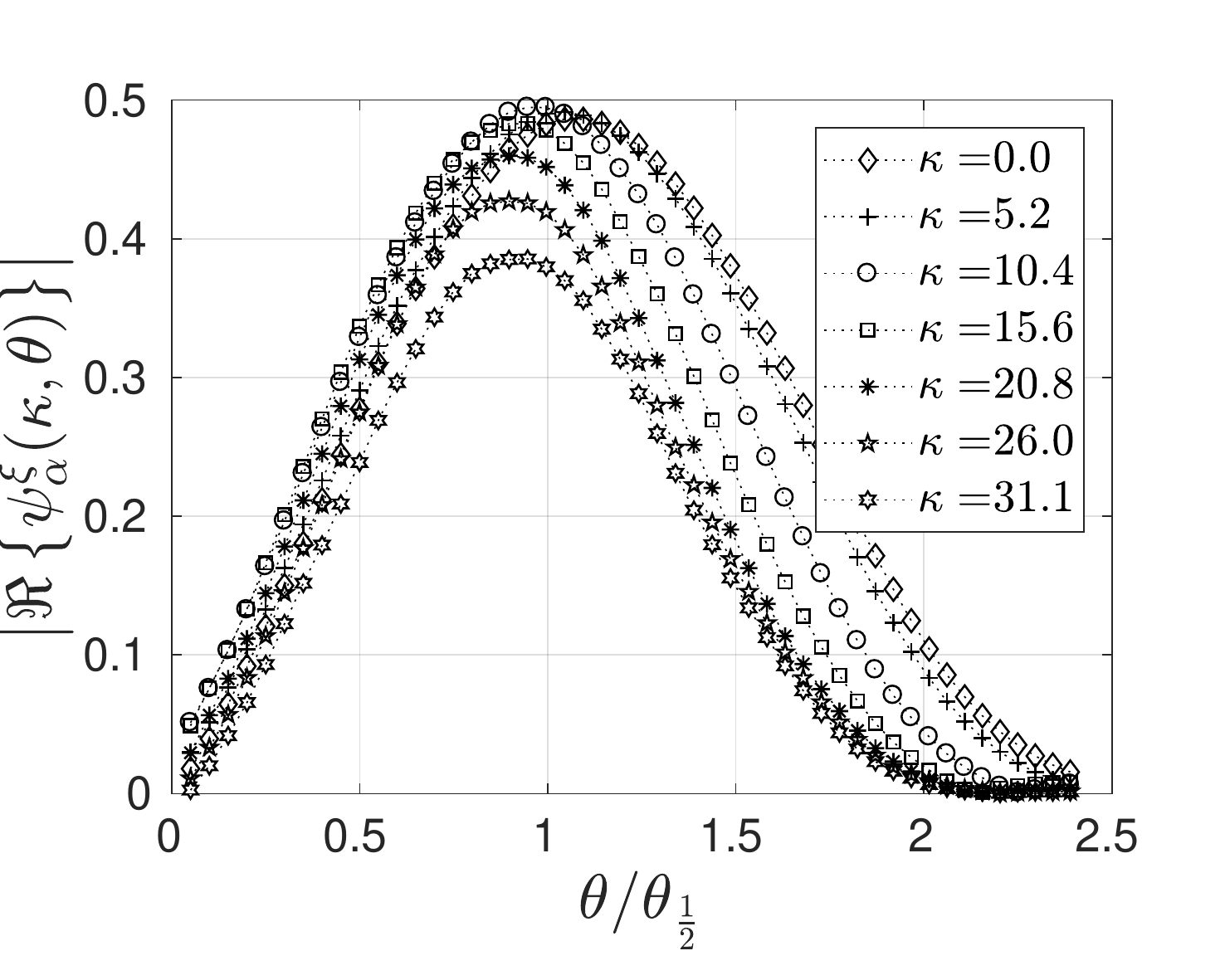}\label{fig:mode_comparison_xi_real}}
\subfloat[]{\includegraphics[width=0.40\linewidth]{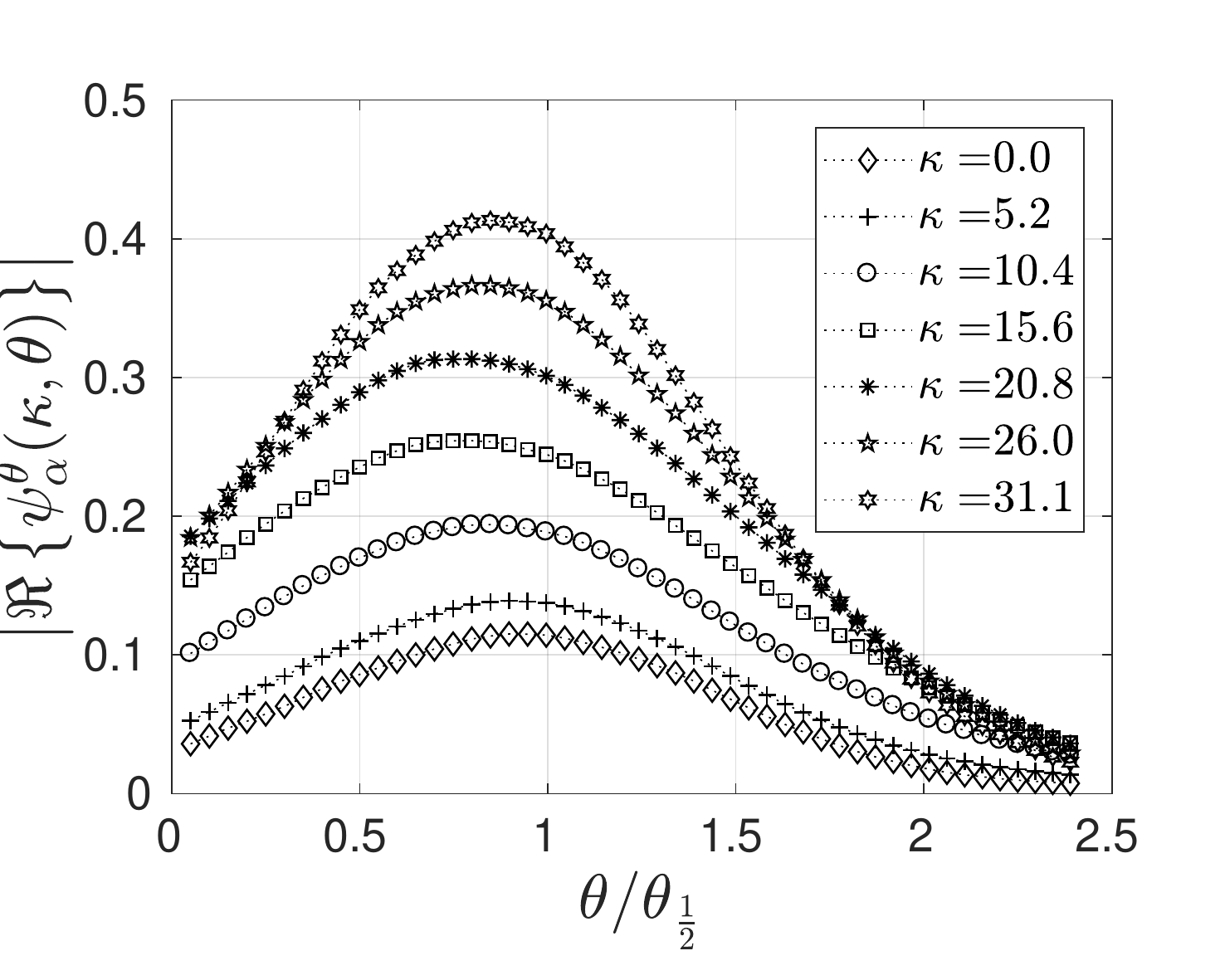}\label{fig:mode_comparison_theta_real}}\\
\subfloat[]{\includegraphics[width=0.40\linewidth]{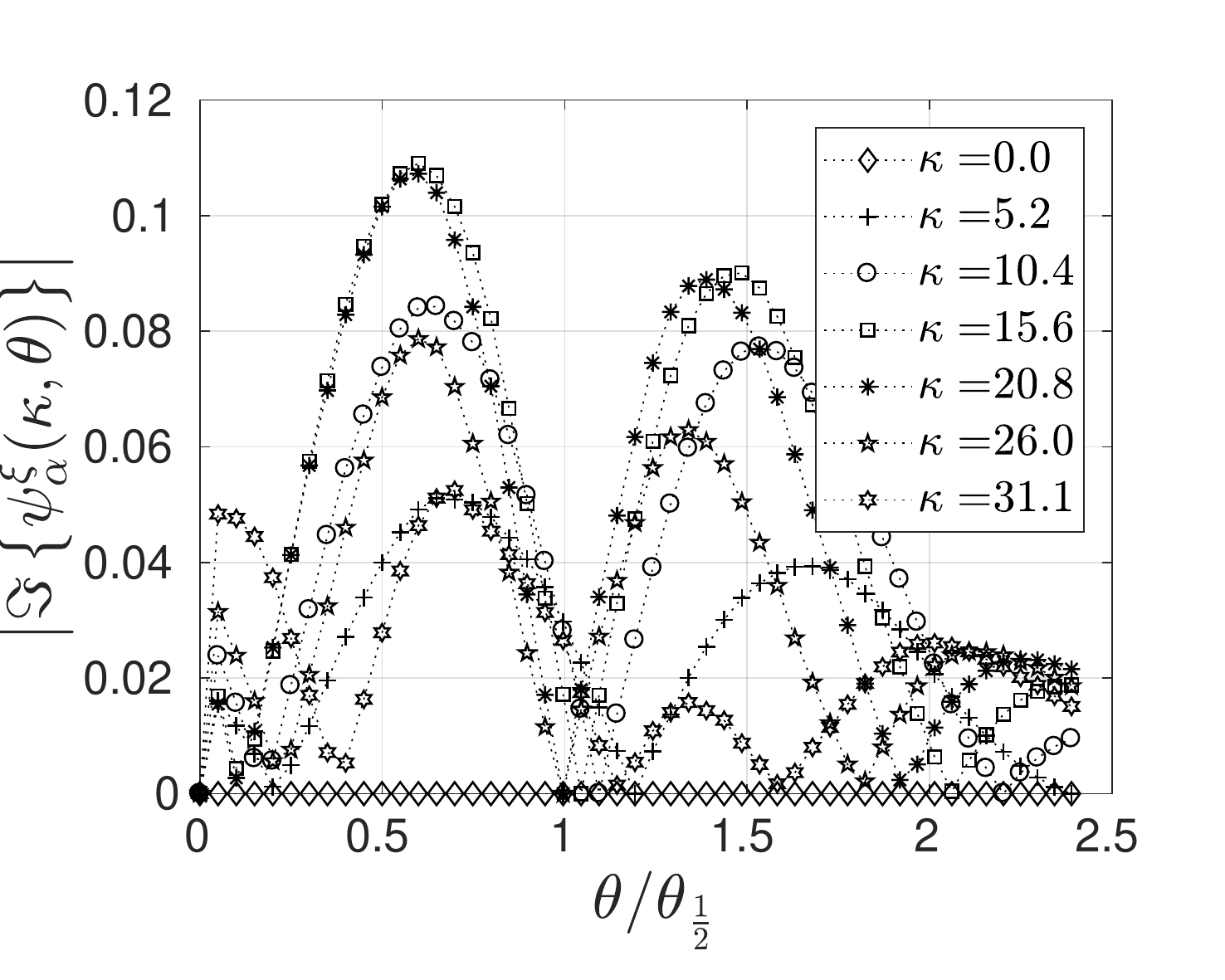}\label{fig:mode_comparison_xi_imag}}
\subfloat[]{\includegraphics[width=0.40\linewidth]{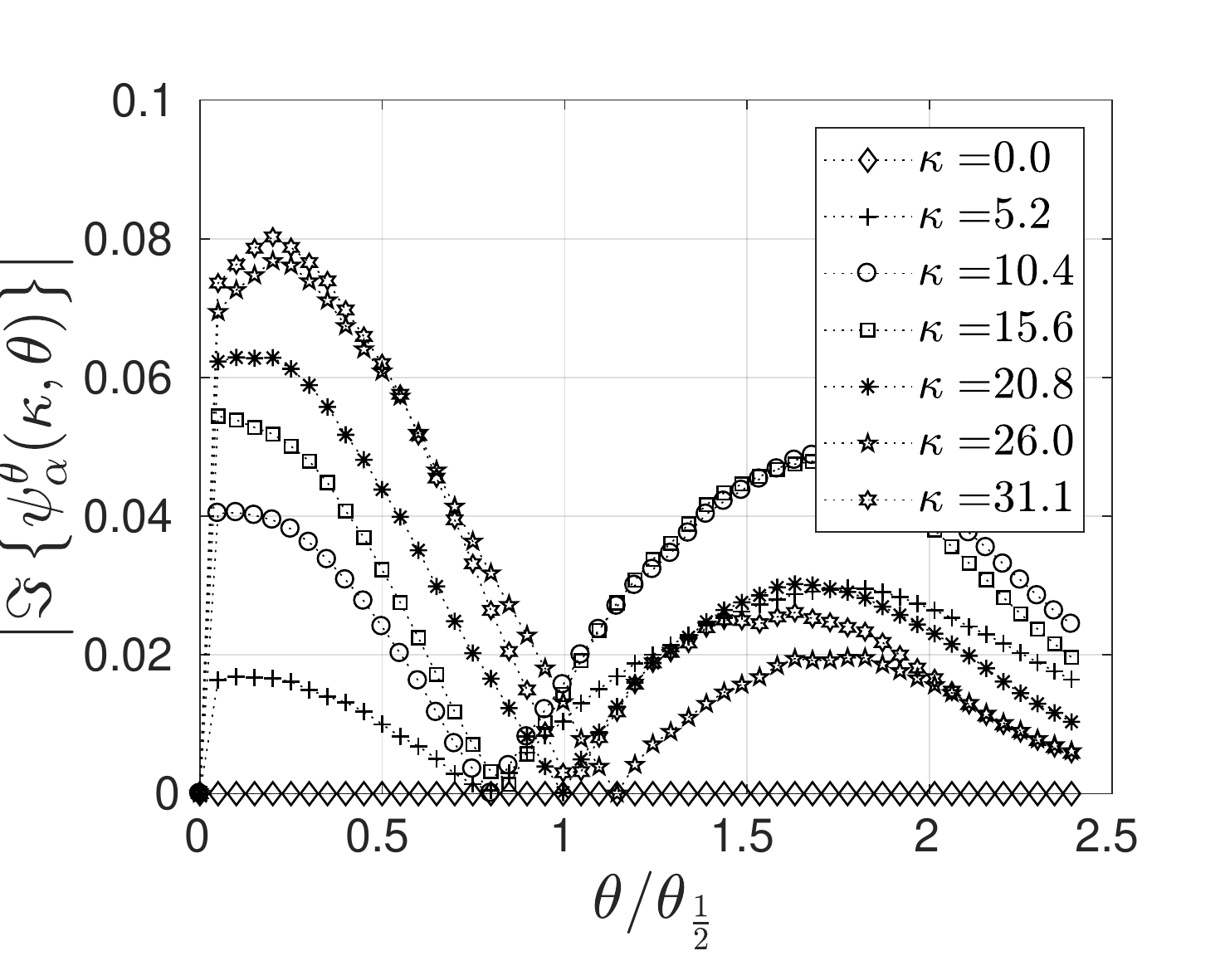}\label{fig:mode_comparison_theta_imag}}
\caption{Visual comparison of modes, $\psi^\xi_\alpha\left(\kappa,\theta\right)$ and $\psi^\theta_\alpha\left(\kappa,\theta\right)$ across $\kappa$ for the mode number $\alpha=1$. Absolute values of the real parts of the $\xi$- and $\theta$-components are shown in (a) and (b) and the imaginary parts are shown in (c) and (d). \label{fig:mode_comparison}}
\end{figure}
\FloatBarrier
\noindent
where $\widetilde{\psi}^{i\alpha}=\widetilde{\psi}^{i\alpha}\left(\theta,\kappa_1\right)$, $\widetilde{\psi}_{i\beta}=\widetilde{\psi}_{i\beta}\left(\theta,\kappa_2\right)$ and $\theta_\text{max}$ is the upper bound of the $\theta$-domain. For $\alpha=\beta$ and $\kappa_1=\kappa_2$ \eqref{eq:parallelity} produces unity whilst for $\alpha\neq\beta$ and $\kappa_1=\kappa_2$ it is zero (see Appendix C in \cite{Hodzic2019_part1}). Nevertheless, \eqref{eq:parallelity} enables us to quantify the parallelity of any pair of numerical components of the LD modes since the orthogonality of $\widetilde{\psi}^{i\alpha}$ and $\widetilde{\psi}_{i\beta}$ with respect to \eqref{eq:parallelity} is not ensured for $\kappa_1\neq\kappa_2$ - even for numerical eigenfunctions that share the same mode number i.e. $\alpha=\beta$. Note that this does not contradict \eqref{eq:orthogonality_eigenfunctions} since the latter is defined for eigenfunctions $\overline{\Phi}_\alpha=\varphi^{j}_\alpha\overline{z}_j$, where $\varphi^j_\alpha$ is defined as \eqref{eq:eigenfunctions} (see Appendix C in \cite{Hodzic2019_part1}) whilst \eqref{eq:parallelity} is defined for the normalized numerical LD components. From a different perspective, it can be said that the loss of orthogonality with respect to \eqref{eq:parallelity} between $\widetilde{\psi}^{i\alpha}$ and $\widetilde{\psi}_{i\beta}$ for $\kappa_1\neq\kappa_2$ is a consequence of the eigenvalue problem being solved separately for each $\kappa$.

Since the visual similarities between any pair of functions in $L^2$ tend to decrease with increasing deviation from orthogonality, it means that functions that are not orthogonal are more likely to share similar traits than orthogonal ones. This is demonstrated by figure \ref{fig:mode_comparison} where the real- and imaginary eigenfunction components for $\alpha=1$ are compared across $\kappa$-values. It is seen from figure \ref{fig:mode_comparison} that the parallelity- and thereby the apparent self-similarity between $\widetilde{\psi}^i_\alpha\left(\kappa_1,\theta\right)$ and $\widetilde{\psi}^i_\alpha\left(\kappa_2,\theta\right)$ decreases for increasing differences between $\kappa_1$- and $\kappa_2$. We can in fact make the statement that modal self-similarity occurs due to the non-orthogonality of $\widetilde{\psi}^{i\alpha}\left(\kappa_1,\theta\right)$ and $\widetilde{\psi}_{i\alpha}\left(\kappa_2,\theta\right)$ with respect to \eqref{eq:parallelity}.

%
\begin{figure}[h]
\centering      
\subfloat[]{\includegraphics[width=0.30\linewidth]{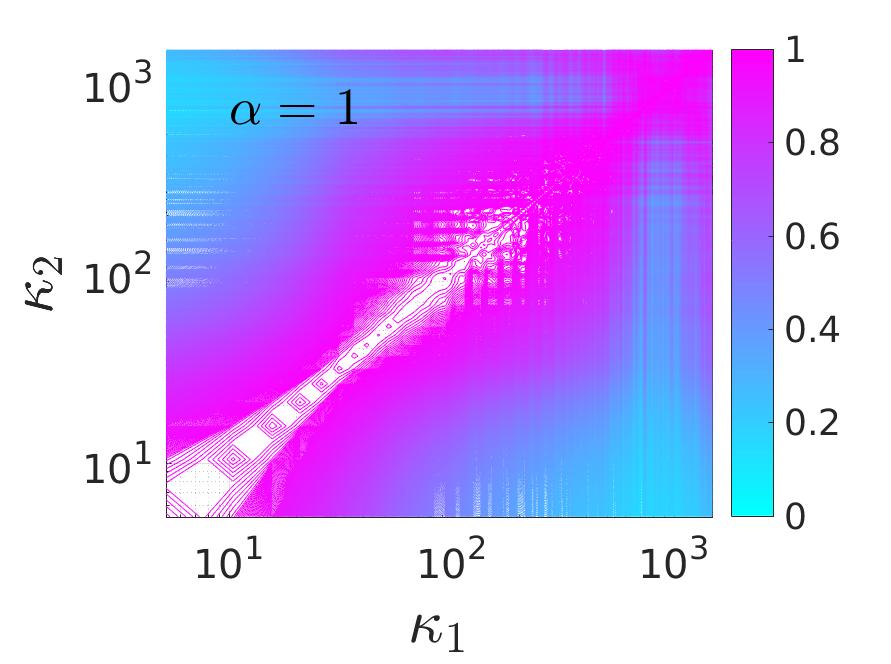}\label{fig:parallelity_kappa_SSC1}}
\subfloat[]{\includegraphics[width=0.30\linewidth]{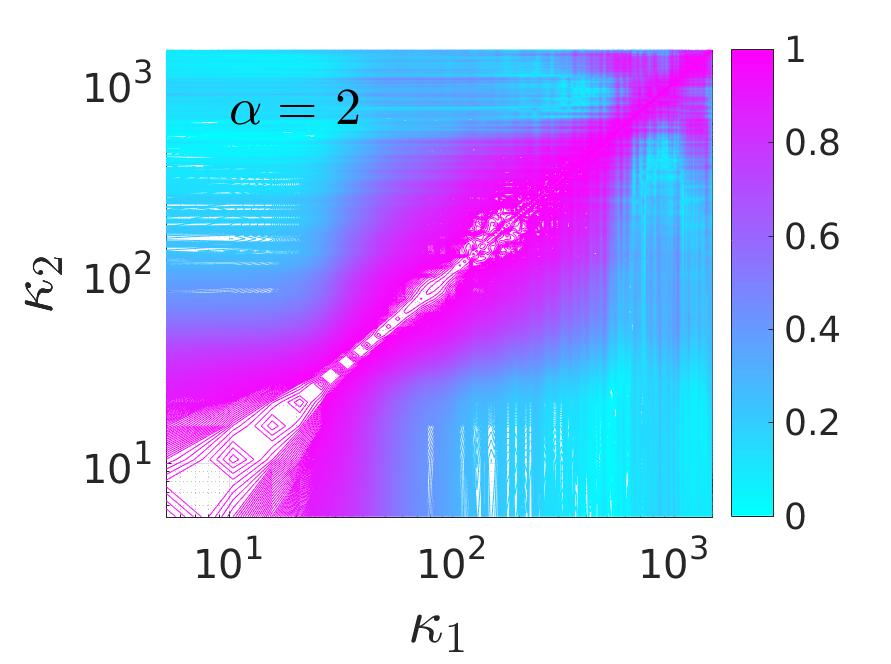}\label{fig:parallelity_kappa_SSC2}}
\subfloat[]{\includegraphics[width=0.30\linewidth]{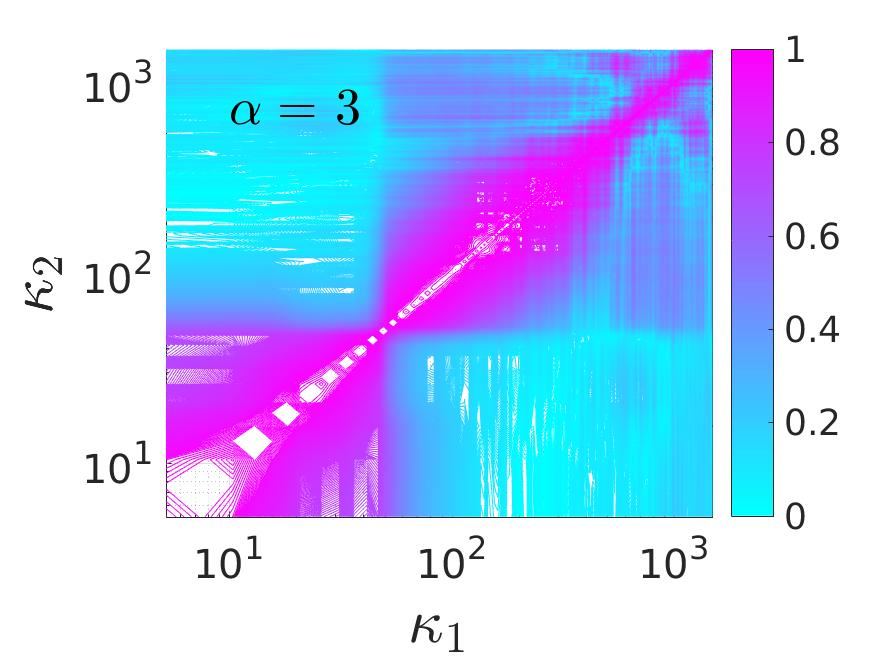}\label{fig:parallelity_kappa_SSC3}}\\
\subfloat[]{\includegraphics[width=0.30\linewidth]{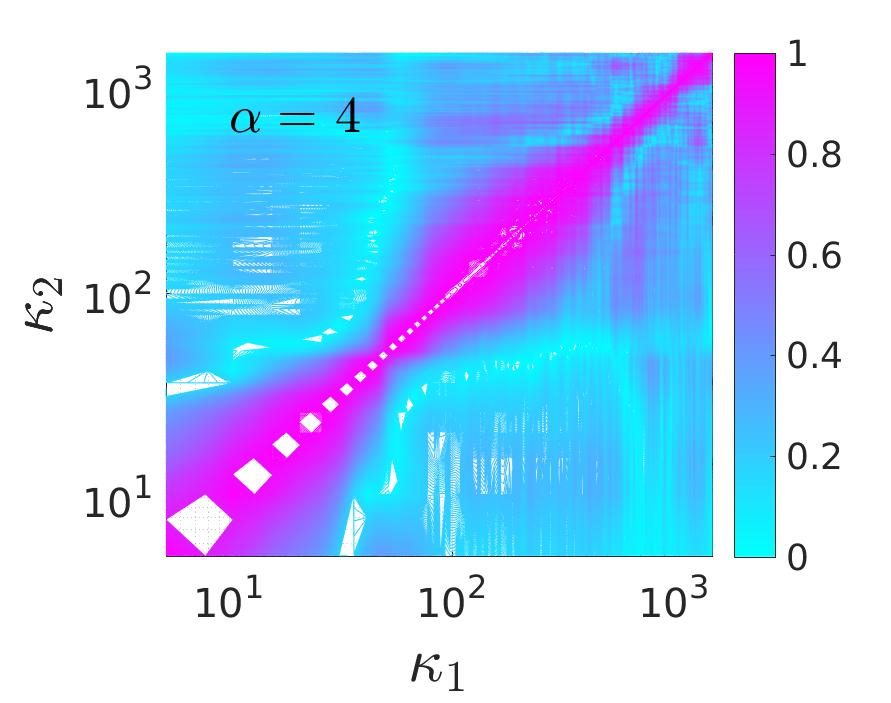}\label{fig:parallelity_kappa_SSC4}}
\subfloat[]{\includegraphics[width=0.30\linewidth]{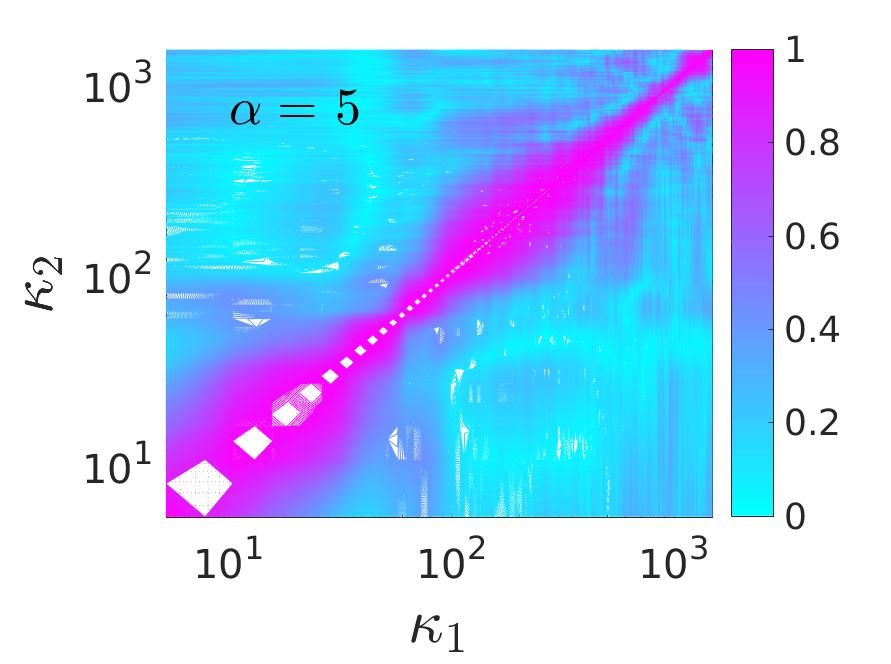}\label{fig:parallelity_kappa_SSC5}}
\subfloat[]{\includegraphics[width=0.30\linewidth]{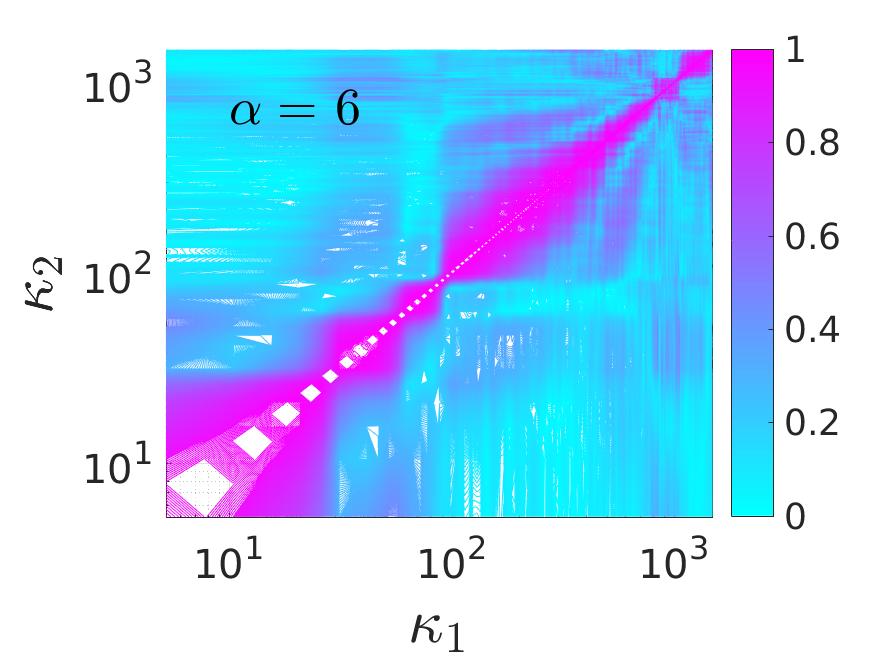}\label{fig:parallelity_kappa_SSC6}}\\
\subfloat[]{\includegraphics[width=0.30\linewidth]{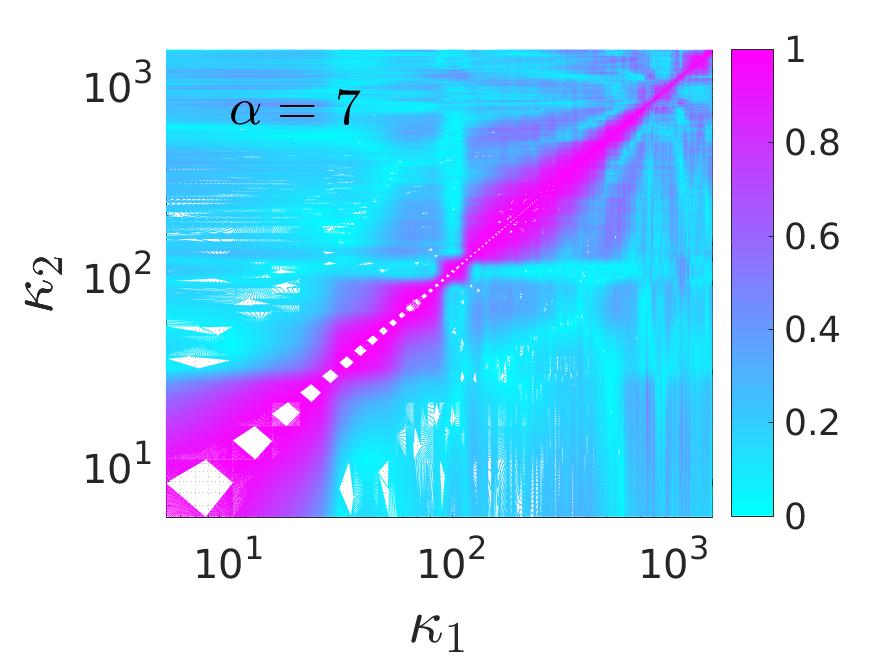}\label{fig:parallelity_kappa_SSC7}}
\subfloat[]{\includegraphics[width=0.30\linewidth]{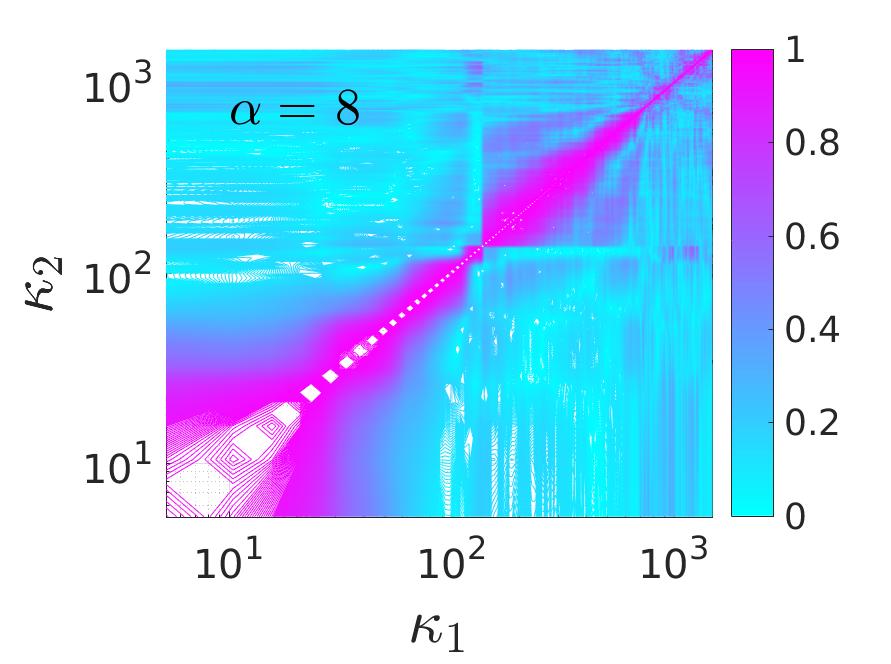}\label{fig:parallelity_kappa_SSC8}}
\subfloat[]{\includegraphics[width=0.30\linewidth]{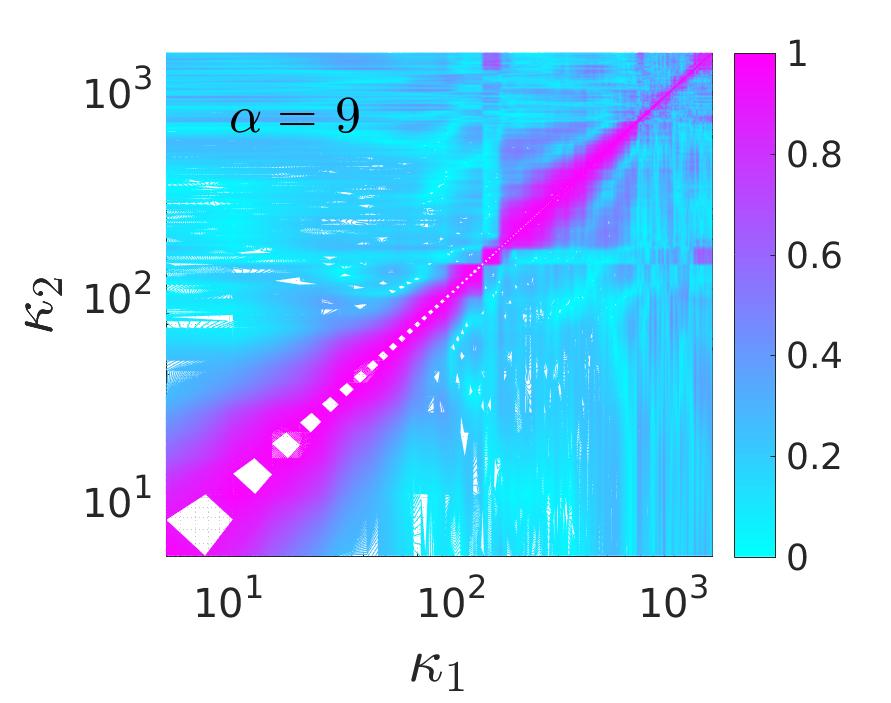}\label{fig:parallelity_kappa_SSC9}}\\
\subfloat[]{\includegraphics[width=0.30\linewidth]{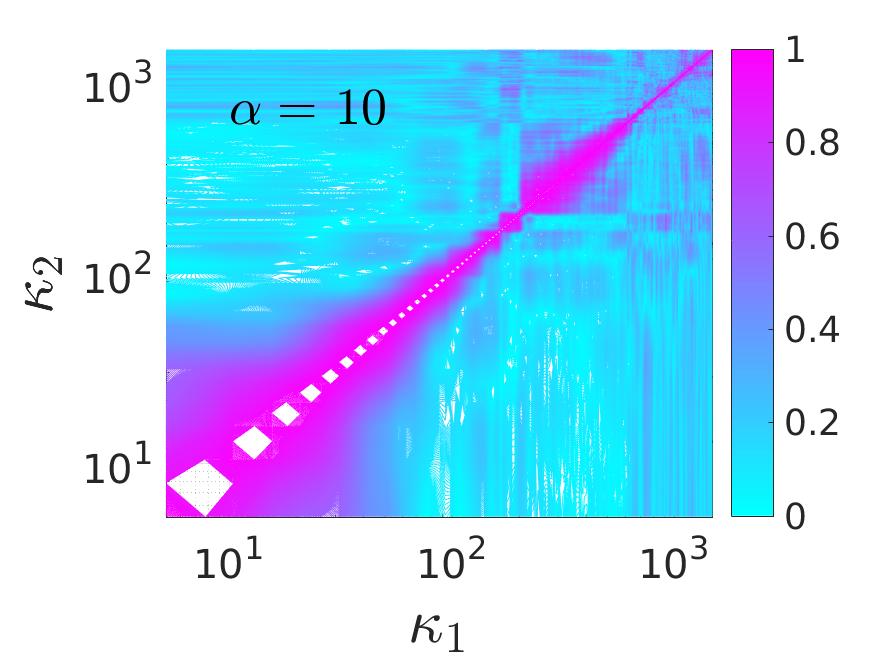}\label{fig:parallelity_kappa_SSC10}}
\subfloat[]{\includegraphics[width=0.30\linewidth]{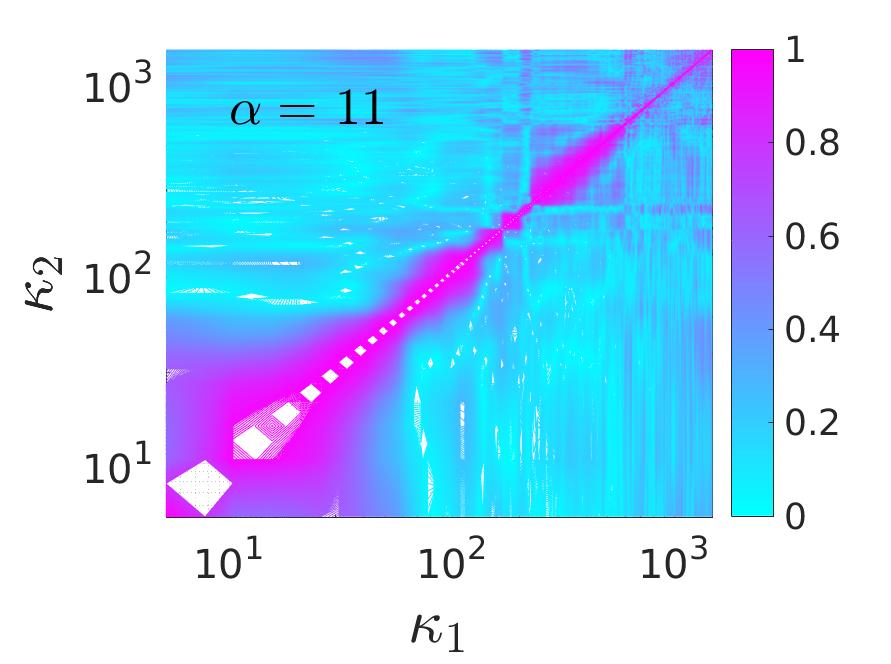}\label{fig:parallelity_kappa_SSC11}}
\subfloat[]{\includegraphics[width=0.30\linewidth]{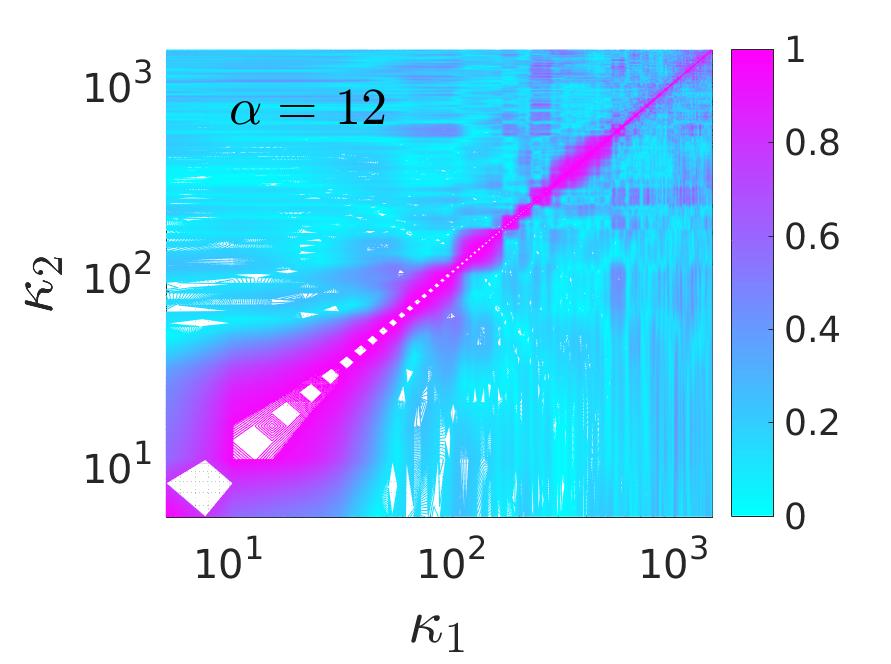}\label{fig:parallelity_kappa_SSC12}}\\
\subfloat[]{\includegraphics[width=0.30\linewidth]{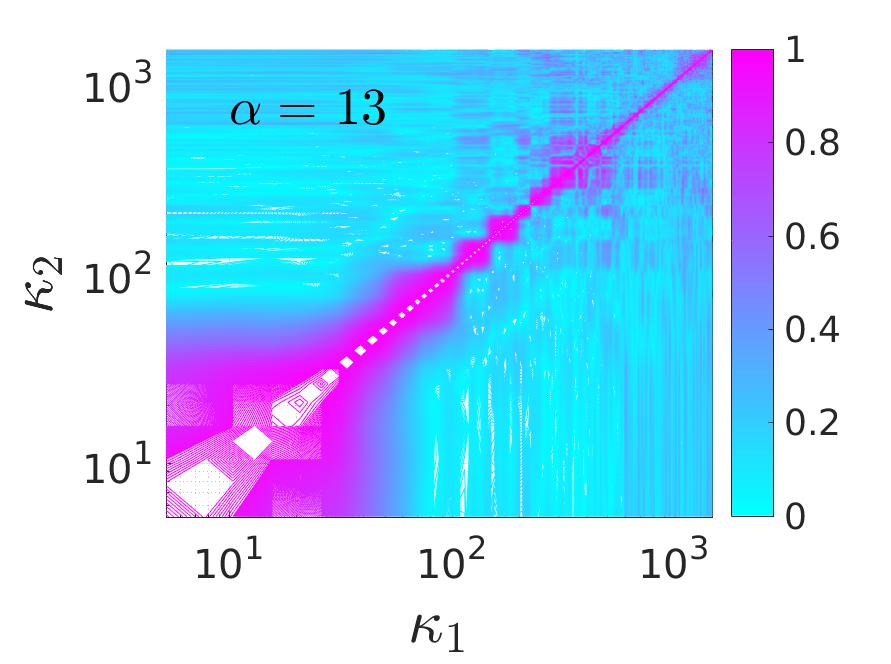}\label{fig:parallelity_kappa_SSC13}}
\subfloat[]{\includegraphics[width=0.30\linewidth]{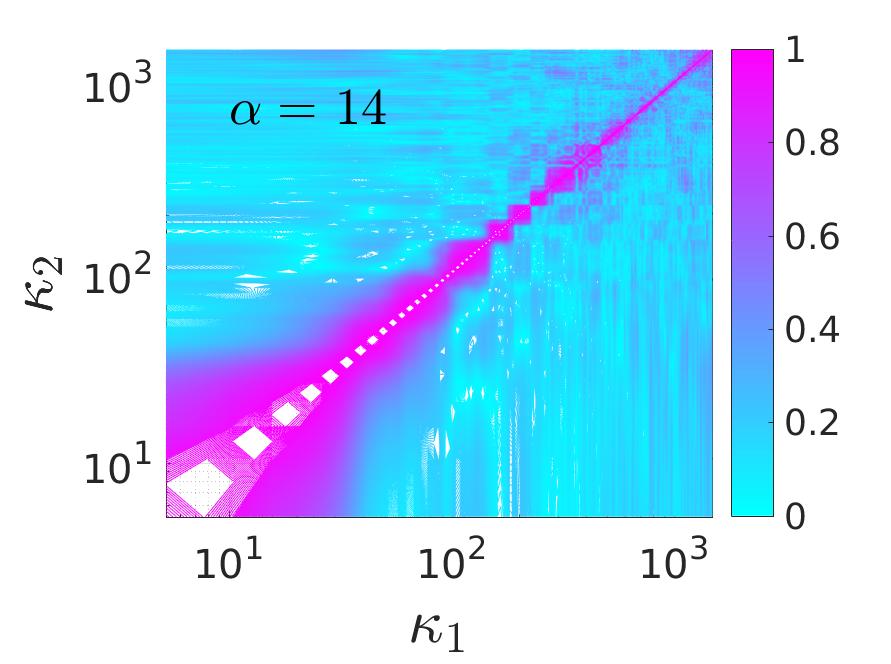}\label{fig:parallelity_kappa_SSC14}}
\subfloat[]{\includegraphics[width=0.30\linewidth]{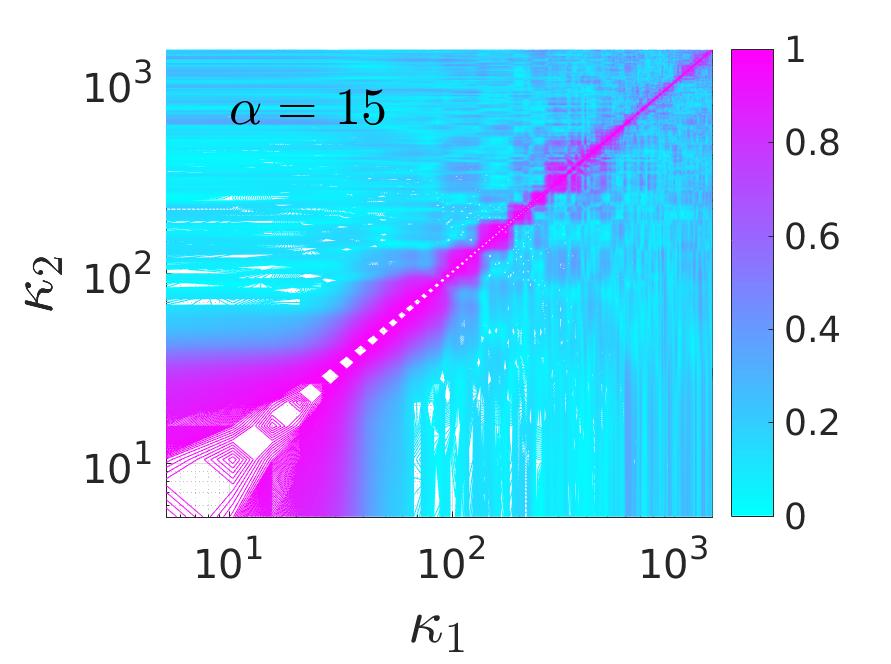}\label{fig:parallelity_kappa_SSC15}}
\caption{Modal self-similarity quantification of eigenfunctions,  $\widetilde{\psi}^\alpha(\theta,\kappa_1)$ and $\widetilde{\psi}_\beta(\theta,\kappa_2)$, for $\alpha=\beta=[1:15]$ across all combinations of $\kappa$-pairs, $\kappa_1$ and $\kappa_2$. The non-orthogonality between $\widetilde{\psi}^{i\alpha}(\theta,\kappa_1)$ and $\widetilde{\psi}_{i\beta}(\theta,\kappa_2)$ is manifested by the size of the off-diagional values of the contour plots. The larger the off-diagonal values the more self-similar the eigenfunctions are likely to appear for different pairs of $\kappa_1$ and $\kappa_2$.\label{fig:parallelity}}
\end{figure}
\FloatBarrier
\noindent

Figure \ref{fig:parallelity} shows the absolute value of \eqref{eq:parallelity} for $\alpha=\beta=[1:15]$ where it is seen that the modal self-similarity generally decreases with increasing $\alpha$. The most distinct "jump" occurs between $\alpha=3$ and $\alpha=4$ after which the contours of figure \ref{fig:parallelity} converge. For each $\alpha$ in figure \ref{fig:parallelity} it is seen that the self-similarity band varies with the size of $\kappa$, demonstrating that the manifestation of the phenomenon is more profound in certain wavenumber ranges than others. For instance, figure \ref{fig:parallelity_kappa_SSC1} shows that a relatively wide range of $\kappa_1,\kappa_2$-combinations are self-similar to each other for $\alpha=1$. Figure \ref{fig:parallelity} then demonstrates that the higher the $\kappa$-values the more localized the similarities of the eigenfunctions are, which is evident from the needle-shaped contours of $\alpha\geq 13$ in figure \ref{fig:parallelity}.
\FloatBarrier
\subsection{Energy production analysis}\label{sec:Energy_production_analysis}
In the following an in-depth energy production analysis will be performed focusing on the local energy production of the modes. The analysis will test the extent to which the hypothesis posed in \cite{Wanstrom2009} holds, namely that multiple modes obtain significant and relatively constant parts of their energy directly from the mean flow. This will lead to a quantification of the energy-normalized production of each mode.

The term \textit{local equilibrium} is generally applied in relation to the averaged turbulence spectra, in particular in wavenumber regions where these exhibit the characteristic $-5/3$-slope predicted by the classical Kolmogorov theory. In fact also the dissipation spectrum is hypothesized by the Kolmogorov theory to have a universal similarity form. 
\begin{figure}[htp]
\centering      
\subfloat[]{\includegraphics[width=0.40\linewidth]{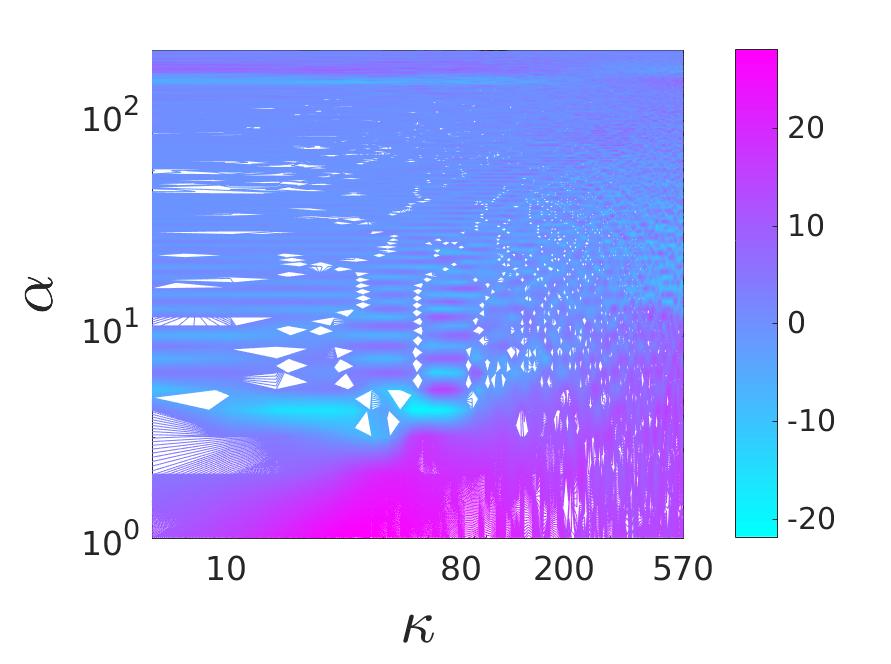}\label{fig:E_prop_real}}
\subfloat[]{\includegraphics[width=0.40\linewidth]{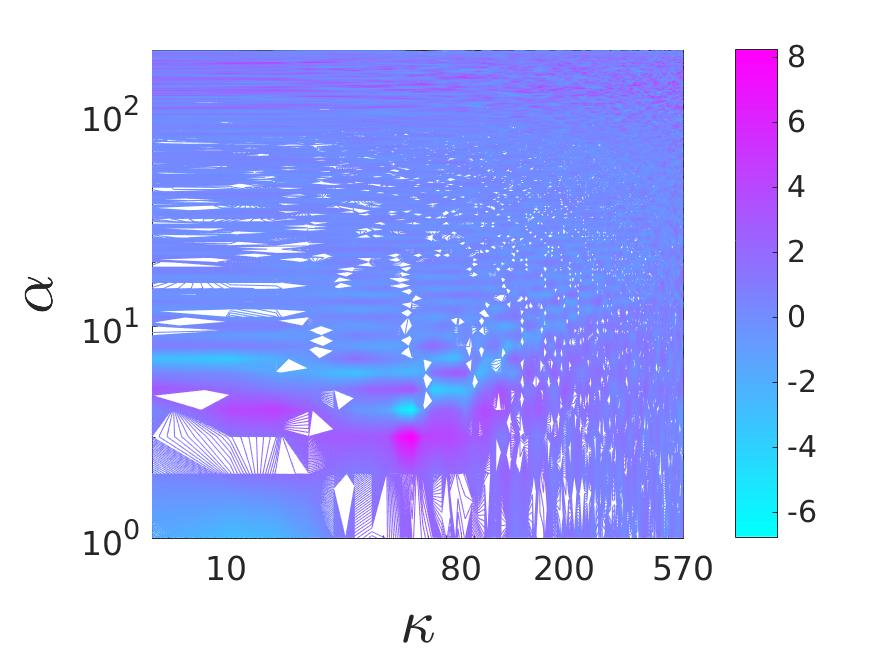}\label{fig:E_prop_imag}}\\
\subfloat[]{\includegraphics[width=0.40\linewidth]{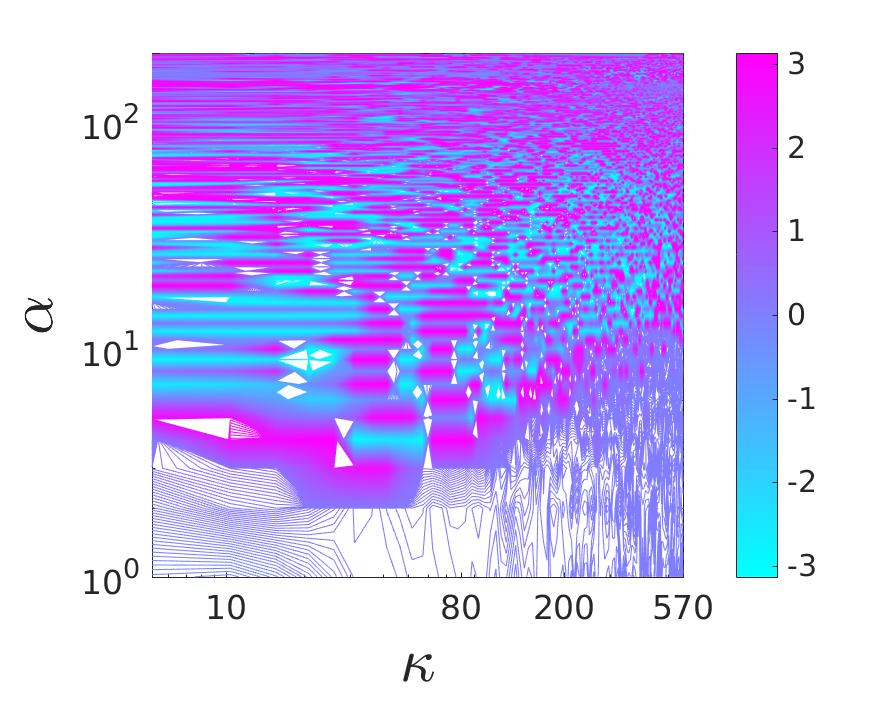}\label{fig:E_prop_arg}}
\subfloat[]{\includegraphics[width=0.40\linewidth]{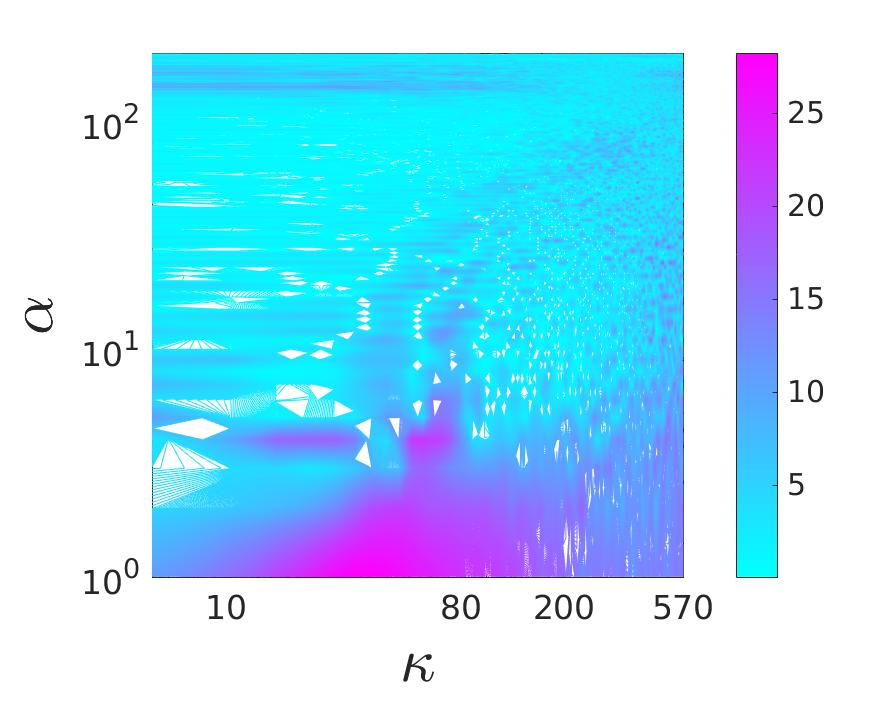}\label{fig:E_prop_abs}}
\caption{Energy-normalized production spectra, $\mathcal{P}_{\rho\lambda}$. (a): $\Re\left\{\mathcal{\mathcal{P}_{\rho\lambda}}\right\}$, (b): $\Im\left\{\mathcal{\mathcal{P}_{\rho\lambda}} \right\}$, (c): $Arg\left(\mathcal{\mathcal{P}_{\rho\lambda}}\right)$ (d): $\left|\mathcal{\mathcal{P}_{\rho\lambda}}\right|$. \label{fig:E_prop}}
\end{figure}
\noindent
%
%
\begin{figure}[h]
    \centering      
\subfloat[$\alpha=1$]{\includegraphics[width=0.3\linewidth]{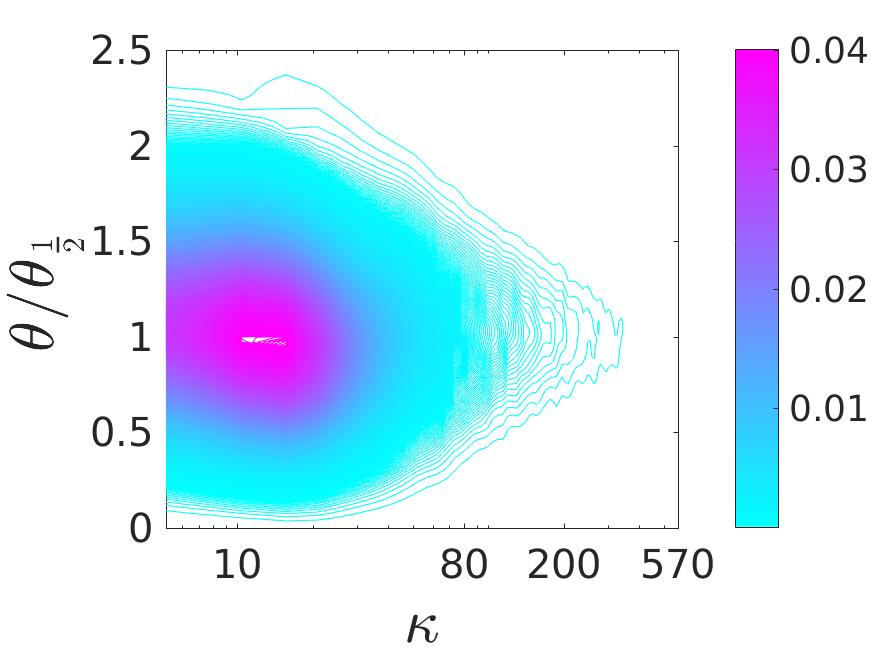}\label{fig:term_II_single_1}}
\subfloat[$\alpha=2$]{\includegraphics[width=0.3\linewidth]{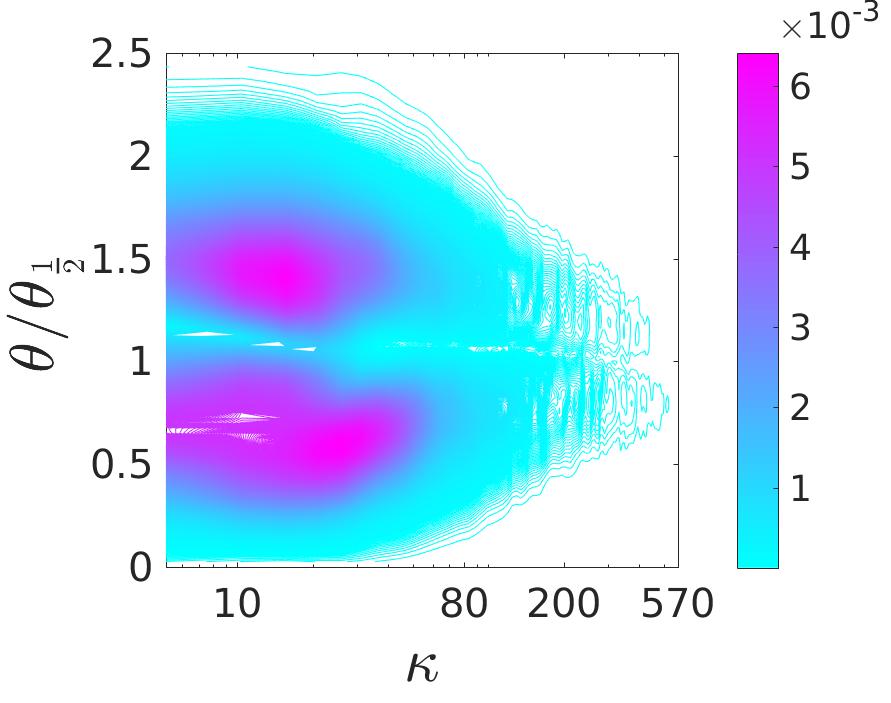}\label{fig:term_II_single_2}}
\subfloat[$\alpha=3$]{\includegraphics[width=0.3\linewidth]{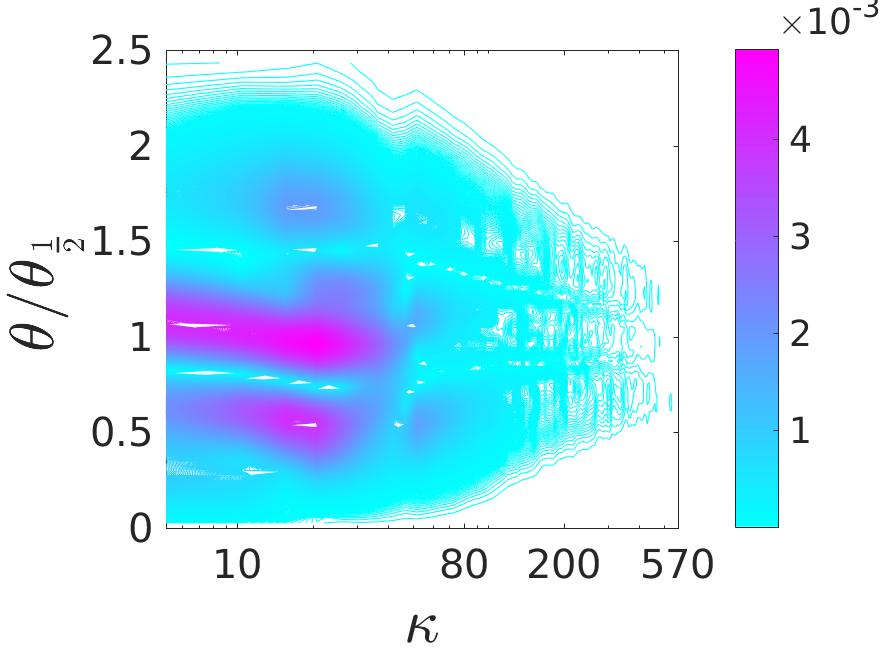}\label{fig:term_II_single_3}}\\
\subfloat[$\alpha=4$]{\includegraphics[width=0.3\linewidth]{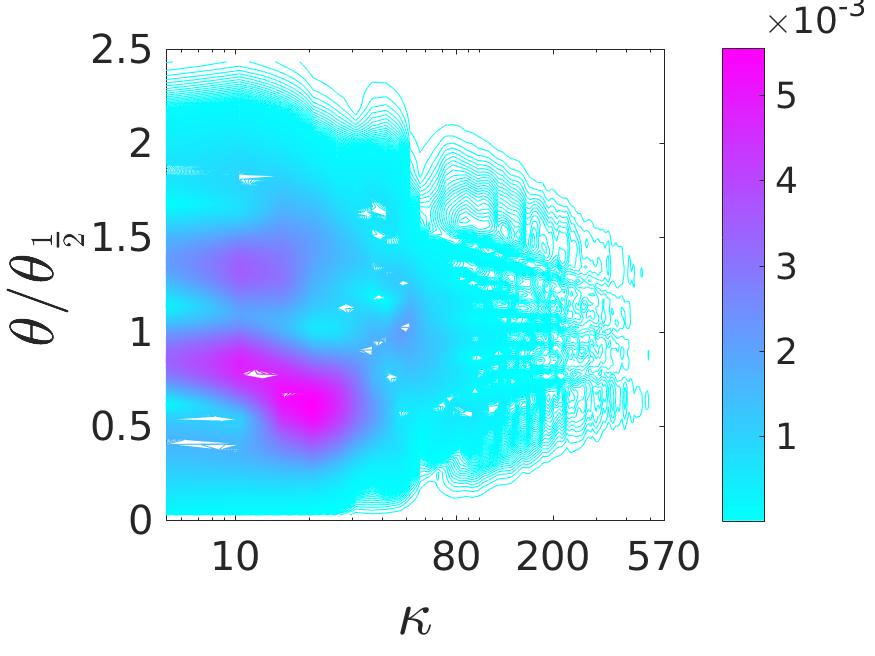}\label{fig:term_II_single_4}}
\subfloat[$\alpha=5$]{\includegraphics[width=0.3\linewidth]{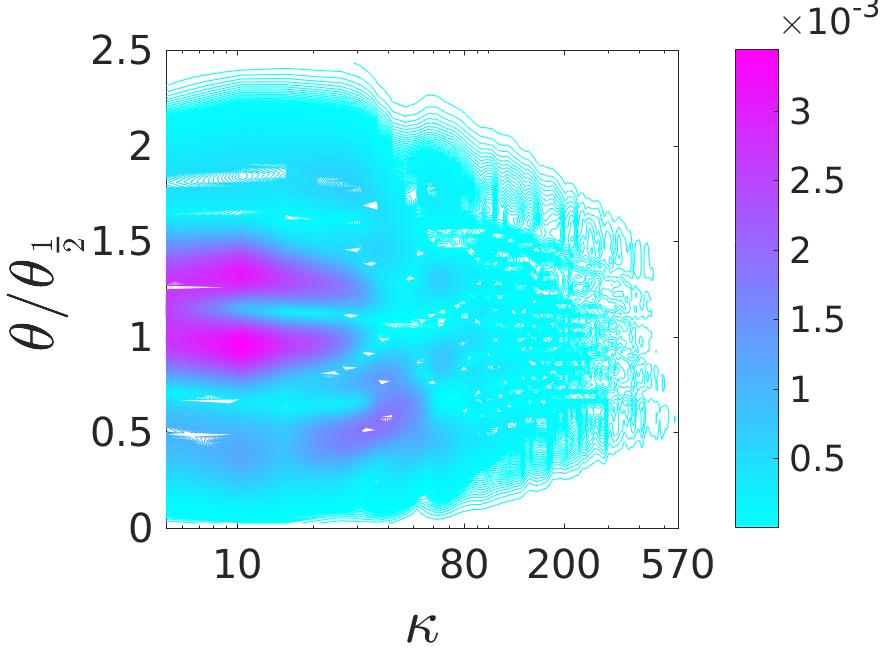}\label{fig:term_II_single_5}}
\subfloat[$\alpha=6$]{\includegraphics[width=0.3\linewidth]{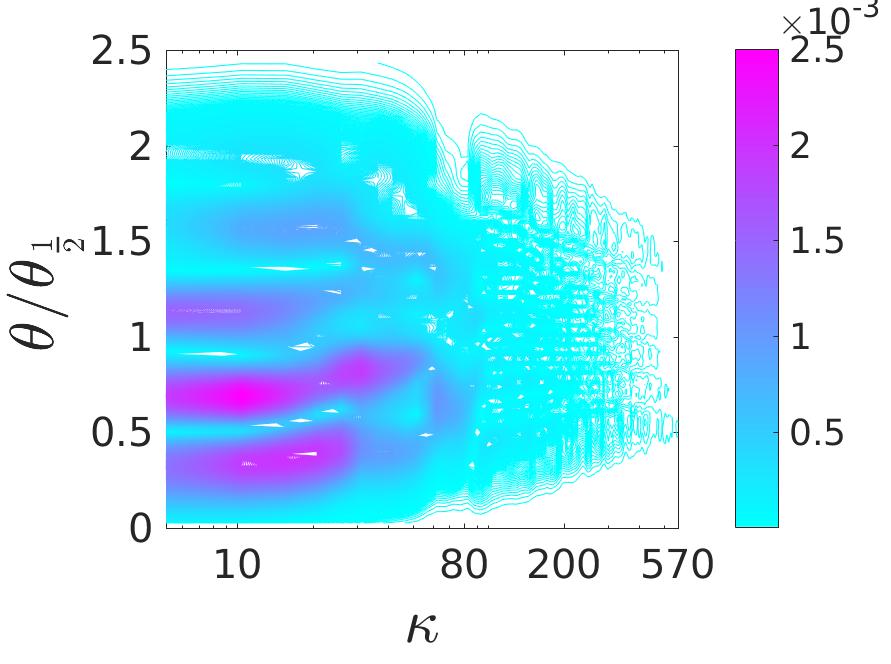}\label{fig:term_II_single_6}}\\
\subfloat[$\alpha=7$]{\includegraphics[width=0.3\linewidth]{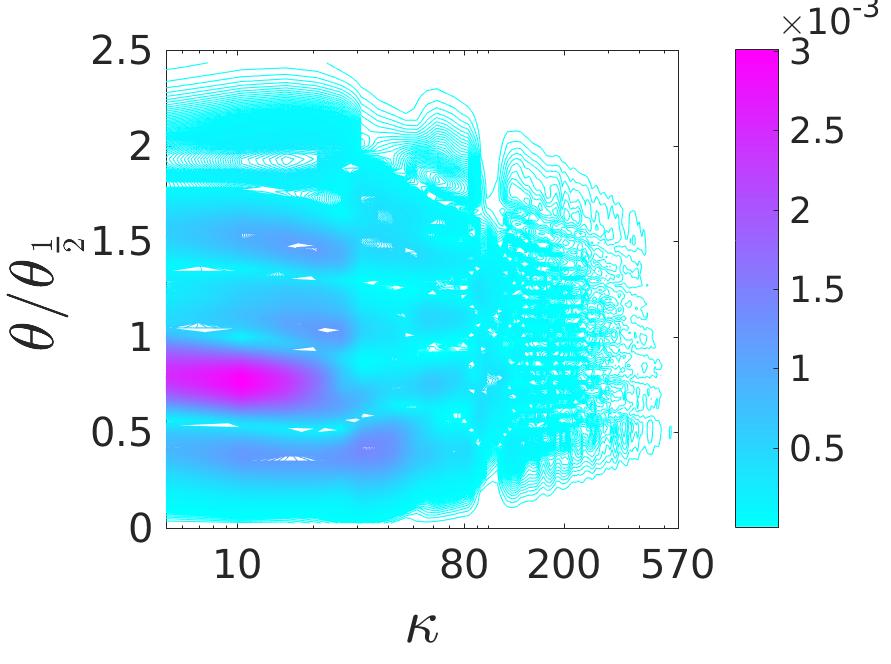}\label{fig:term_II_single_7}}
\subfloat[$\alpha=8$]{\includegraphics[width=0.3\linewidth]{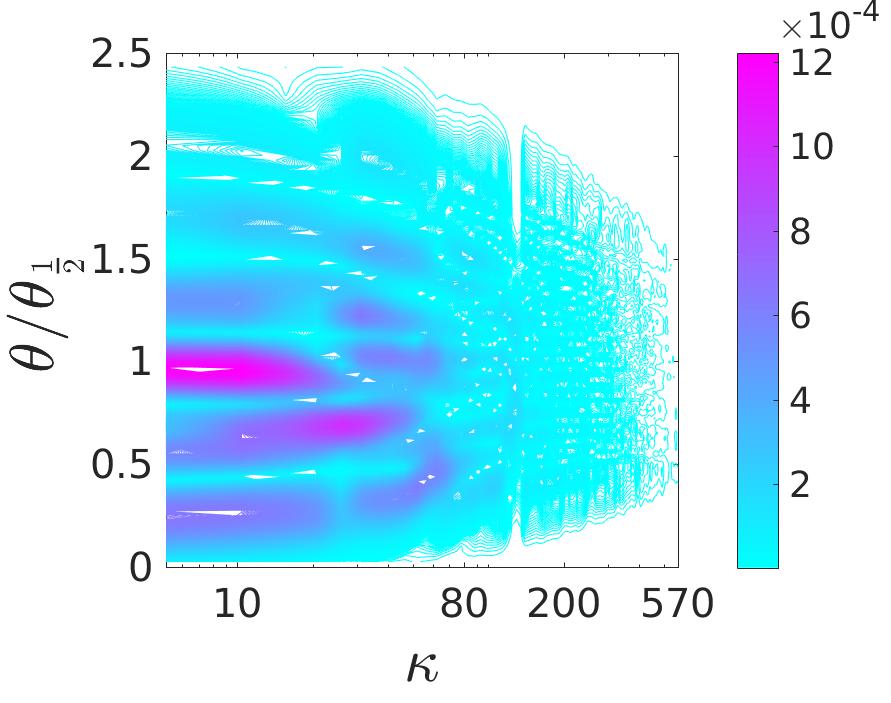}\label{fig:term_II_single_8}}
\subfloat[$\alpha=9$]{\includegraphics[width=0.3\linewidth]{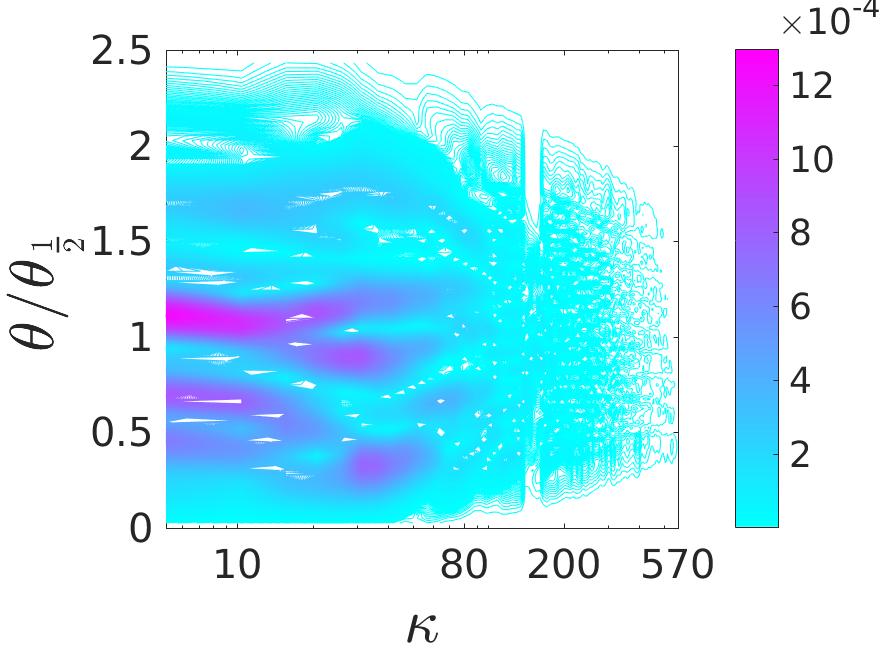}\label{fig:term_II_single_9}}\\
\subfloat[$\alpha=10$]{\includegraphics[width=0.3\linewidth]{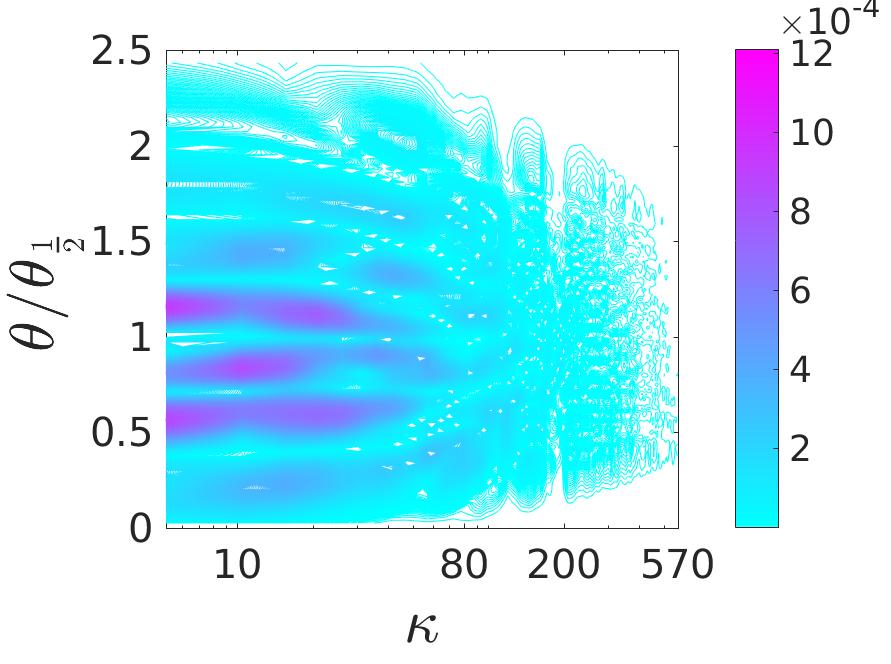}\label{fig:term_II_single_10}}
\subfloat[$\alpha=11$]{\includegraphics[width=0.3\linewidth]{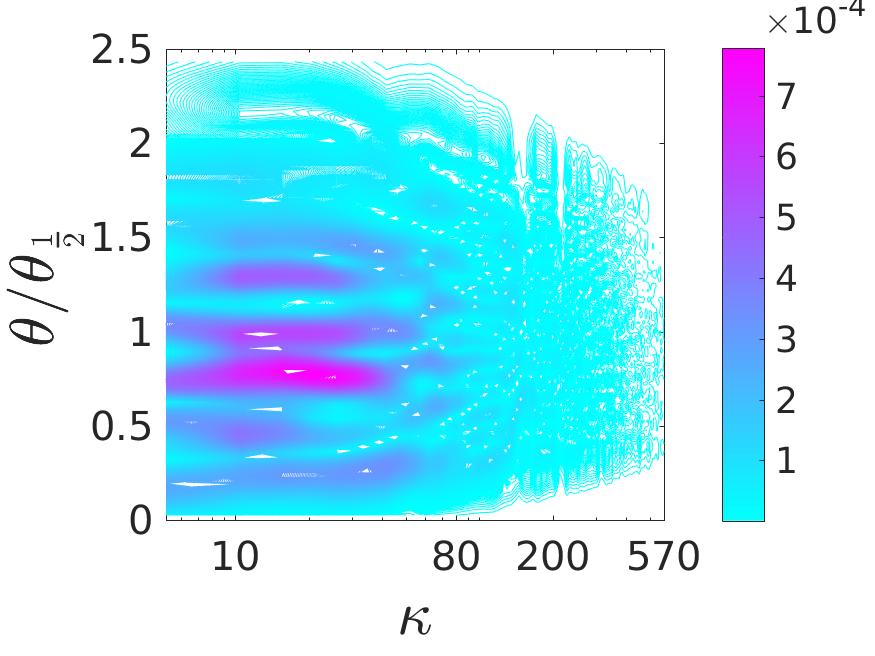}\label{fig:term_II_single_11}}
\subfloat[$\alpha=12$]{\includegraphics[width=0.3\linewidth]{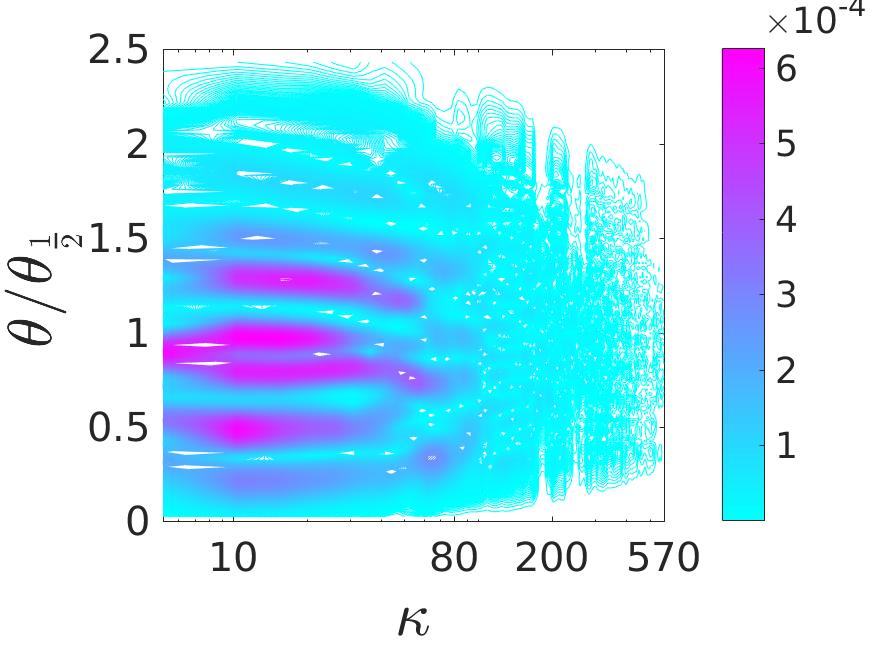}\label{fig:term_II_single_12}}\\
\subfloat[$\alpha=13$]{\includegraphics[width=0.3\linewidth]{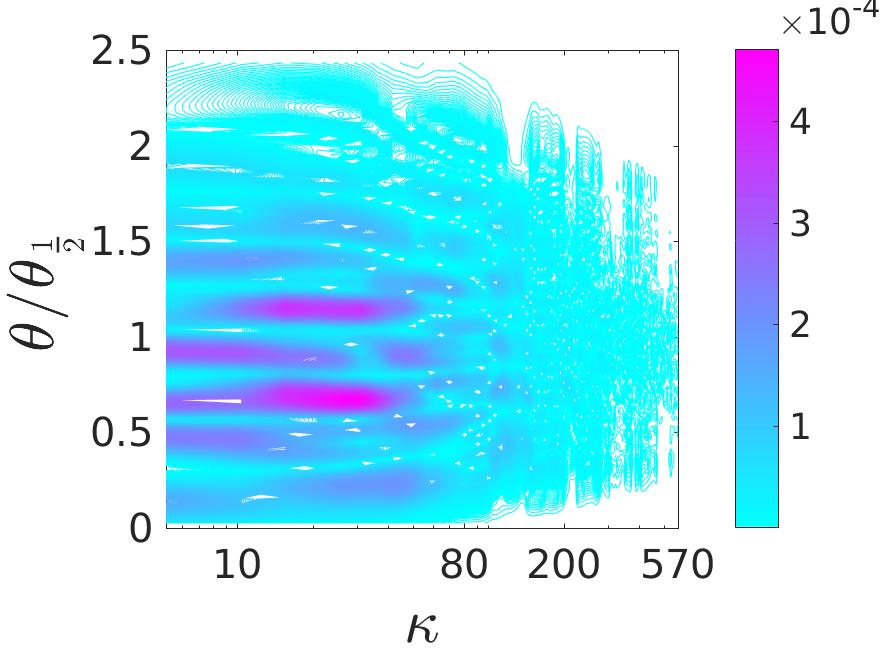}\label{fig:term_II_single_13}}
\subfloat[$\alpha=14$]{\includegraphics[width=0.3\linewidth]{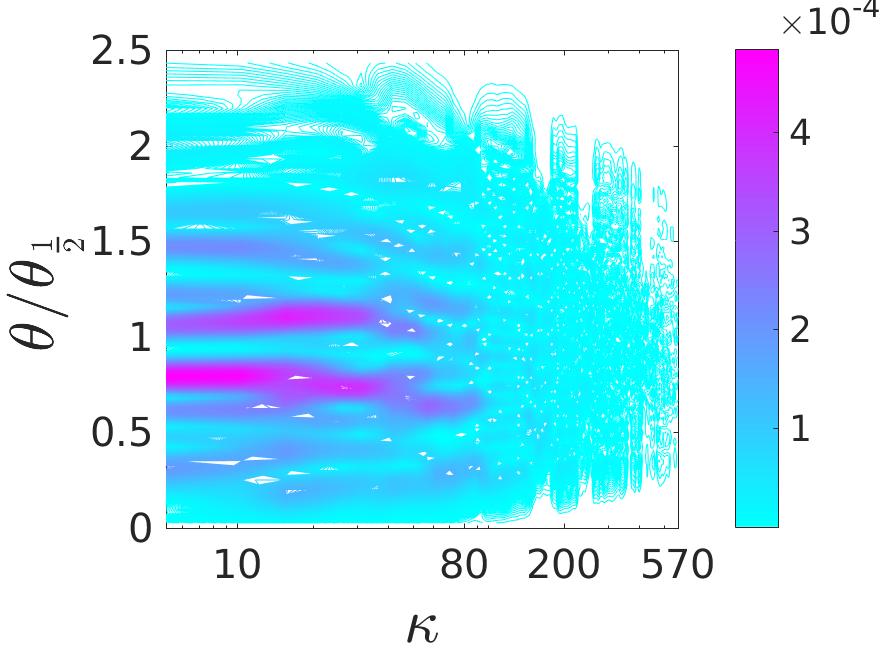}\label{fig:term_II_single_14}}
\subfloat[$\alpha=15$]{\includegraphics[width=0.3\linewidth]{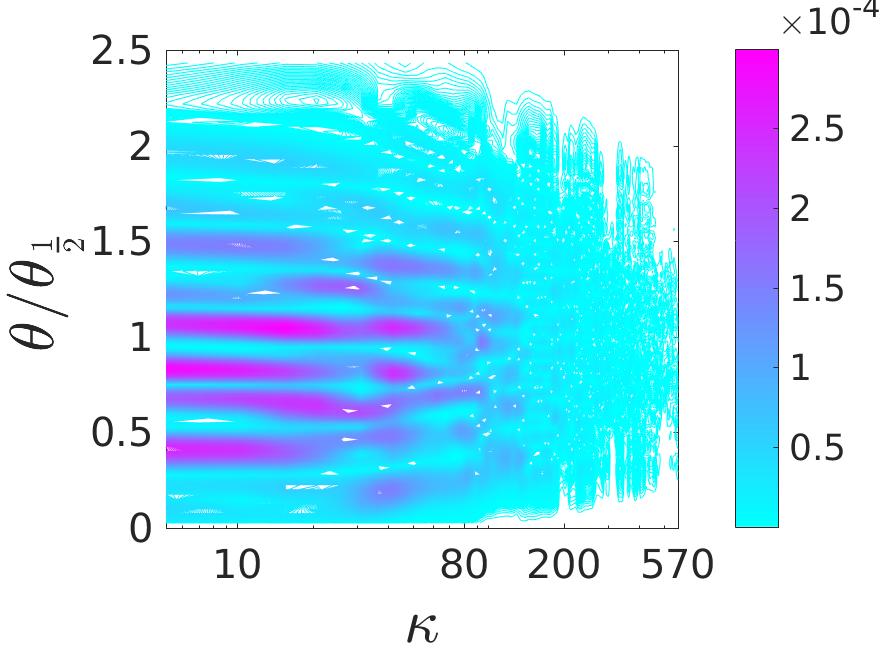}\label{fig:term_II_single_15}}
\caption{Modal components of the energy-normalized production (ENP), $\mathcal{P}_{\text{tot},\theta}$, for~${\alpha=1:15}$. \label{fig:term_II_single}}
\end{figure}
\FloatBarrier
\noindent
It is worth keeping in mind, however, that the LD is in fact itself a statistical decomposition, and the modes (eigenfunctions) represent component parts of a probability density function (pdf), which describes the dynamics of the entire field. The fact that the modes are the building blocks of this pdf is seen directly from the orthogonality criterion of \eqref{eq:orthogonality_eigenfunctions}. The criterion implies that the integrand of the $L^2_w(\Omega,\mathbb{C}^3)$-norm of any eigenfunction, $\overline{\Phi}_\alpha$, is a pdf. This is due to the fact that any integral equaling unity directly implies that the integrand can be perceived as a pdf. Since the energy spectra are reconstructed from these eigenfunctions, the components of the spectral reconstruction should be viewed as a more detailed dynamic representation of the spectrum, namely one that is weighted by the degree of correlation between SADFM coefficients across the span of the jet. 

The energy production term, $II$, is the only term in \eqref{eq:galerkin_energy_terms} for which the full spatial reconstruction is possible to perform from the current data. 
%
%
From $II$ the total averaged turbulent kinetic energy production rate ($\mathrm{W} = \mathrm{J/s}$) is obtained by
\begin{equation}
\mathcal{P}_{\text{tot}} = \int_\Omega\rho\mathcal{P} d\mu = 2\pi\rho\int_0^{\theta_{\text{max}}}\int_{\xi_1}^{\xi_2}\mathcal{P}\sqrt{Z}d\xi d\theta,\label{eq:production_total}
\end{equation}
where $\mathcal{P} = II$ in \eqref{eq:galerkin_production_term}. By scaling \eqref{eq:production_total} with the energy, $\rho\lambda$, one obtains the \textit{energy-normalized production} (ENP)
\begin{equation}
\mathcal{P}_{\rho\lambda}=2\pi\int_0^{\theta_{\text{max}}}\int_{\xi_1}^{\xi_2}\varphi_i^\alpha\varphi^{j*}_\alpha\nabla_j\left\langle V^i\right\rangle \sqrt{Z}d\xi d\theta,\label{eq:PROD_integrated}
\end{equation}
which is the relative energy production distribution in terms of LD mode number and wavenumber with units $\mathrm{W/J = s^{-1}}$. The reduced forms of \eqref{eq:production_total} and \eqref{eq:PROD_integrated} can be found in Appendix \ref{app:galerkin_of_II}. Considered separately for each $\alpha-\kappa$-combination \eqref{eq:PROD_integrated} represents the amount of turbulence kinetic energy produced relative to the energy contained by a given eigenfunction. It therefore represents the ENP related to each mode. Since \eqref{eq:PROD_integrated} is complex-valued the real-, and imaginary parts, the arguments and the absolute values of these can be computed. These are shown in figures \ref{fig:E_prop_real}, \ref{fig:E_prop_imag}, \ref{fig:E_prop_arg}, and \ref{fig:E_prop_abs}, respectively. It is seen from \eqref{eq:PROD_relative} that any imaginary parts are directly affiliated with energy production due to shear-stresses, as the production related to normal stresses is entirely real - for each $\alpha$ we have that $\varphi^{i\alpha}\varphi^{*}_{j\alpha} =|\varphi^\alpha_i|^2$ for $i=j$. The real parts in figure \ref{fig:E_prop_real} therefore must contain the aggregate contributions from both shear- and normal stresses (the normal stresses were shown in \cite{Hodzic2019_part1} to have a non-negligible contribution to the total production), while the imaginary parts exclusively contain the information related to shear stresses. Figure \ref{fig:E_prop_abs} shows that the energy production is concentrated around low wave- and mode numbers. In the most energy productive region of the spectrum the production is characterized by real modes. This is seen from $Arg\left(\mathcal{P}_{\rho\lambda}\right)$ in figure \ref{fig:E_prop_arg}, since these are negligible for the corresponding $\alpha$-$m$ combinations. It is profound to note that ENP levels are nearly constant over a relatively wide range of $\alpha$ and $\kappa$, as hypothesized by \cite{Wanstrom2009}. Figure \ref{fig:E_prop} therefore quantifies which modes produce TKE in proportion to their energy content. For the individual modes, e.g. $\alpha=1$ which exhibits the $-5/3$-slope for~${20<\kappa<300}$ (see figure \ref{fig:cumulative_spectra_vv_reconstructed_SSC} and also Appendix \ref{app:individual_modal_components_spectra} for additional individual contributions to the spectra), it is observed from figure \ref{fig:E_prop_abs} that the ENP values diminish for increasing $\kappa$. This indicates that modes related to higher wavenumbers play a more central part in the transport of energy than the modes related to lower $\kappa$. Nevertheless, significant ENP levels are observed even for higher $\kappa$-values in the $-5/3$-range. This means that even these modes obtain significant fractions of their energy directly from the mean flow and are thus less dependent on an energy cascade process from more energetic eigenfunctions, than what might initially be thought.
%
%
\begin{figure}[h]
\centering      
\subfloat[$\alpha=1$]{\includegraphics[width=0.3\linewidth]{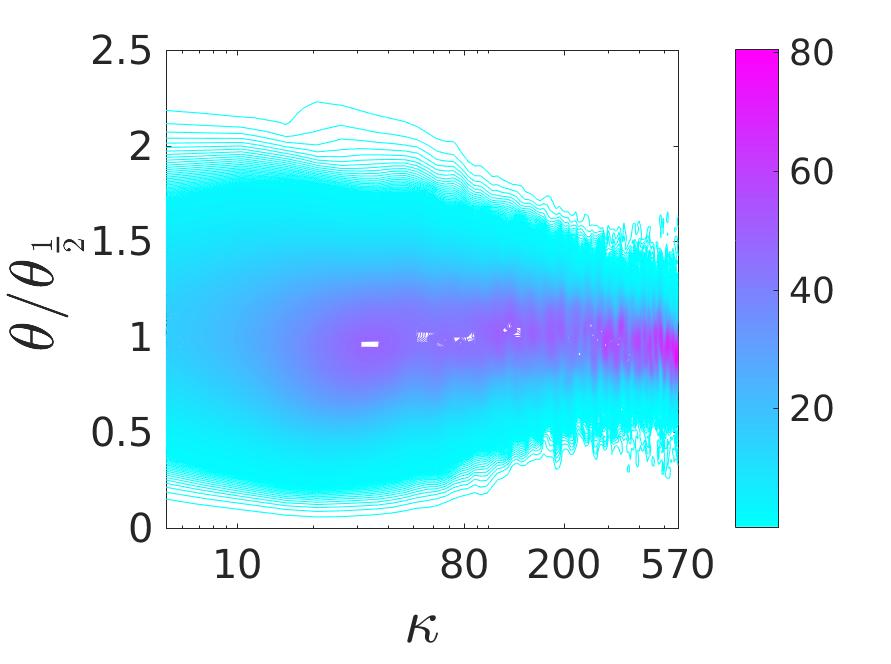}\label{fig:term_II_single_scaled_1}}
\subfloat[$\alpha=2$]{\includegraphics[width=0.3\linewidth]{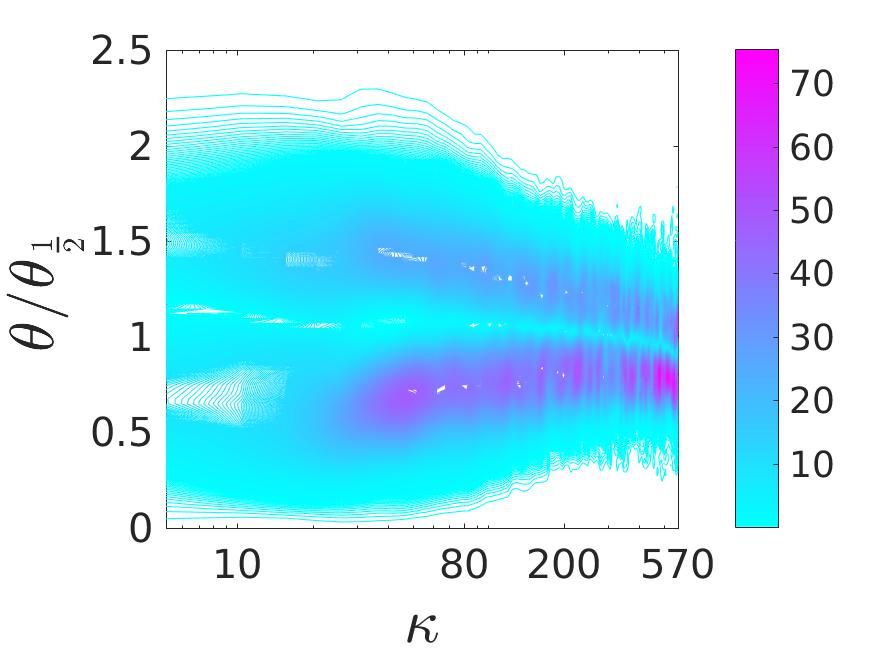}\label{fig:term_II_single_scaled_2}}
\subfloat[$\alpha=3$]{\includegraphics[width=0.3\linewidth]{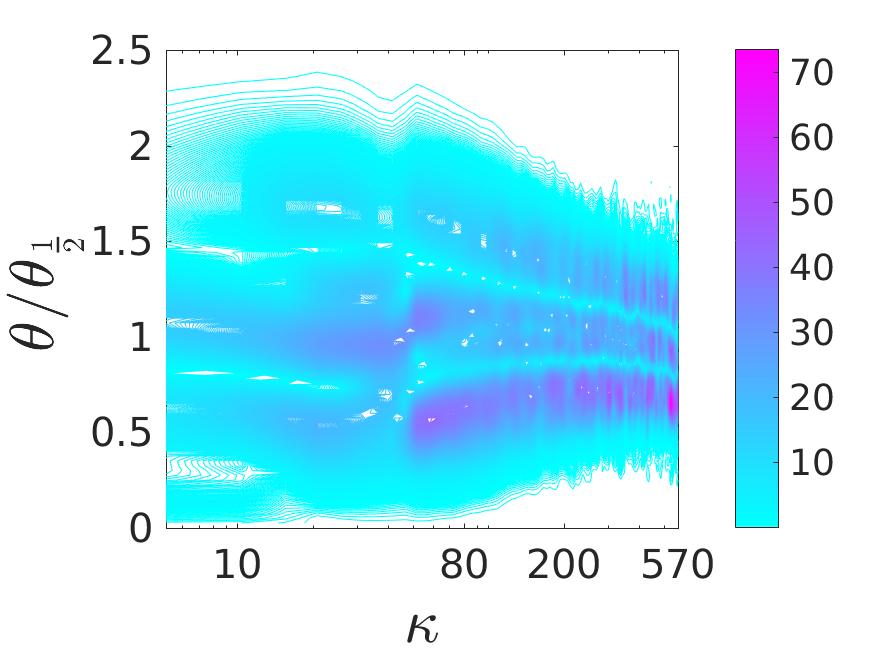}\label{fig:term_II_single_scaled_3}}\\
\subfloat[$\alpha=4$]{\includegraphics[width=0.3\linewidth]{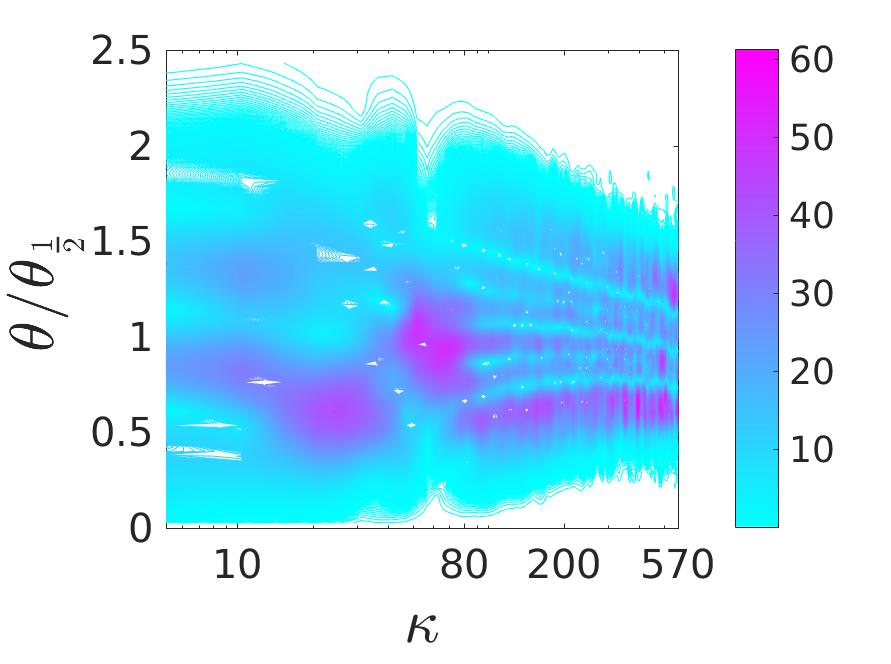}\label{fig:term_II_single_scaled_4}}
\subfloat[$\alpha=5$]{\includegraphics[width=0.3\linewidth]{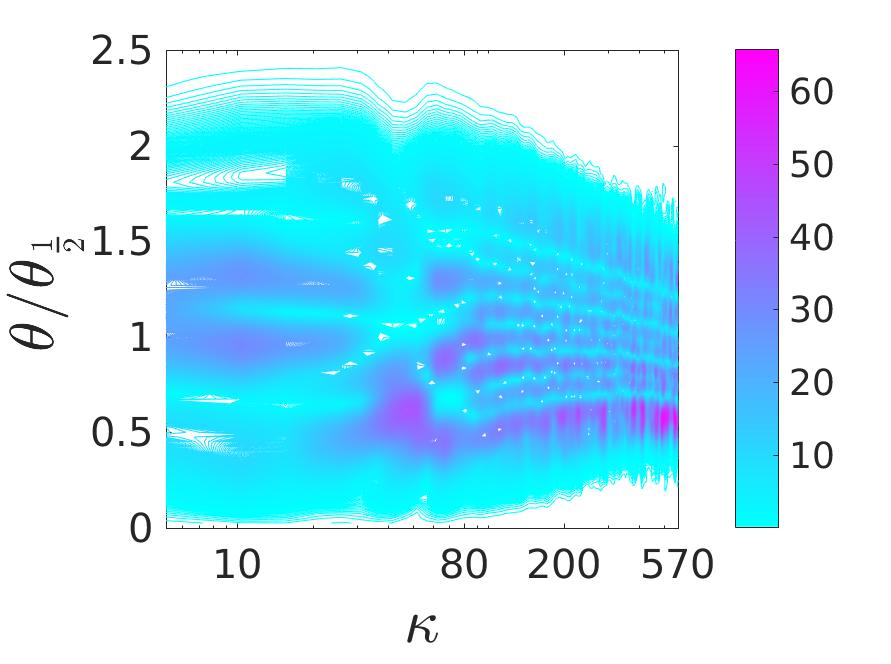}\label{fig:term_II_single_scaled_5}}
\subfloat[$\alpha=6$]{\includegraphics[width=0.3\linewidth]{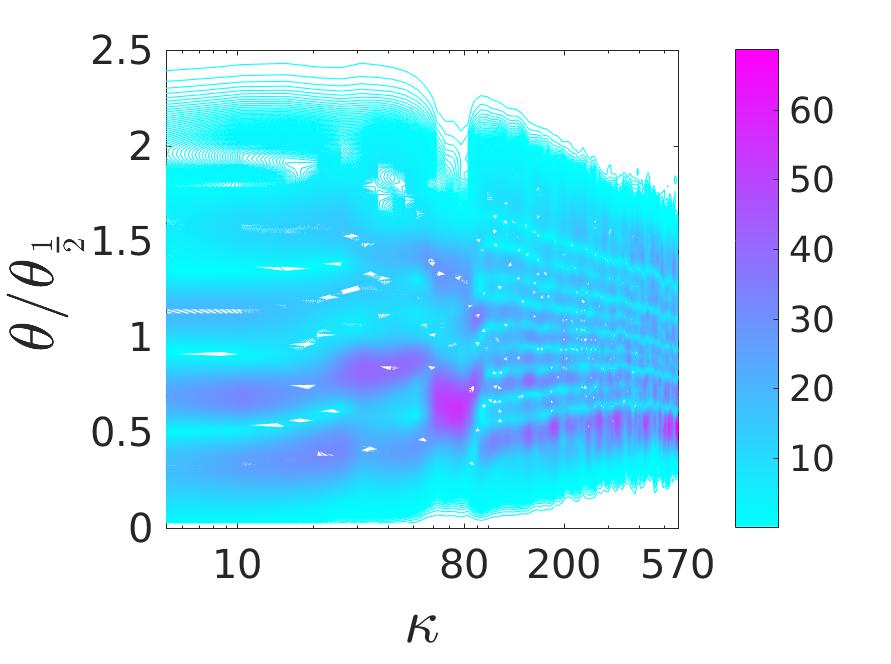}\label{fig:term_II_single_scaled_6}}\\
\subfloat[$\alpha=7$]{\includegraphics[width=0.3\linewidth]{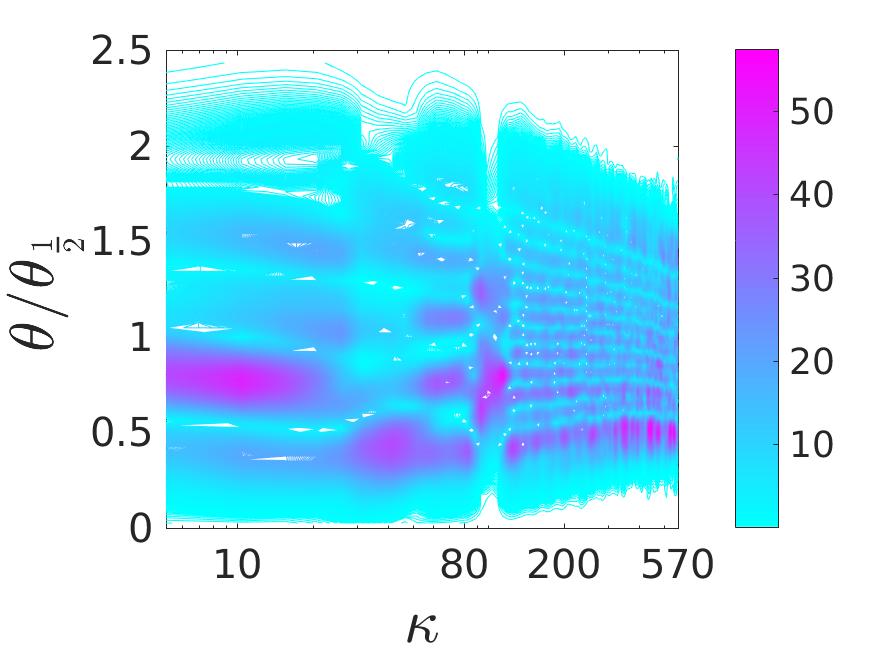}\label{fig:term_II_single_scaled_7}}
\subfloat[$\alpha=8$]{\includegraphics[width=0.3\linewidth]{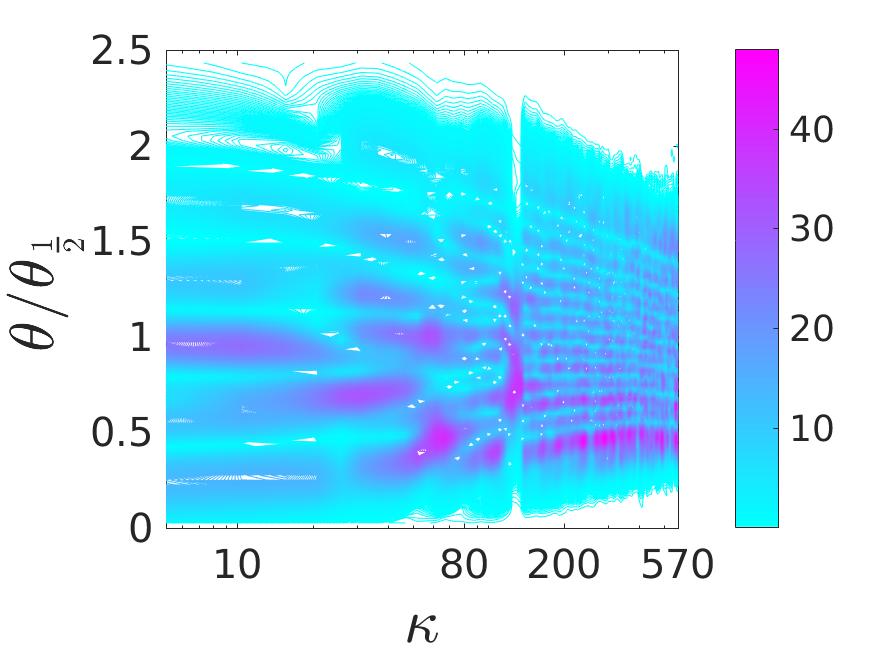}\label{fig:term_II_single_scaled_8}}
\subfloat[$\alpha=9$]{\includegraphics[width=0.3\linewidth]{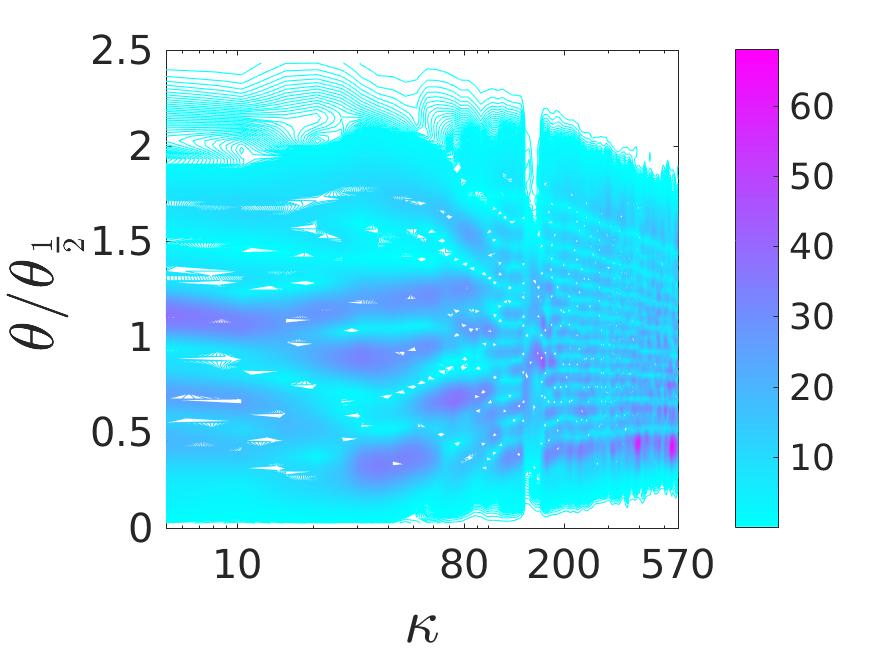}\label{fig:term_II_single_scaled_9}}\\
\subfloat[$\alpha=10$]{\includegraphics[width=0.3\linewidth]{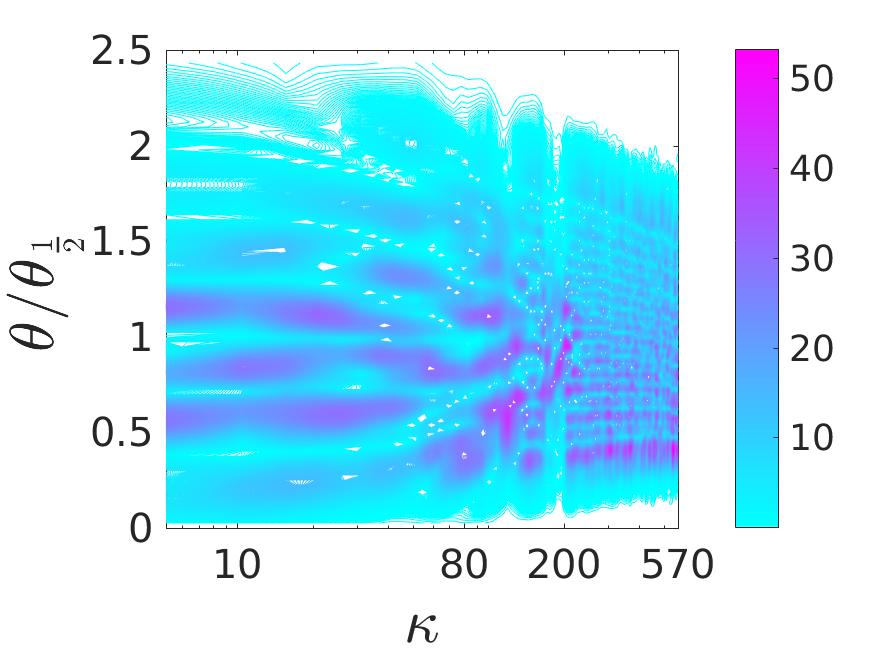}\label{fig:term_II_single_scaled_10}}
\subfloat[$\alpha=11$]{\includegraphics[width=0.3\linewidth]{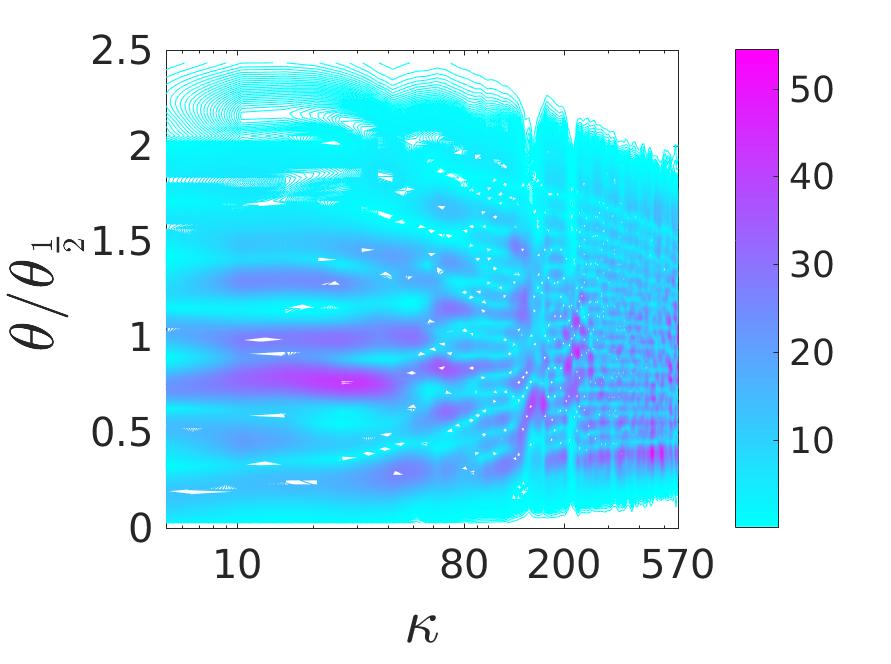}\label{fig:term_II_single_scaled_11}}
\subfloat[$\alpha=12$]{\includegraphics[width=0.3\linewidth]{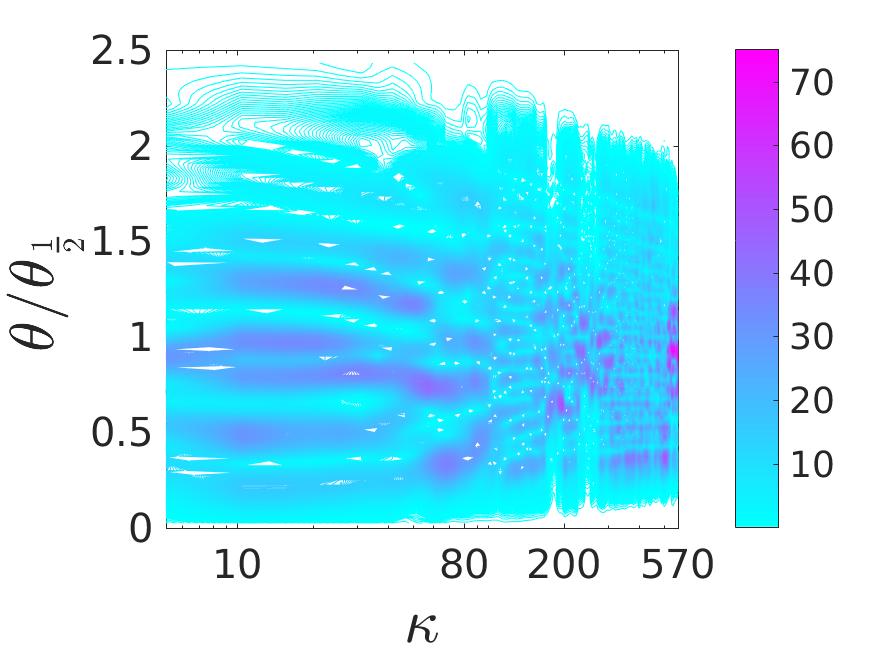}\label{fig:term_II_single_scaled_12}}\\
\subfloat[$\alpha=13$]{\includegraphics[width=0.3\linewidth]{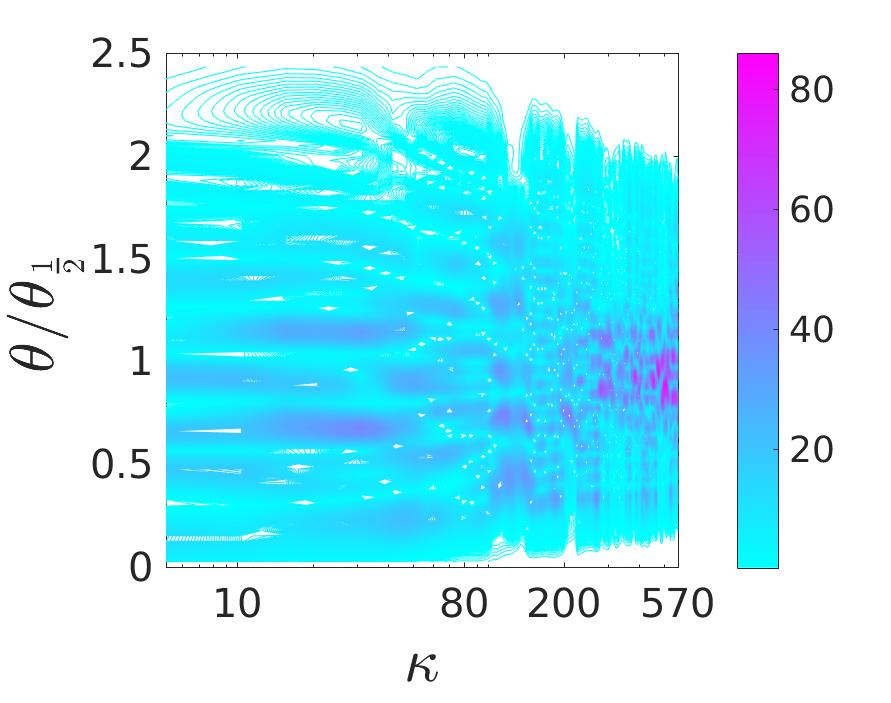}\label{fig:term_II_single_scaled_13}}
\subfloat[$\alpha=14$]{\includegraphics[width=0.3\linewidth]{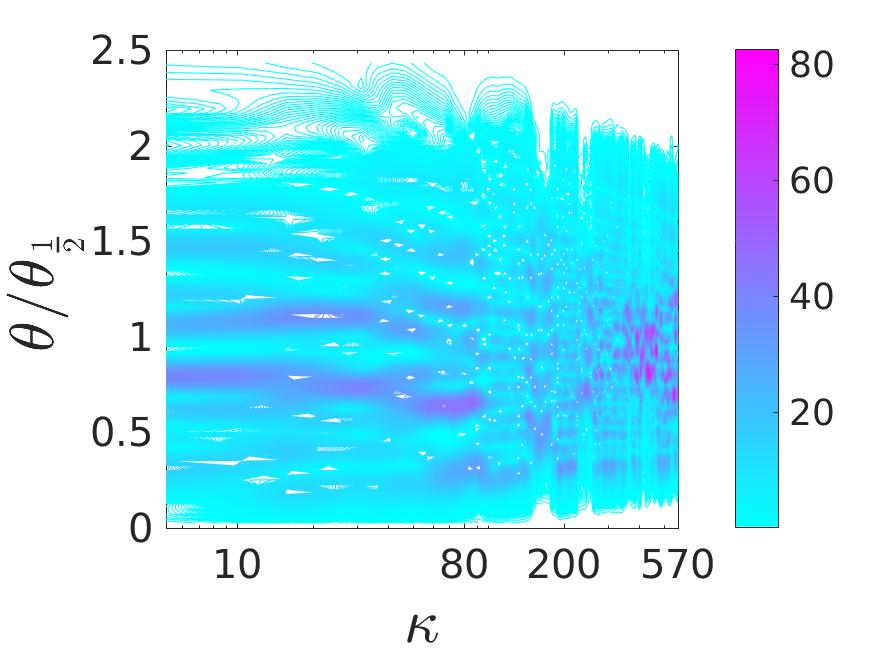}\label{fig:term_II_single_scaled_14}}
\subfloat[$\alpha=15$]{\includegraphics[width=0.3\linewidth]{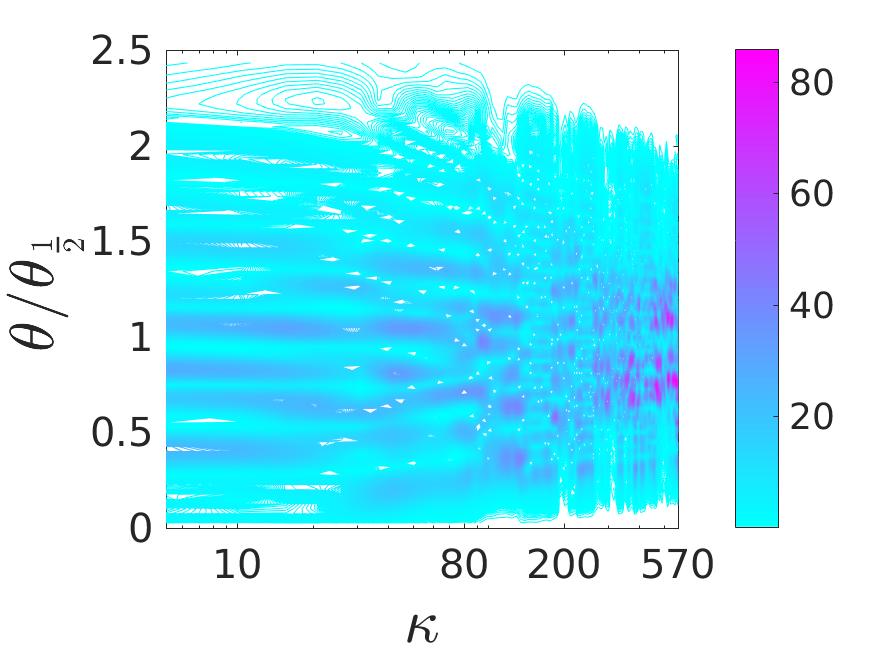}\label{fig:term_II_single_scaled_15}}
\caption{Eigenvalue-normalized modal components, $\mathcal{P}_{\rho\lambda,\theta}$, for $\alpha=1:15$ as a function of wavenumber and span of the jet. These illustrate the turbulence energy production capacity as a function of $\theta$ and $\kappa$ for each LD mode, relative to the local eigenvalue. \label{fig:term_II_single_scaled}}
\end{figure}
\FloatBarrier
\noindent

The total energy production contribution of each mode as a function of $\theta$ is obtained by integrating out the $\xi$- and $\phi$-dependencies of $\rho\mathcal{P}$
\begin{equation}
\mathcal{P}_{\text{tot},\theta} = 2\pi\rho\int_{\xi_1}^{\xi_2}\mathcal{P}\sqrt{Z}d\xi,\label{eq:production_total_theta}
\end{equation}
leaving the dependence on $\theta$. Figure \ref{fig:term_II_single} shows the evaluation of \eqref{eq:production_total_theta}. Figure \ref{fig:term_II_single} exposes the modal contributions to the total energy production as a function of $\theta$. It is seen that most of the absolute energy is produced below $\kappa=100$ around the high mean shear region ($\theta/\theta_{\frac{1}{2}}\approx 1$) at low mode numbers.

Since \eqref{eq:PROD_integrated} is integrated over the spatial domain, it means that regions in the jet which are characterized by negligible mean shear weigh in on the results in figure \ref{fig:E_prop} with insignificant energy production levels - effectively reducing the total ENP estimate. This means that local regions in the integration domain characterized by much higher ENP levels may exist. This can be analyzed by evaluating the relative energy production in terms of mode number, wavenumber, and $\theta$, which is obtained by scaling \eqref{eq:production_total} by the energy, $\rho\lambda$, and integrating out the $\xi$- and $\phi$-dependencies
\begin{equation}
\mathcal{P}_{\rho\lambda,\theta}=2\pi\int_{\xi_1}^{\xi_2}\varphi_i^\alpha\varphi^{j*}_\alpha\nabla_j\left\langle V^i\right\rangle \sqrt{Z}d\xi.\label{eq:PROD_relative}
\end{equation}
Figure \ref{fig:term_II_single_scaled} shows the eigenvalue normalized energy productions, defined by \eqref{eq:PROD_relative}. The evaluation of \eqref{eq:PROD_relative} for $\alpha=1$ is shown in figure \ref{fig:term_II_single_scaled_1}. It is seen that $\alpha=1$ retains most of its ENP for $\theta/\theta_{\frac{1}{2}}=1$, throughout the entire $\kappa$-range shown in figure \ref{fig:term_II_single_scaled}. Upon closer inspection we note that the variations from $\kappa=26$ to $\kappa=300$ have a standard deviation corresponding to about $8.5\%$. We recall that the $-5/3$-region was identified to be in $\kappa\in[20:300]$. Similar tendencies are seen for the remaining modes in figure \ref{fig:term_II_single_scaled}, indicating that this indeed is a general trend for all modes.

Since the half-width region is characterized by high mean shear it is not surprising that even higher modes affiliated to higher $\kappa$-values are able to produce significant levels of their own energy content. This is closely related to the earlier discussions regarding the $-7/3$-region of the cross-spectra and their relation to the production of shear-stresses. We recall that the cross-spectra in the high-shear region were recovered from the first mode alone. From the relatively constant ENP levels observed over a wide range of wavenumbers (including the $-5/3$-region) in figure \ref{fig:term_II_single_scaled_1}, we see that a wide range of modes obtain significant portions of their energy directly from the mean flow - especially in the vicinity of high mean shear regions, namely $\theta/\theta_{\frac{1}{2}}=0.5-1.5$. Outside these $\theta$-ranges energy transfer between LD modes becomes increasingly important in order to facilitate the energy transfer across wavenumbers and eventually down to the dissipative scales. This follows directly from the diminishing ENP for large $\alpha$ and $\kappa$ observed from the data.

We note that this - at least locally - contradicts the classical Kolmogorov theory. The latter states that the $-5/3$-range is characterized exclusively by local energy transfers across adjecent scales, in the spirit of the Richardson cascade, and thereby very little energy is produced. The exclusive energy transfer in the assumed inertial subrange can - based on these results - only potentially hold true in terms of the absolute energy contributions of the individual modes in figure \ref{fig:term_II_single}. Here it is seen that the total turbulence energy production is limited in comparison to low mode and wavenumber regions. However, the \textit{relative} production levels in figure \ref{fig:term_II_single_scaled_1} are shown to be significant especially in high shear regions. This shows that energy is in fact introduced over a wide range of scales and this process is not confined to the low wavenumber range - the so-called energy production range.
\FloatBarrier
\subsection{Non-linear energy transport analysis}
Having concluded that multiple modes produce significant levels of TKE, we are interested in analyzing the non-linear energy transfer within- and across modes, which has proved to be more significant for higher $\kappa$-values from earlier results. 
\begin{figure}[h]
\centering      
\subfloat[]{\includegraphics[width=0.4\linewidth]{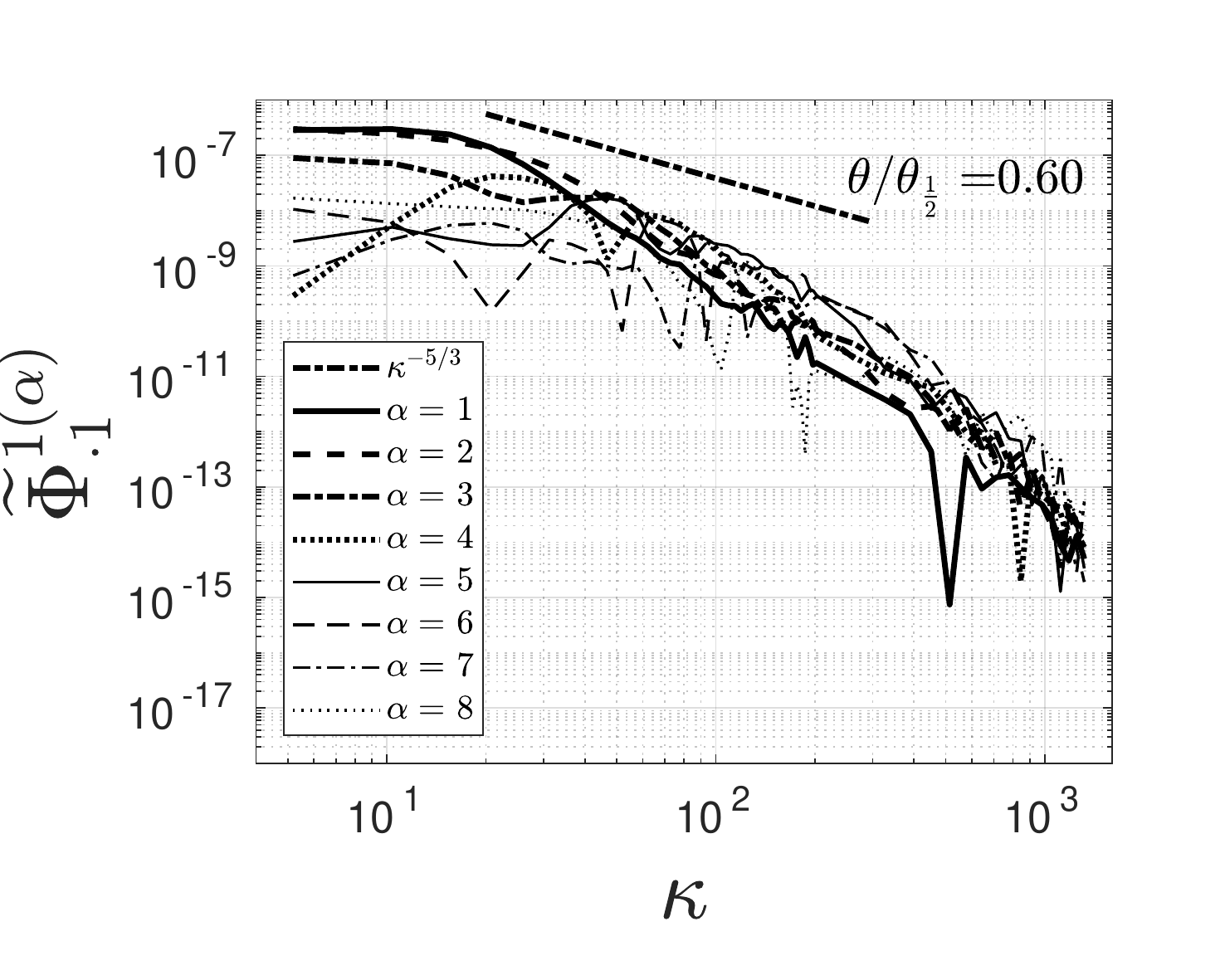}\label{fig:single_spectra_uu_60}}
\subfloat[]{\includegraphics[width=0.4\linewidth]{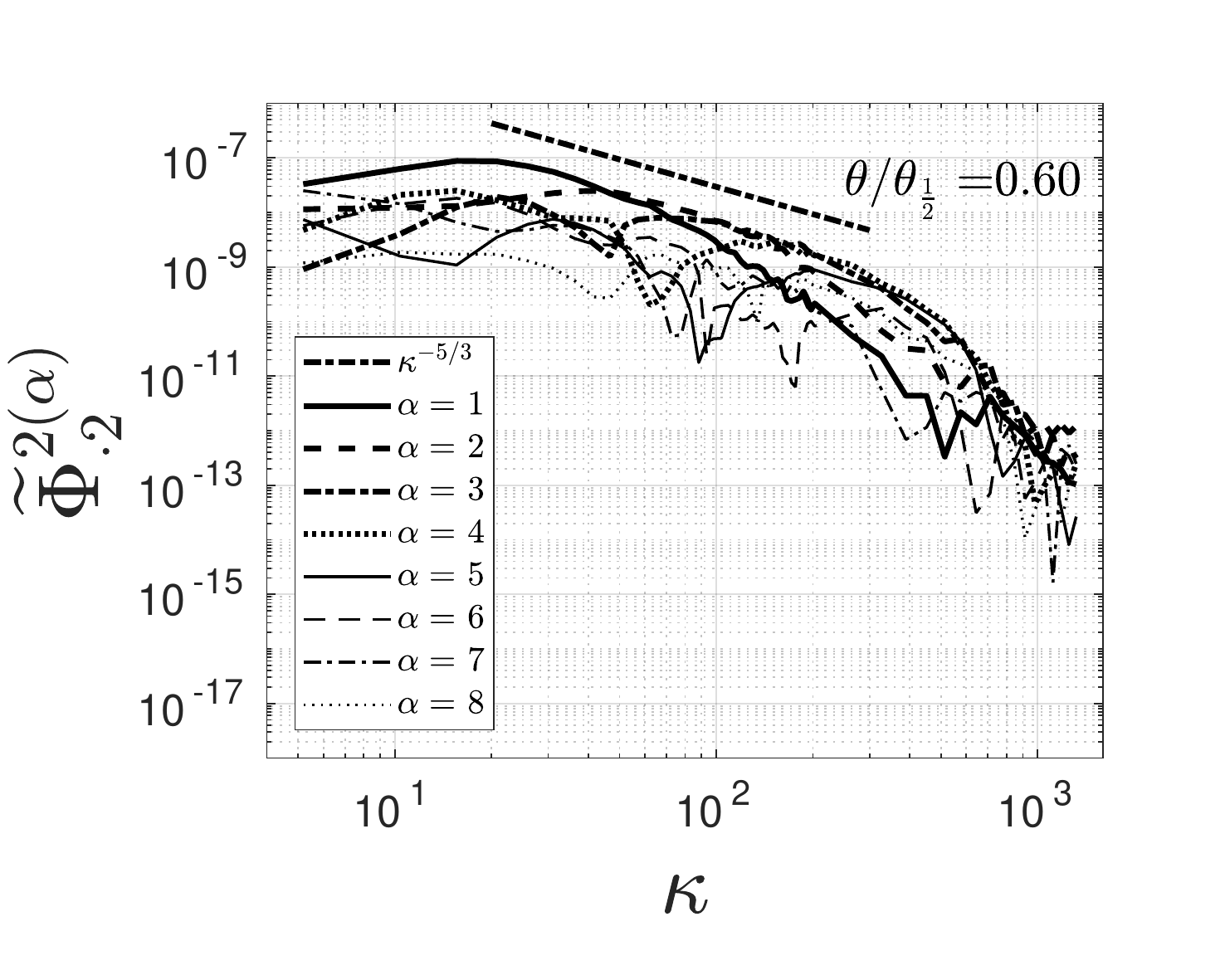}\label{fig:single_spectra_vv_60}}\\
\subfloat[]{\includegraphics[width=0.4\linewidth]{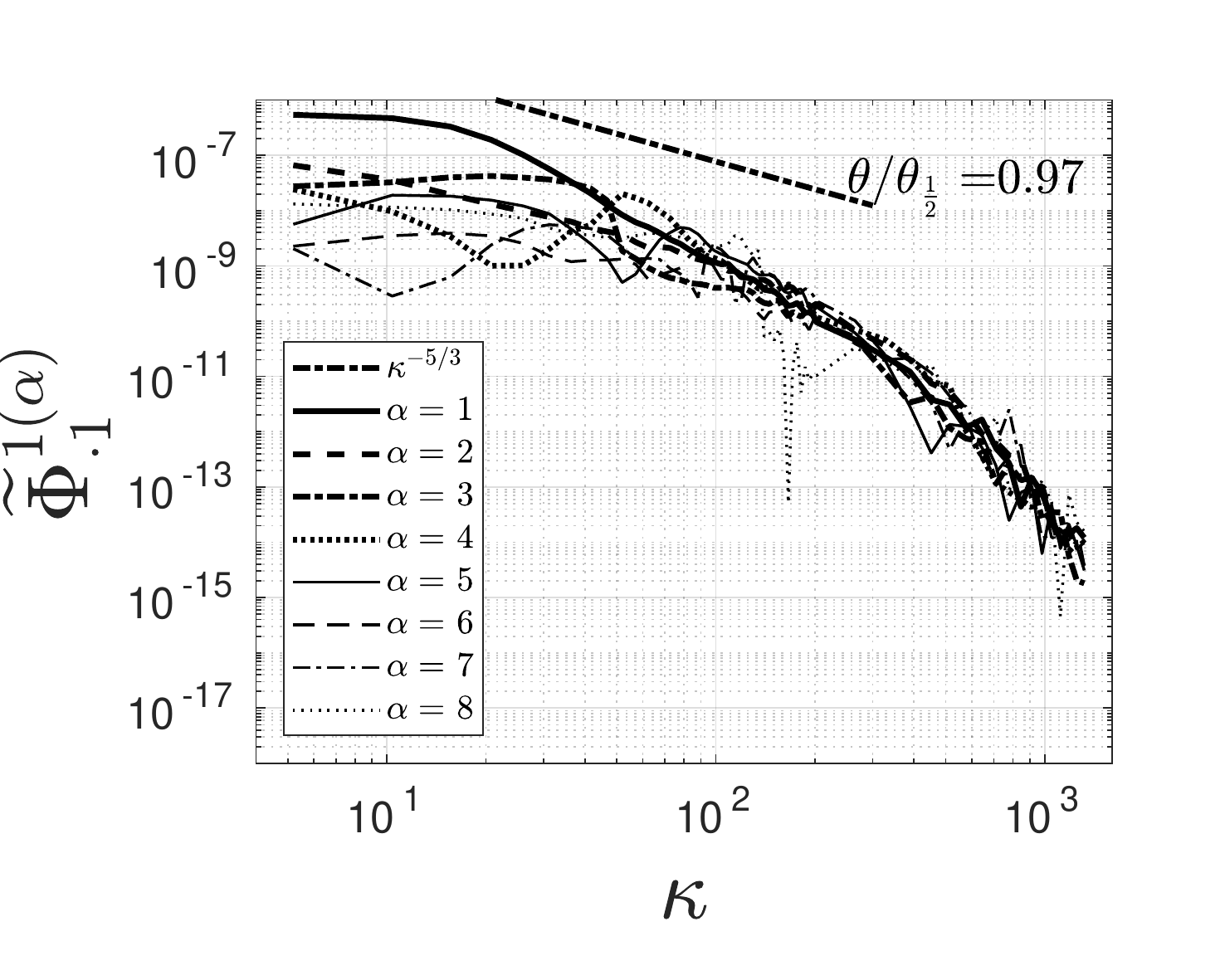}\label{fig:single_spectra_uu_97}}
\subfloat[]{\includegraphics[width=0.4\linewidth]{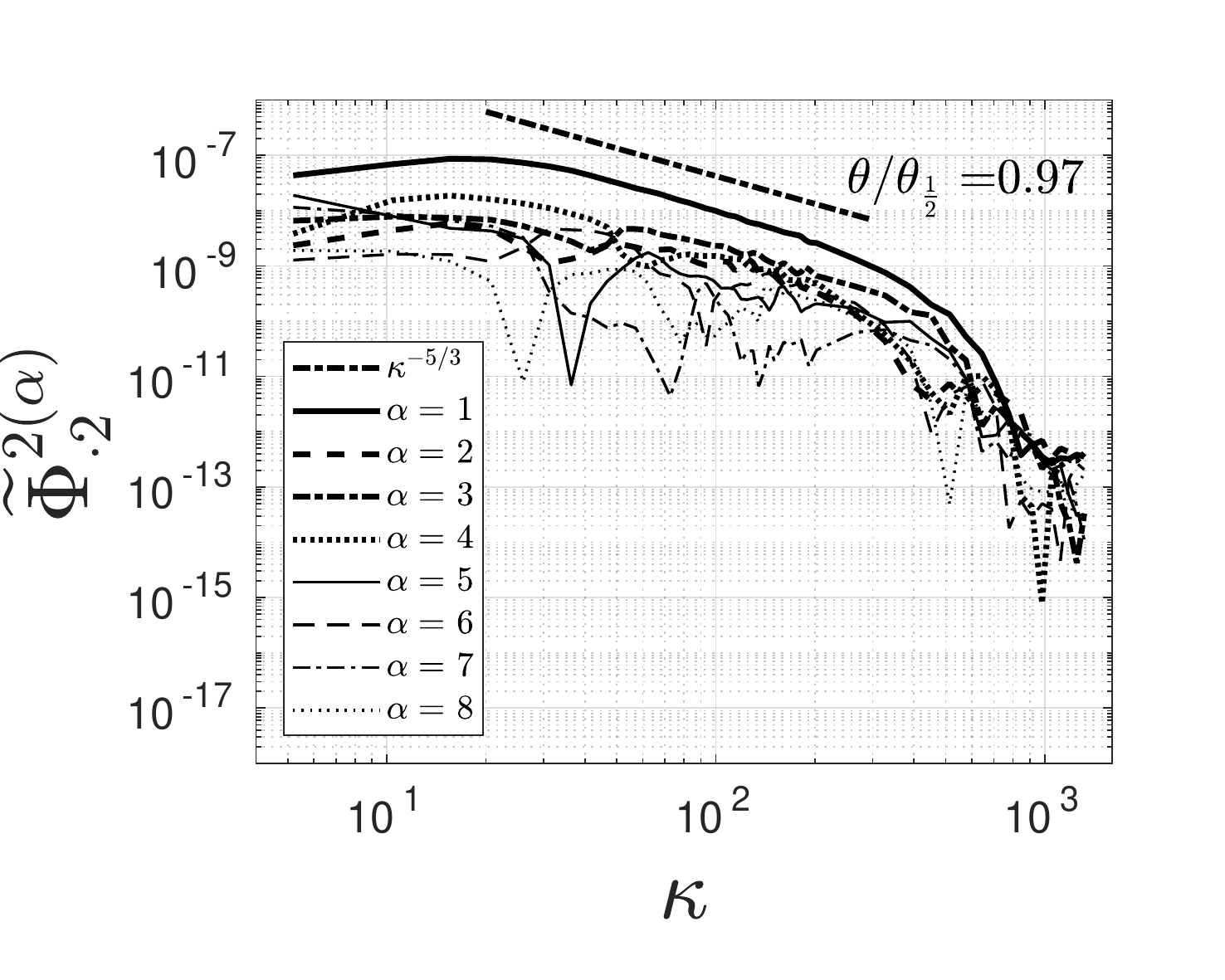}\label{fig:single_spectra_vv_97}}
\caption{Individual modal components of single-point spatial spectra, $\widetilde{\Phi}^{1(\alpha)}_{\cdot,1}$, and $\widetilde{\Phi}^{2(\alpha)}_{\cdot,2}$, ${n=1:8}$ at~${\theta/\theta_{\frac{1}{2}}=0.60}$ and ~${\theta/\theta_{\frac{1}{2}}=0.97}$. Multiple modes reconstruct the $-5/3$-slope for $\widetilde{\Phi}^2_{\cdot 2}$, wheras no modes exhibit the $-5/3$-slope in the reconstruction of $\widetilde{\Phi}^{1(\alpha)}_{\cdot,1}$. \label{fig:single_spectra_reconstructed_SSC}}
\end{figure}
\noindent
With this in mind, the variation of modal components of the spectra seen in figure \ref{fig:single_spectra_reconstructed_SSC} (see also figures \ref{fig:app_single_spectra_uu_reconstructed_SSC} and \ref{fig:app_single_spectra_vv_reconstructed_SSC} in Appendix \ref{app:individual_modal_components_spectra})  can be analyzed in terms of the spectral energy flux. From dimensional grounds, we can infer that the energy spectrum must exhibit a $-5/3$-range on average in regions where global (in terms of wavenumber) production and dissipation can be deemed insignificant. From this follows that the one-dimensional spectra can be modeled as, \cite{Tennekes1972}
\begin{eqnarray}
\widetilde{\Phi}^1_{\cdot 1} &=& \alpha\frac{9}{55}\epsilon_\kappa^{2/3}\kappa^{-5/3},\label{eq:Phi_11-model}\\
\widetilde{\Phi}^2_{\cdot 2} &=& \alpha\frac{12}{55}\epsilon_\kappa^{2/3}\kappa^{-5/3}\label{eq:Phi_22-model},
\end{eqnarray}
in the case of isotropic turbulence, where $\epsilon_\kappa$ is the net energy spectral flux across a given wavenumber, $\kappa$. It is worth noting that \eqref{eq:Phi_11-model} and \eqref{eq:Phi_22-model} assume that only the local spectral flux across wavenumbers is of significance, since the underlying dimensional analysis of the expressions is based on assumptions of insignificant energy production and dissipation in this range. Note that the locality of the spectral flux is indicated by the subscripted $\kappa$ and is therefore an important deviation from the traditional assumption of statistical equilibrium of the turbulence, for which the spectral flux is assumed to be constant across an inertial subrange. 
\begin{figure}[h]
\centering      
\subfloat[]{\includegraphics[width=0.4\linewidth]{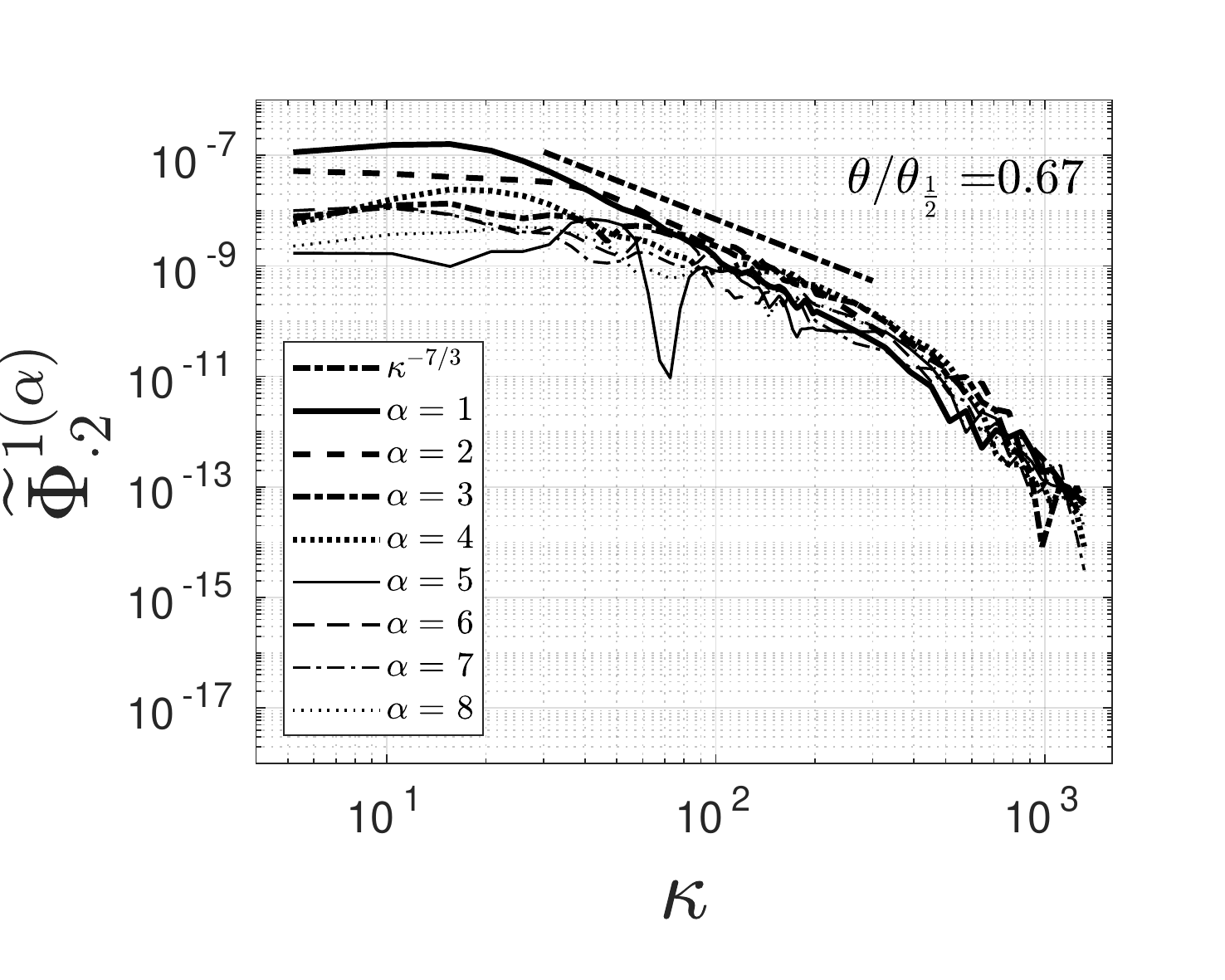}\label{fig:single_spectra_uv_67}}
\subfloat[]{\includegraphics[width=0.4\linewidth]{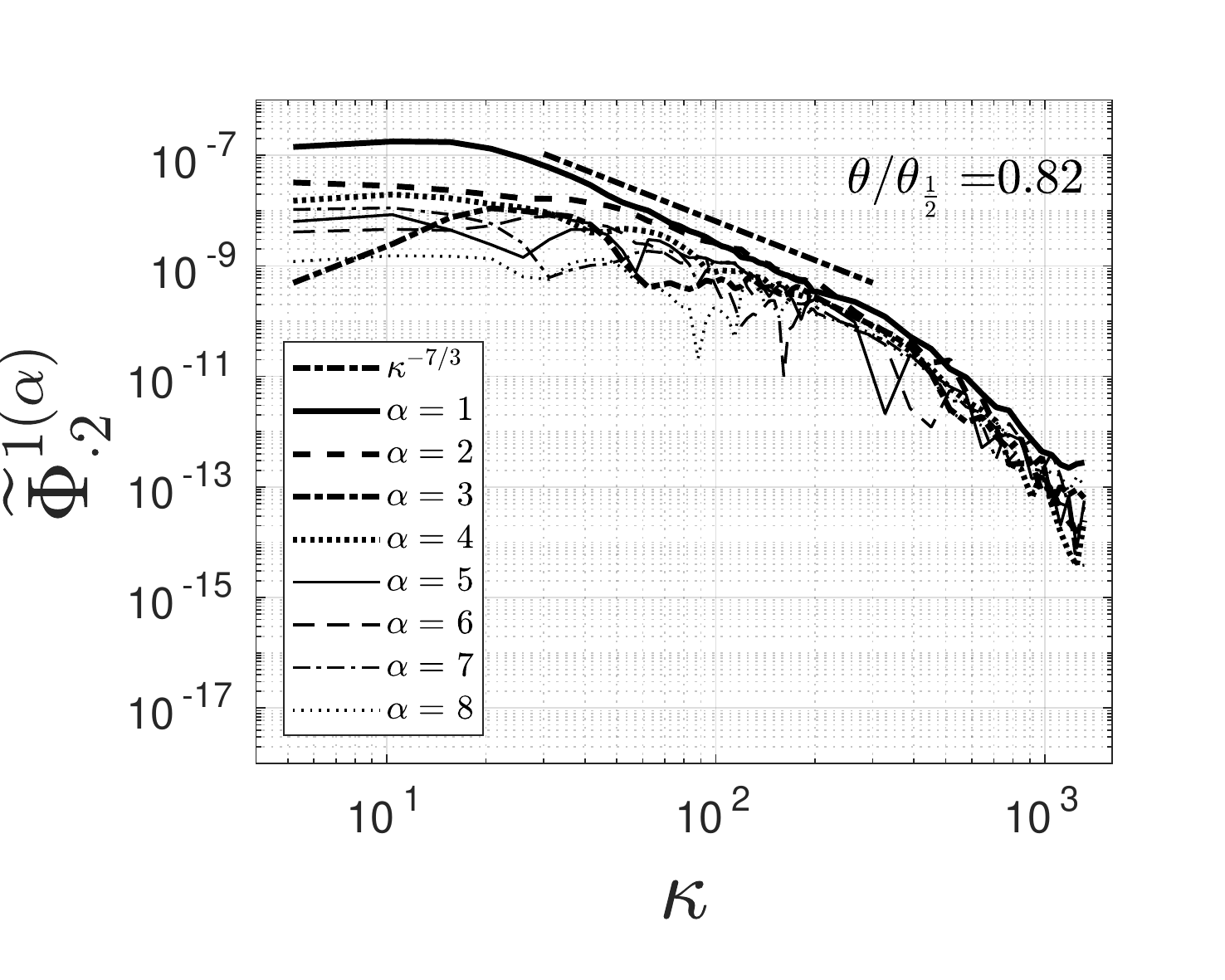}\label{fig:single_spectra_uv_82}}\\
\subfloat[]{\includegraphics[width=0.4\linewidth]{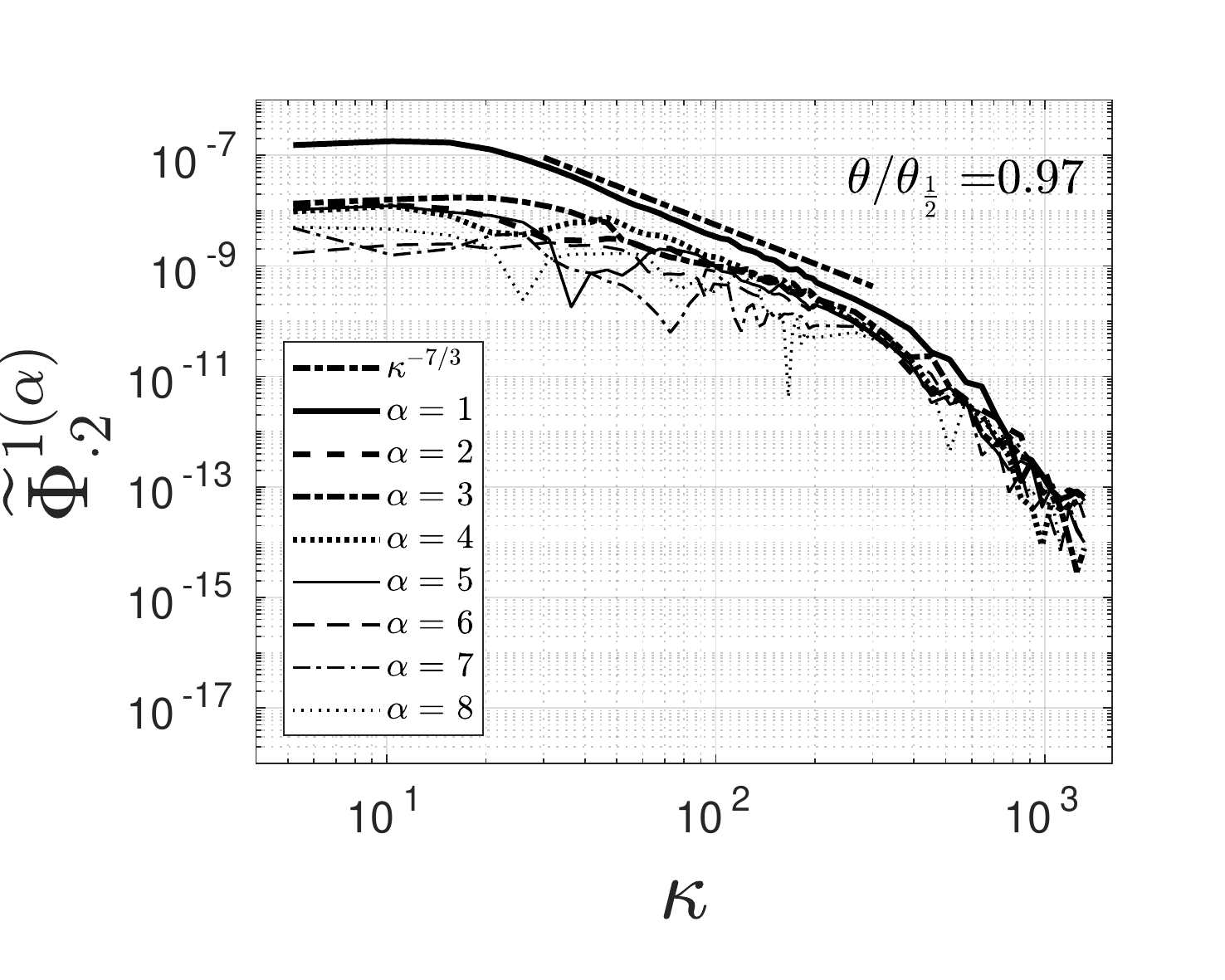}\label{fig:single_spectra_uv_97}}
\subfloat[]{\includegraphics[width=0.4\linewidth]{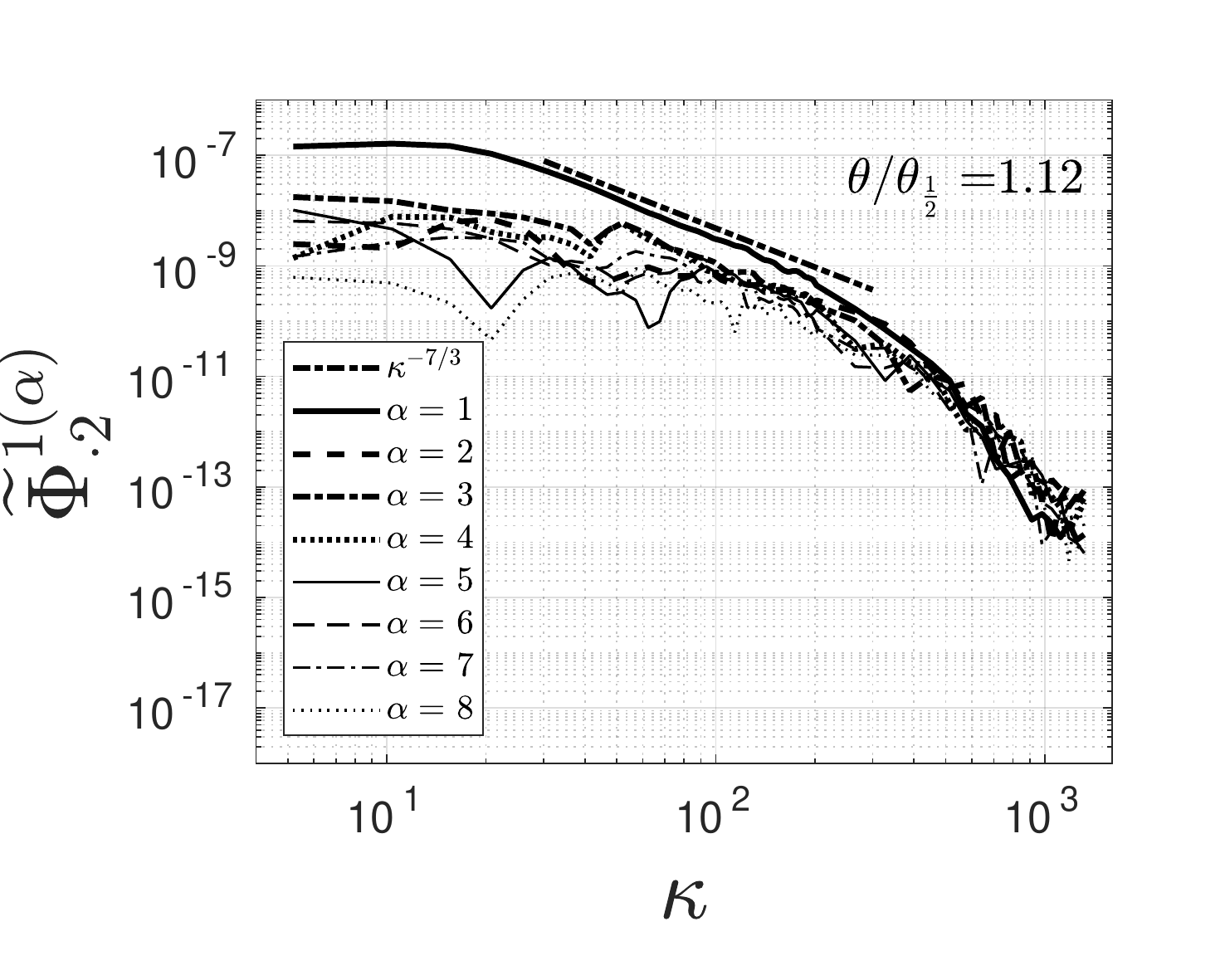}\label{fig:single_spectra_uv_112}}
\caption{Individual modal components of cross-spectra, $\widetilde{\Phi}^{1(\alpha)}_{\cdot,2}$, ${n=1:8}$ at~${\theta/\theta_{\frac{1}{2}}=[0.60,0.82,0.97,1.12]}$. Modes $n=1,2,3$ of the cross-spectrum reconstruction exhibit the $-7/3$-slope. \label{fig:single_spectra_reconstructed_SSC}}
\end{figure}
\noindent

Since the wavenumber gradient of the component energy spectra, \eqref{eq:Phi_11-model} and \eqref{eq:Phi_22-model}, can only deviate from $-5/3$ if $\epsilon_\kappa$ is dependent on the wavenumber the individual modal contributions to the reconstructions of $\widetilde{\Phi}^1_{\cdot 1}$ and $\widetilde{\Phi}^2_{\cdot 2}$ in figures \ref{fig:single_spectra_reconstructed_SSC} reveal that most modal contributions are in fact characterized by varying spectral fluxes. This is seen from \eqref{eq:Phi_11-model} and \eqref{eq:Phi_22-model} where only the local spectral flux, $\epsilon_\kappa$, remains as a variable. 
\begin{figure}[h]
    \centering      
    \subfloat[]{\includegraphics[width=0.40\linewidth]{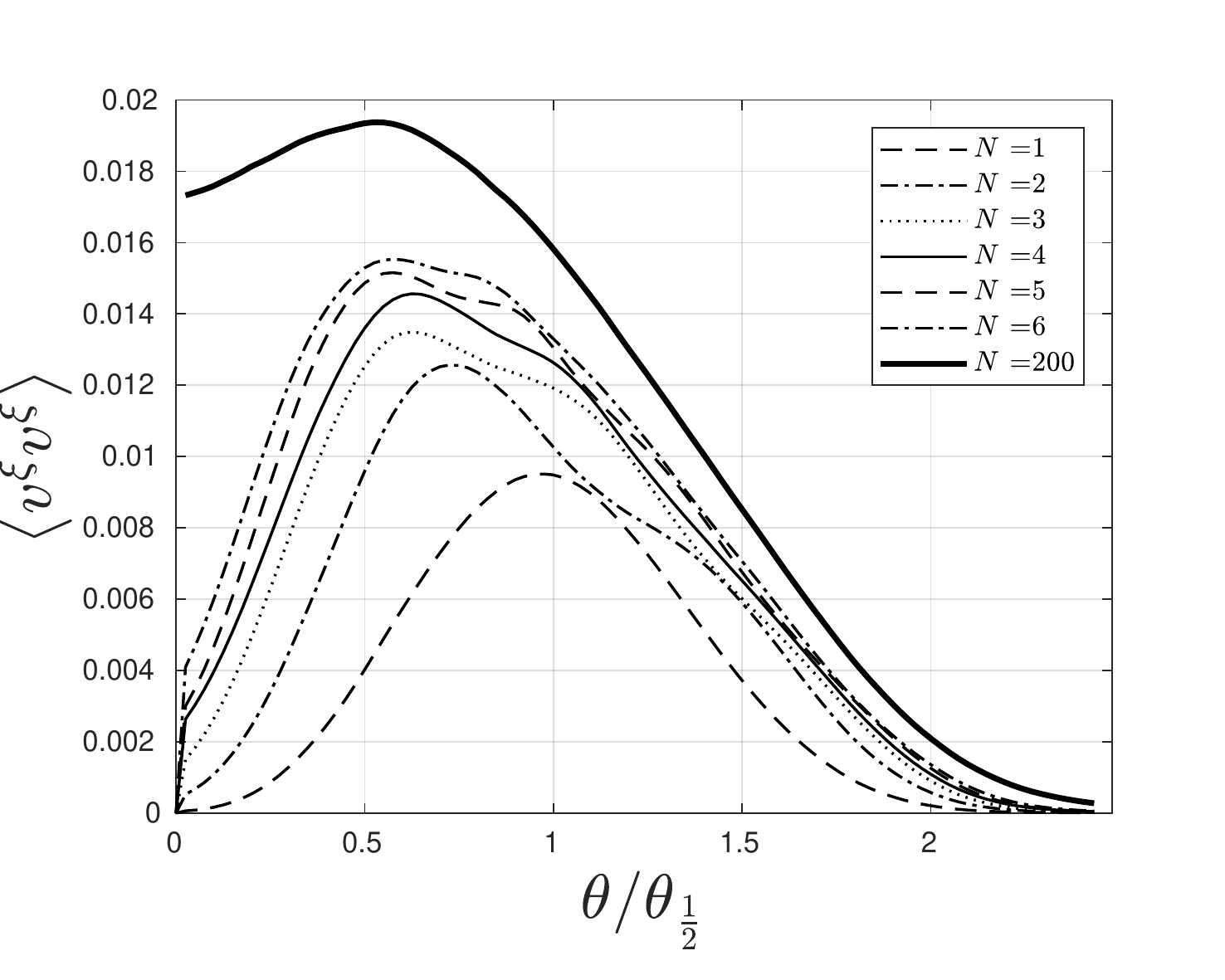}\label{fig:v_xi_v_xi_SSC}}
    \subfloat[]{\includegraphics[width=0.40\linewidth]{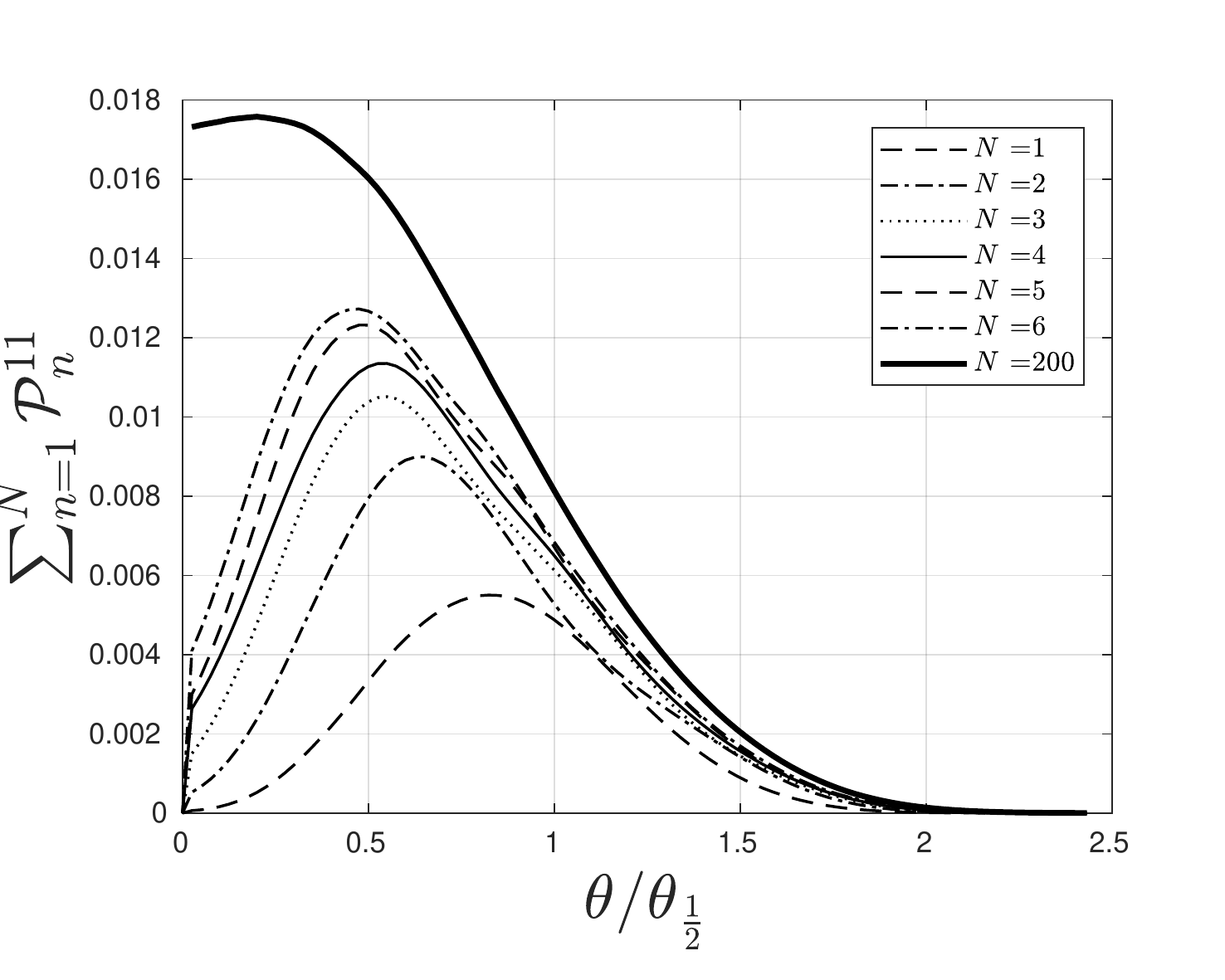}\label{fig:recon_production_uu_SSC}}\\
    \subfloat[]{\includegraphics[width=0.40\linewidth]
    {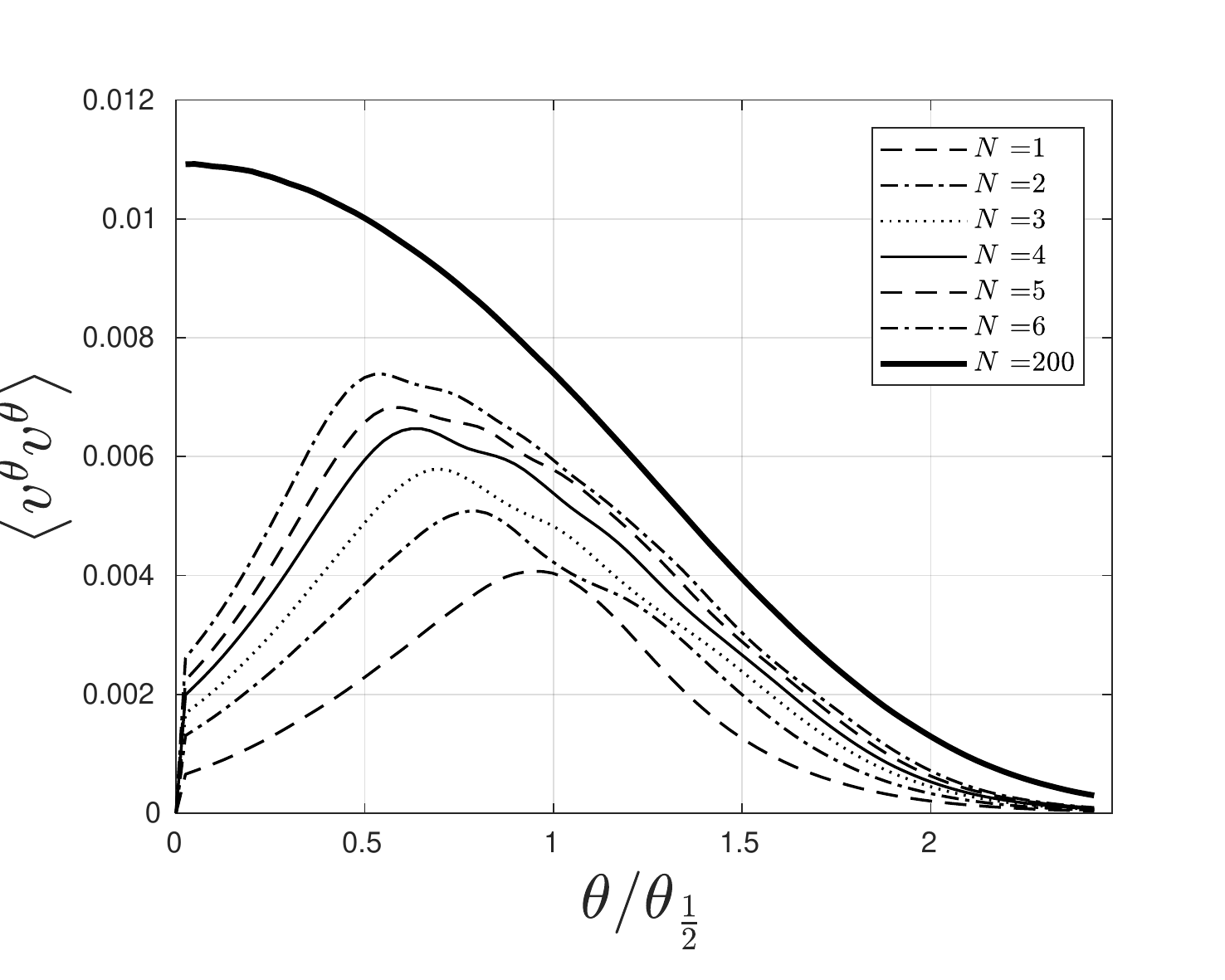}\label{fig:v_theta_v_theta_SSC}}
    \subfloat[]{\includegraphics[width=0.40\linewidth]
    {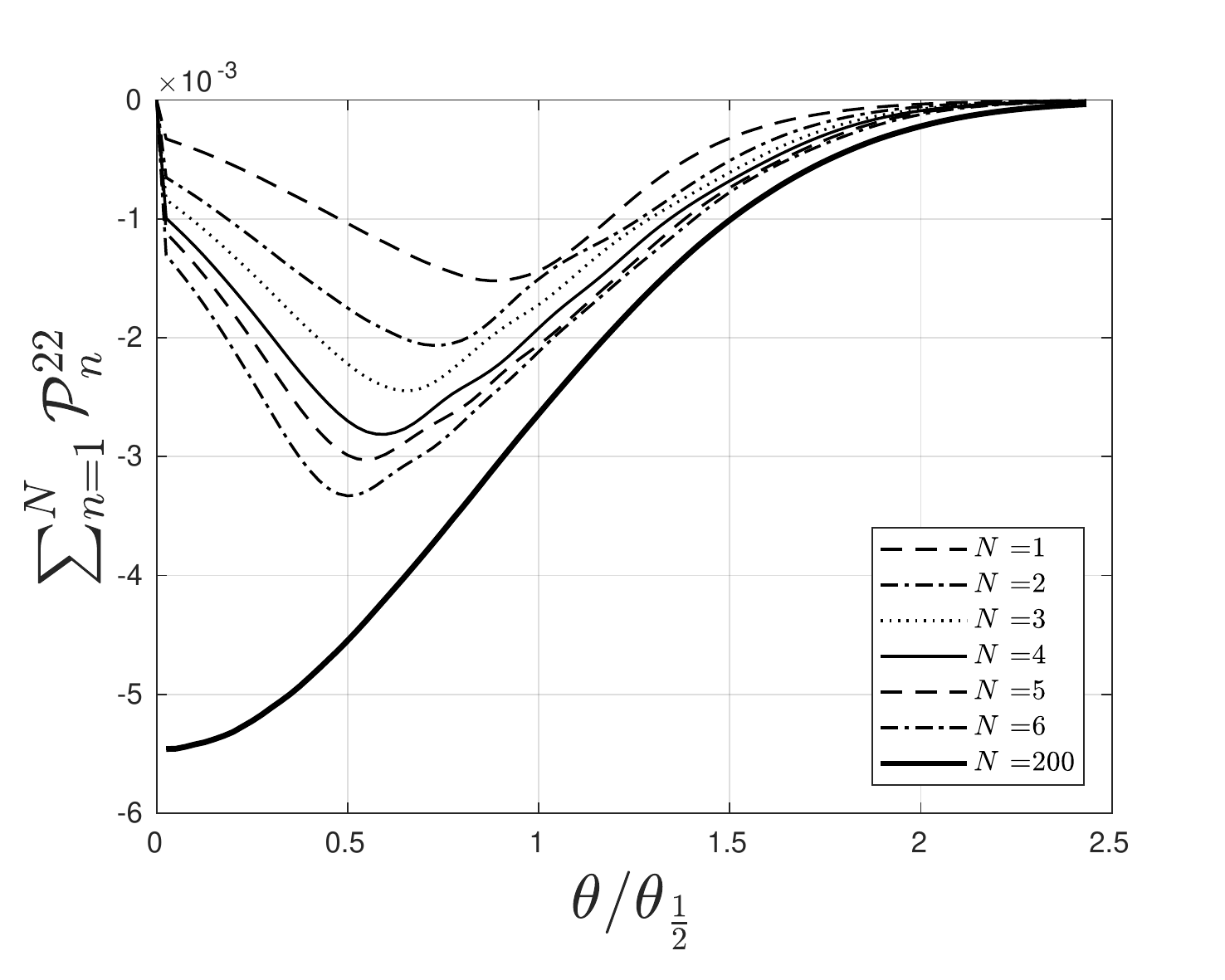}\label{fig:recon_production_vv_SSC}}\\
    \subfloat[]{\includegraphics[width=0.40\linewidth]{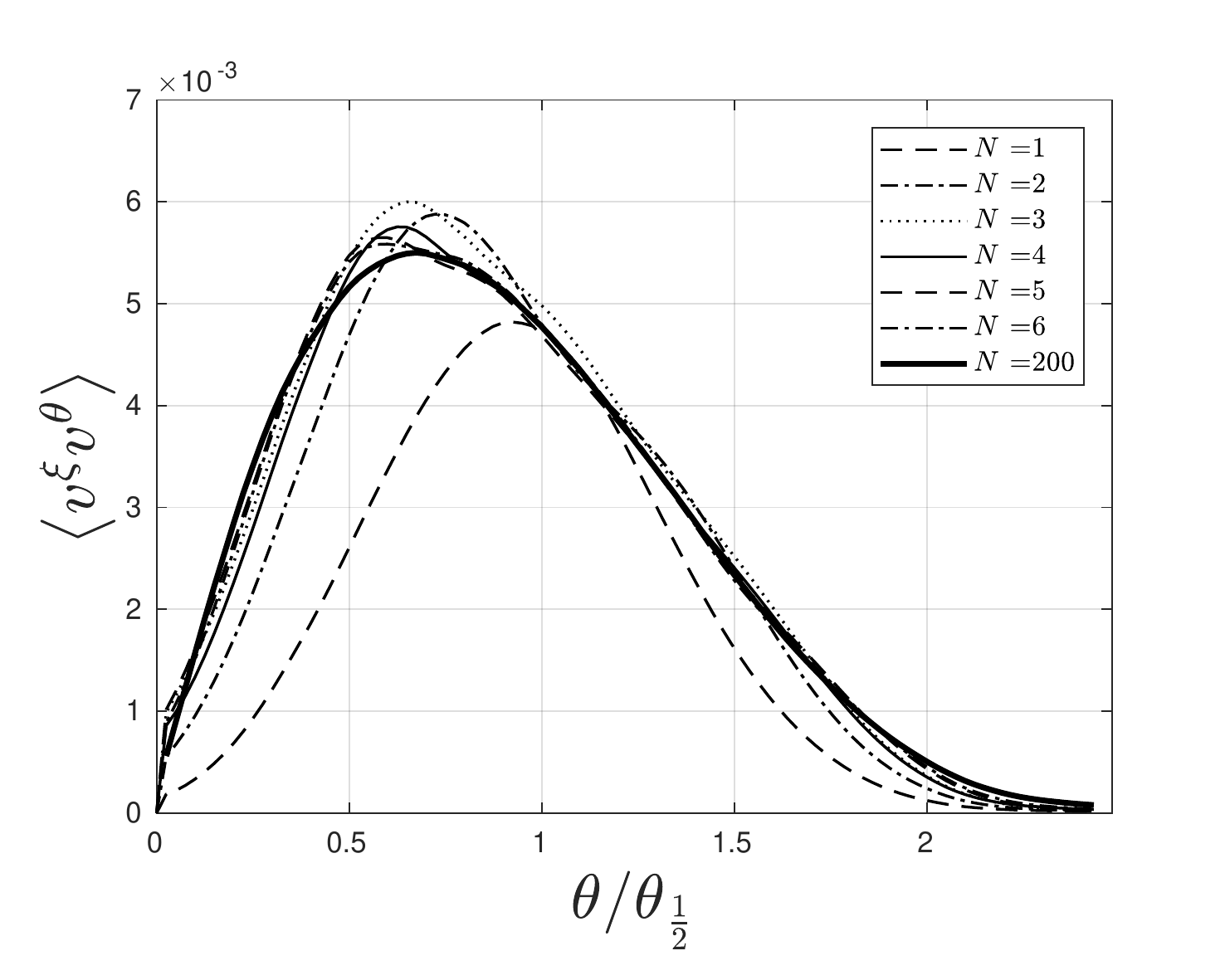}\label{fig:v_xi_v_theta_SSC}}
    \subfloat[]{\includegraphics[width=0.40\linewidth]{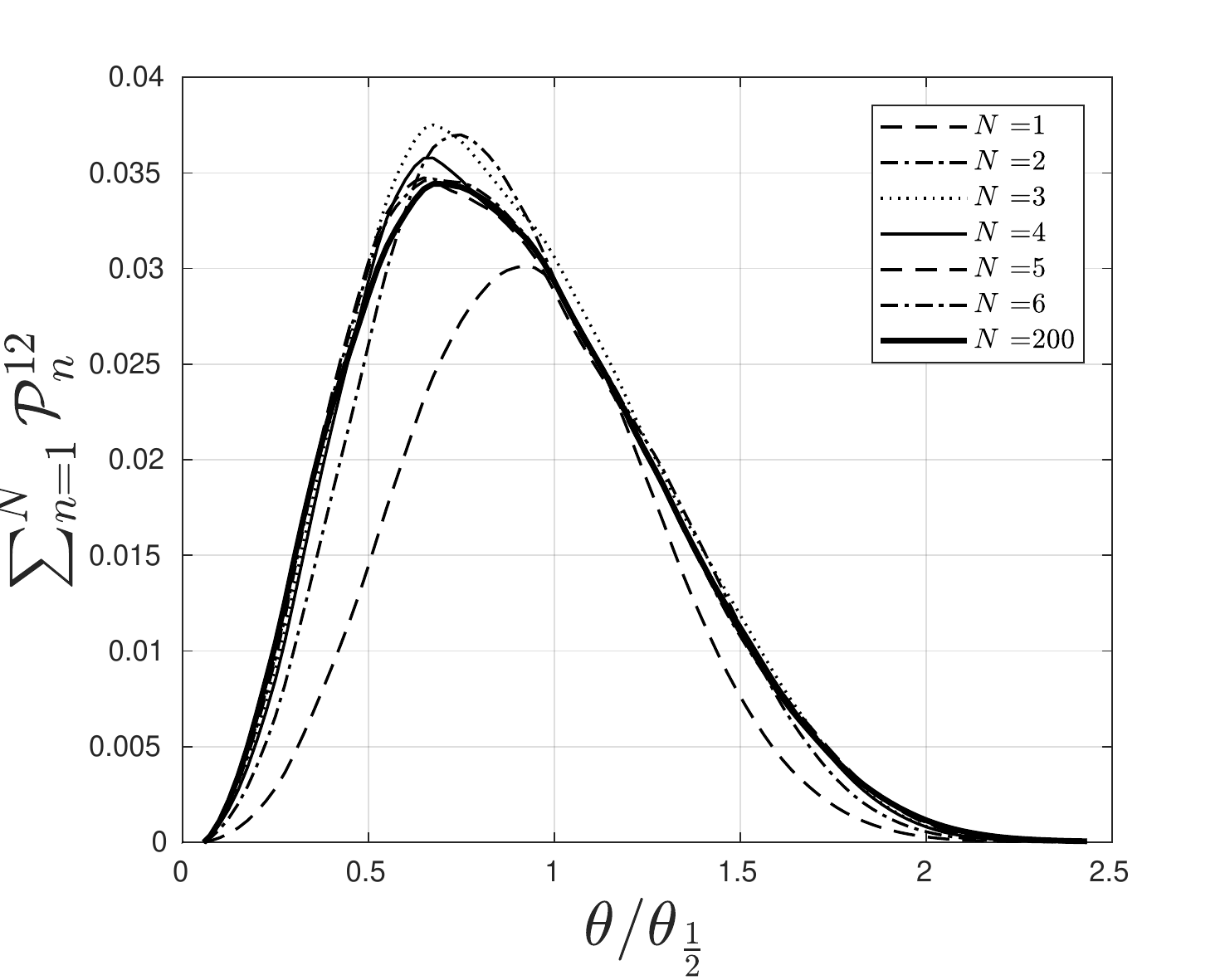}\label{fig:recon_production_uv_SSC}}\\
     \subfloat[]{\includegraphics[width=0.40\linewidth]{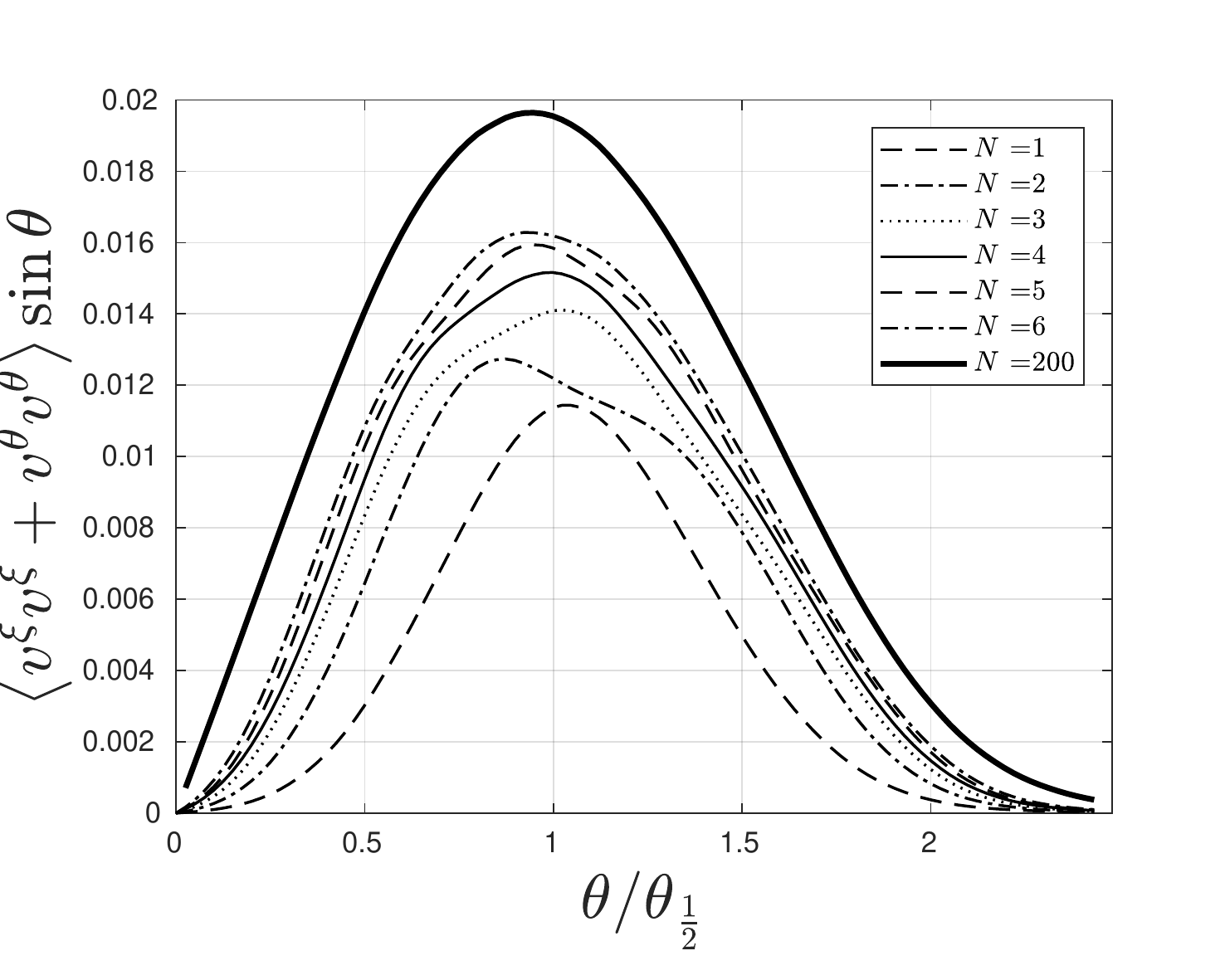}\label{fig:TKE_SSC}}
\subfloat[]{\includegraphics[width=0.40\linewidth]{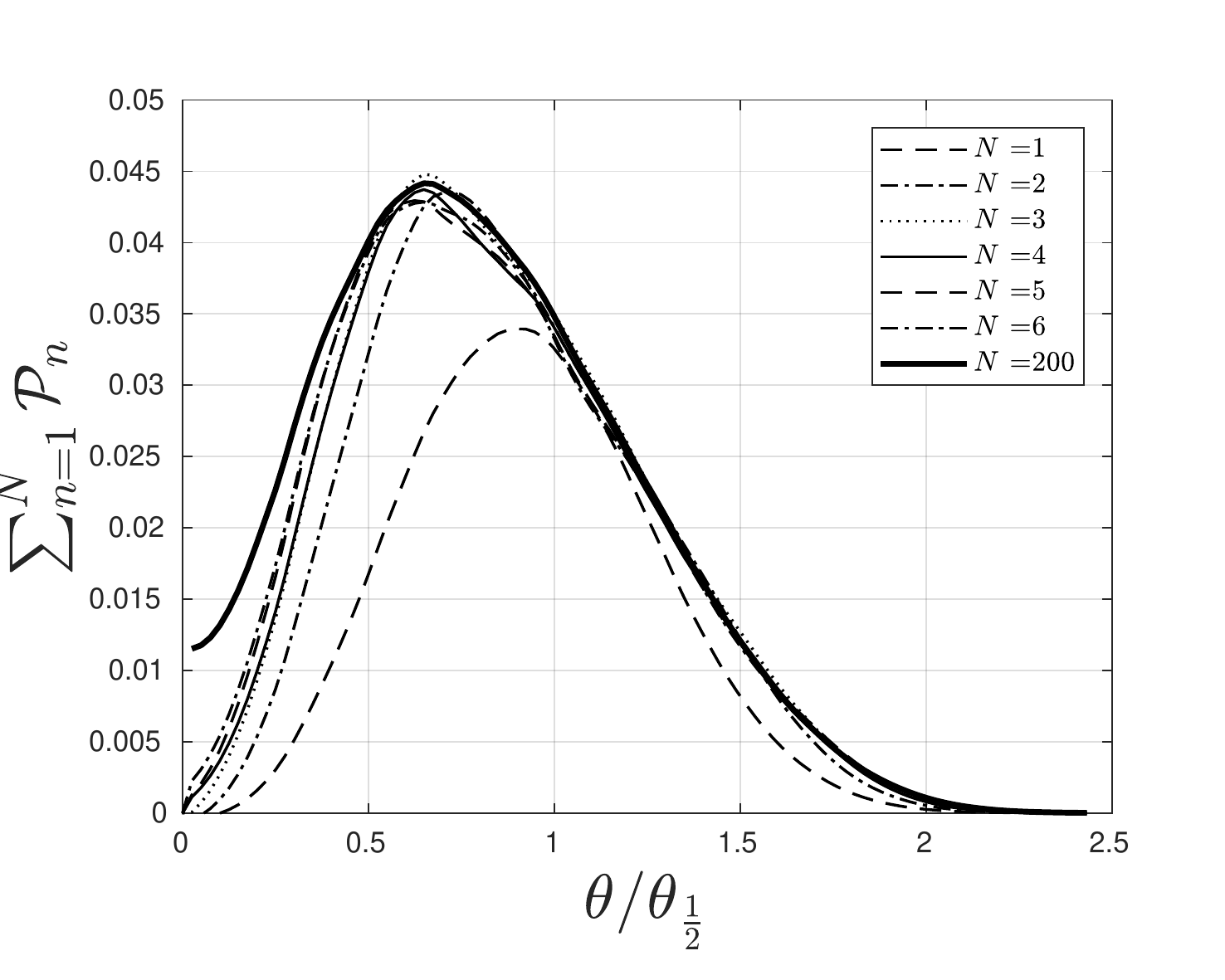}\label{fig:recon_production_SSC}}
\caption{The cumulutive sum of modal building blocks of various single-point statistics. Figures (a), (c), and (e) show the reconstruction of Reynolds stresses, (b) and (d), show the reconstruction of the $\xi$- and $\theta$-components of the energy production, respectively, (f) shows the shear-stress-related components of the energy production. \label{fig:single-point_statistics_reconstructed_SSC}}
\end{figure}
\FloatBarrier
\noindent
The net fluxes have a tendency to change signs over the same wavenumber range where the mean spectrum exhibits the characteristic $-5/3$-gradient, demonstrating negative as well as positive net spectral fluxes across wavenumbers. It is only the aggregate of the modal contributions that reproduces the $-5/3$-range in the energy spectra - at least away from regions where the mean shear is significant. The spectral reconstruction produces spectra characterized by component contributions in which only the minority are defined by a constant energy flux across wavenumbers in the $-5/3$-range. 

Although comprising a minority, these spectral components do exists in regions where mean shear dominates. 
Several of the most energetic modal contributions of $\widetilde{\Phi}^{2}_{\cdot,2}$ exhibit a constant spectral energy flux across wavenumbers, particularly around $\theta/\theta_{\frac{1}{2}}=1$ as figure \ref{fig:single_spectra_reconstructed_SSC} shows. This is seen from their preservation of the $-5/3$-gradients. Similarly, the rapid reconstructions of the $-7/3$-range, also indicate a constant spectral flux across wavenumbers for modes $1-3$ approximately. 

The modal reconstruction of $\widetilde{\Phi}^{2}_{\cdot,2}$ is characterized by the $-5/3$-range due to the key role that $v^\theta$ plays in the transport of $v^\xi$ across the jet. These results indicate that for statistically stationary high mean shear flows, there exists a statistical equilibrium within the spectral energy transport for individual LD modes. This means that within individual LD modes there exists a constant energy flux across $\kappa$-values. While the detailed nature of these energy transfers is not entirely clear in terms of the ratio between the energy transport across LD modes and the energy flux within individual modes, the current data does indicate that the most energetic modes are in a state of equilibrium around the half-width region.

From the $\widetilde{\Phi}^{1}_{\cdot,1}$ reconstruction, it is seen that the constant spectral flux is not exhibited in any wavenumber range for the first eight modes in the jet. A general decrease in wavenumber gradients with increasing mode number in the $-5/3$-region indicates that the net negative energy fluxes across wavenumbers are decreasing as the mode number increases - even around $\theta/\theta_{\frac{1}{2}}=1$. This stands in contrast to what was observed from the same modes in the reconstruction of $\widetilde{\Phi}^{2}_{\cdot,2}$. It indicates that the energy transfer across mode numbers must play a considerable role as no signs of spectral equilibrium across wavenumbers are observed in individual modes. 

%

The contributions of the less energetic modes, $4-8$, are weighted towards increasing the energy levels at higher wavenumbers, such that the higher end of the spectrum approaches the $-5/3$-slope. This is more clearly seen by the individual modal contributions to the $\widetilde{\Phi}^{2}_{\cdot,2}$-spectra in figures \ref{fig:app_single_spectra_uu_30_SSC}-\ref{fig:app_single_spectra_uu_90_SSC} in Appendix \ref{app:individual_modal_components_spectra}. The spectrum is reconstructed progressively from low- to high wavenumbers for increasing mode numbers. This indicates that for higher wavenumbers the energy contributions of higher mode numbers are dominant, and support the idea of an energy cascade from lower- to higher mode numbers in the $-5/3$-region.

The reconstruction of single-point statistics can be seen in figure \ref{fig:single-point_statistics_reconstructed_SSC}. Note that a Parzen window was applied on the data before the spectral analysis, resulting in a reduction of turbulence kinetic energy in figure \ref{fig:single-point_statistics_reconstructed_SSC}, compared to the scaled profiles, \cite{Hodzic2019_part1}. In order to avoid this, the LD analysis could be performed on the data before application of the Parzen window. But since the windowing effect would have a more profound effect on the spectra and the LD modes, the current approach was chosen. Nevertheless, the lower energy does not affect our interpretation of the reconstructed single-point statistics. 

Figures \ref{fig:v_xi_v_xi_SSC}, \ref{fig:v_theta_v_theta_SSC}, and \ref{fig:v_xi_v_theta_SSC} show the reconstruction of the Reynolds stresses, whereas figures \ref{fig:recon_production_uu_SSC}, \ref{fig:recon_production_vv_SSC}, and \ref{fig:recon_production_uv_SSC} show the reconstruction of the corresponding $\mathcal{P}^{ij}$-terms in \eqref{eq:galerkin_production_term}. The shear-stresses in figure \ref{fig:v_xi_v_theta_SSC} are seen to converge to their final form rapidly. In fact, only two LD modes reconstruct the shear-stress profile. The normal stresses are seen to converge much slower, and as figures \ref{fig:v_xi_v_xi_SSC} and \ref{fig:v_theta_v_theta_SSC} show, the convergence rate is slowest around the centerline. Since the modes are determined due to their energy content, it is expected that the energy content of the modes is highest away from the centerline, due to the increasing weighing of the Jacobian moving away from the centerline. Near the centerline the modal contribution to the energy reconstruction is very limited due to this fact. 

Since the shear-stresses are responsible for most of the production, unless significant energy transport is dominating, the TKE will have a similar profile. This is substantiated by the total turbulence kinetic energy production profile shown in figure \ref{fig:recon_production_SSC}. The profile is nearly fully reconstructed after the contribution of only two modes. Furthermore, multiplying the normal stresses by $\sin\theta$ yields the cumulative modal contributions to the turbulence kinetic energy integrand, which is shown in figure \ref{fig:v_xi_v_theta_SSC}. From this it becomes apparent that the LD optimizes the modes in terms of maximizing their total turbulence kinetic energy representation. This can be concluded since every contribution to the reconstruction is a self-similar profile to the preceding one, and each bell-shaped contribution is centered around $\theta/\theta_{\frac{1}{2}}\approx 1$. 
\FloatBarrier
\section{Conclusions}
A Galerkin projection was combined with tensor calculus in order to be able to expand the governing equations by the eigenfunctions in curvilinear coordinates. The applied methodology provided the possibility of easily performing modal analysis of each term of the energy equation in a curvilinear coordinate systems in order to rigorously quantify the various processes, herein the energy production, energy dissipation, and the non-linear transport in terms of the LD eigenfunctions. With this formalism the current work has paved the way for future investigations of the turbulence kinetic energy transport equation in the turbulent jet by means of modal expansions in general coordinate systems. 

The modal reconstruction of the energy spectra obtained from the SADFM identified the modal self-similarity in the jet far-field. This shows that the energy spectra from fundamentally different flow-types show similar traits, and may prove to be highly significant for turbulence modeling purposes. Furthermore, the current work has focused on testing the hypothesis posed in \cite{Wanstrom2009}, namely that multiple modes can extract a constant and significant portion of their relative energies from the mean flow. The analysis of the Galerkin projection of the turbulence kinetic energy transport equation substantiated this possibility and the results showed that a wide range of modes indeed obtain a substantial part of their energy directly from the mean flow.

The analysis has revealed that the energy production capacity of a given mode is varying across the flow domain as well as over wave- and mode numbers. A wide range of modes are observed to obtain significant levels of their energy directly from the mean flow, even at wavenumbers for which the averaged spectra exhibit the characteristic $-5/3$-region. This is particularly shown to be true in high mean shear regions. The reconstruction of the energy spectra revealed that the spectral flux of the individual modal building blocks of the spectra is varying across the $-5/3$-region (in the averaged spectrum) and can be seen to change signs across wavenumbers. This indicates that a more complex energy transport mechanism than the classical one stated by Richardson underlies the averaged spectra.
\section*{Acknowledgments}
The authors would also like to thank Dr. Maja W{\"a}nstr{\"o}m for sharing data.

The authors gratefully acknowledge the long-term support of the Department of Mechanical Engineering at the Technical University of Denmark. All the experimental work of this long research program was carried out in these laboratories. This generous support has made these investigations and their continuation possible. A considerable part of the current work is obtained from the PhD-dissertation of Azur Hod\v zi\' c, \cite{Hodzic2018b}.

Last, but not least, the final stages of the project would not have been possible without the support of the European Research Council: This project has received funding from the European Research Council (ERC) under the European Union’s Horizon 2020 research and innovation program (grant agreement No 803419).
\FloatBarrier

\bibliographystyle{jfm}
\appendix
\clearpage
\section{Evaluation of the Galerkin projection of $II$}\label{app:galerkin_of_II}
The production term, $II$, in \eqref{eq:galerkin_energy_terms} is defined as 
\begin{equation}
\mathcal{P} = \lambda\varphi_i^\alpha\varphi^{j*}_\alpha\nabla_j\left\langle V^i\right\rangle,
\end{equation}
from which the total reconstructed energy production, $\mathcal{P}_{\text{tot}}$, is obtained by multiplying by $\rho$ and integrating over the domain, $\Omega\subset\mathbb{R}^3$
\begin{equation}
\mathcal{P}_{\text{tot}} = \int_\Omega\rho\mathcal{P}d\mu.
\end{equation}
This reduces to the following expression for the eigenfunctions defined in \eqref{eq:eigenfunctions}
\begin{eqnarray}
\mathcal{P}_{\text{tot}} &=& \frac{\rho\lambda B\sqrt{M_0}}{L_\xi C^2}\left(1-e^{-L_\xi}\right)\int_0^{\theta_{\text{max}}}f\left(\theta\right)d\theta ,\nonumber\\
&=& \frac{\rho\lambda\left(U_c\left(0\right)-U_c\left(L_\xi\right)\right)}{L_\xi C}\int_0^{\theta_{\text{max}}}f\left(\theta\right)d\theta\label{eq:app_P_tot},
\end{eqnarray}
where $U_c\left(0\right)$ and $U_c\left(L_\xi\right)$, respectively, are the centerline velocities at the lower- and upper bounds of the domain in $\xi$-coordinates, where $U_c=B\sqrt{M_0}/\left(Ce^\xi\right)$, and 
\begin{eqnarray}
f\left(\theta\right) = -\psi^{\xi\alpha }\psi^{\xi*}_\alpha\left\langle V^\xi\right\rangle &+&\psi^{\xi\alpha }\psi^{\theta*}_\alpha\left(\frac{\partial\left\langle V^\xi\right\rangle}{\partial \theta}-\left\langle V^\theta\right\rangle\right)\nonumber \\
&-&\psi^{\theta\alpha }\psi^{\xi*}_\alpha\left\langle V^\theta\right\rangle 
+ \psi^{\theta\alpha }\psi^{\theta *}_\alpha\left(\frac{\partial\left\langle V^\theta\right\rangle}{\partial \theta}+\left\langle V^\xi\right\rangle\right),
\end{eqnarray}
where $\widetilde{\psi}^1_\alpha = \psi^\xi_\alpha$, and $\widetilde{\psi}^2_\alpha = \psi^\theta_\alpha$, following the notation of \cite{Hodzic2019_part1} . Normalizing \eqref{eq:app_P_tot} by the local energy, $\rho\lambda$, yields 
\begin{equation}
\mathcal{P}_{\rho\lambda} = \frac{U_c\left(0\right)-U_c\left(L_\xi\right)}{L_\xi C}\int_0^{\theta_{\text{max}}}f\left(\theta\right)d\theta. \label{eq:app_P_rholambda}
\end{equation}
Suppressing the integrals in \eqref{eq:app_P_tot} and \eqref{eq:app_P_rholambda} we obtain the total production density over $\theta$
\begin{equation}
\mathcal{P}_{\text{tot},\theta} = \frac{\rho\lambda\left(U_c\left(0\right)-U_c\left(L_\xi\right)\right)}{L_\xi C}f\left(\theta\right)\label{eq:app_P_tot_theta},
\end{equation}
and the energy-normalized production density as a function of $\theta$
\begin{equation}
\mathcal{P}_{\rho\lambda,\theta} = \frac{U_c\left(0\right)-U_c\left(L_\xi\right)}{L_\xi C}f\left(\theta\right), \label{eq:app_P_rhotheta}
\end{equation}
which reveal the local contributions to the reconstruction of the production term over the span of the jet. 
\FloatBarrier
\section{Individual modal components of spectral densities}\label{app:individual_modal_components_spectra}
\begin{figure}[h]
    \centering      
\subfloat[]{\includegraphics[width=0.40\linewidth]{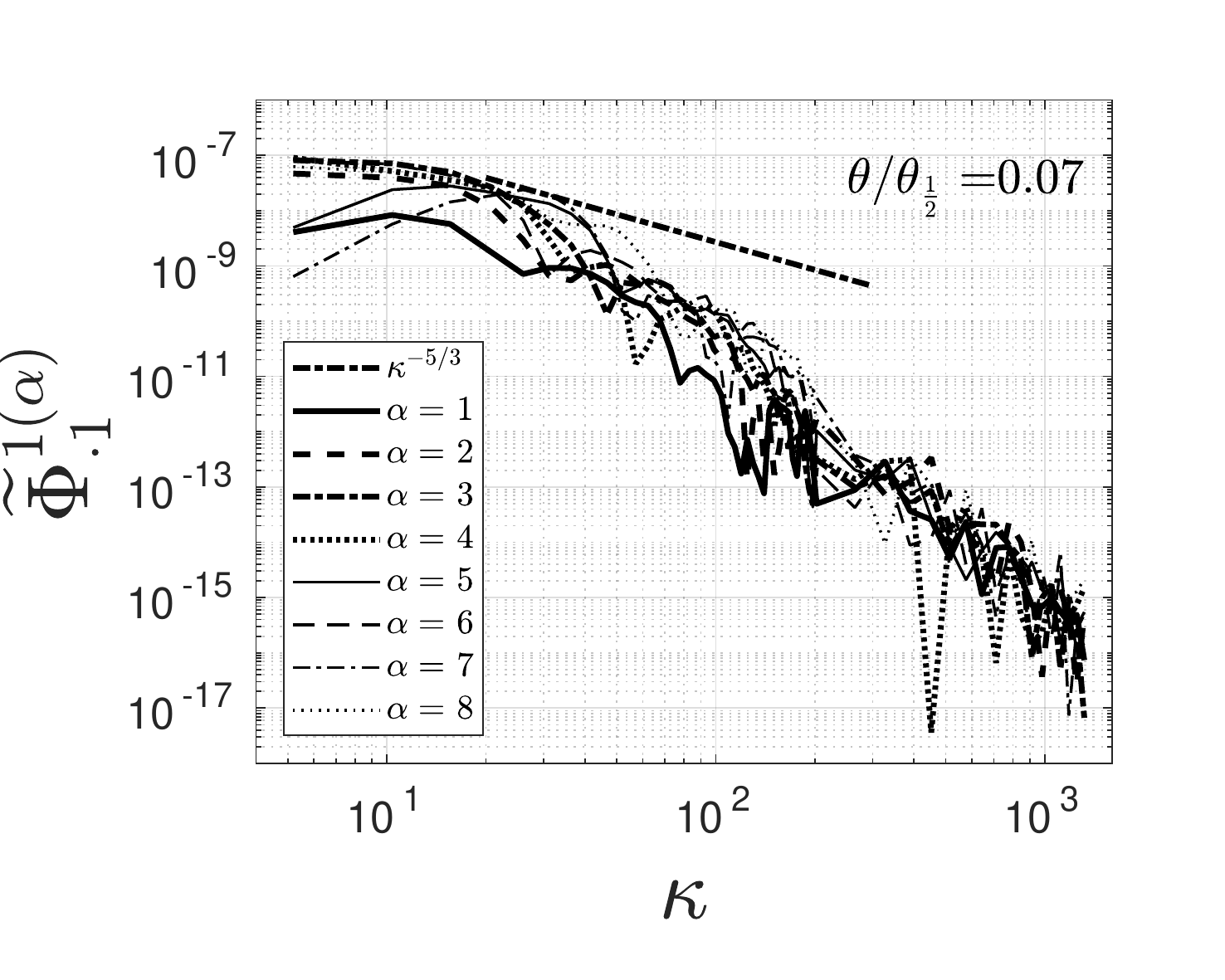}\label{fig:app_single_spectra_uu_7_SSC}}
\subfloat[]{\includegraphics[width=0.40\linewidth]{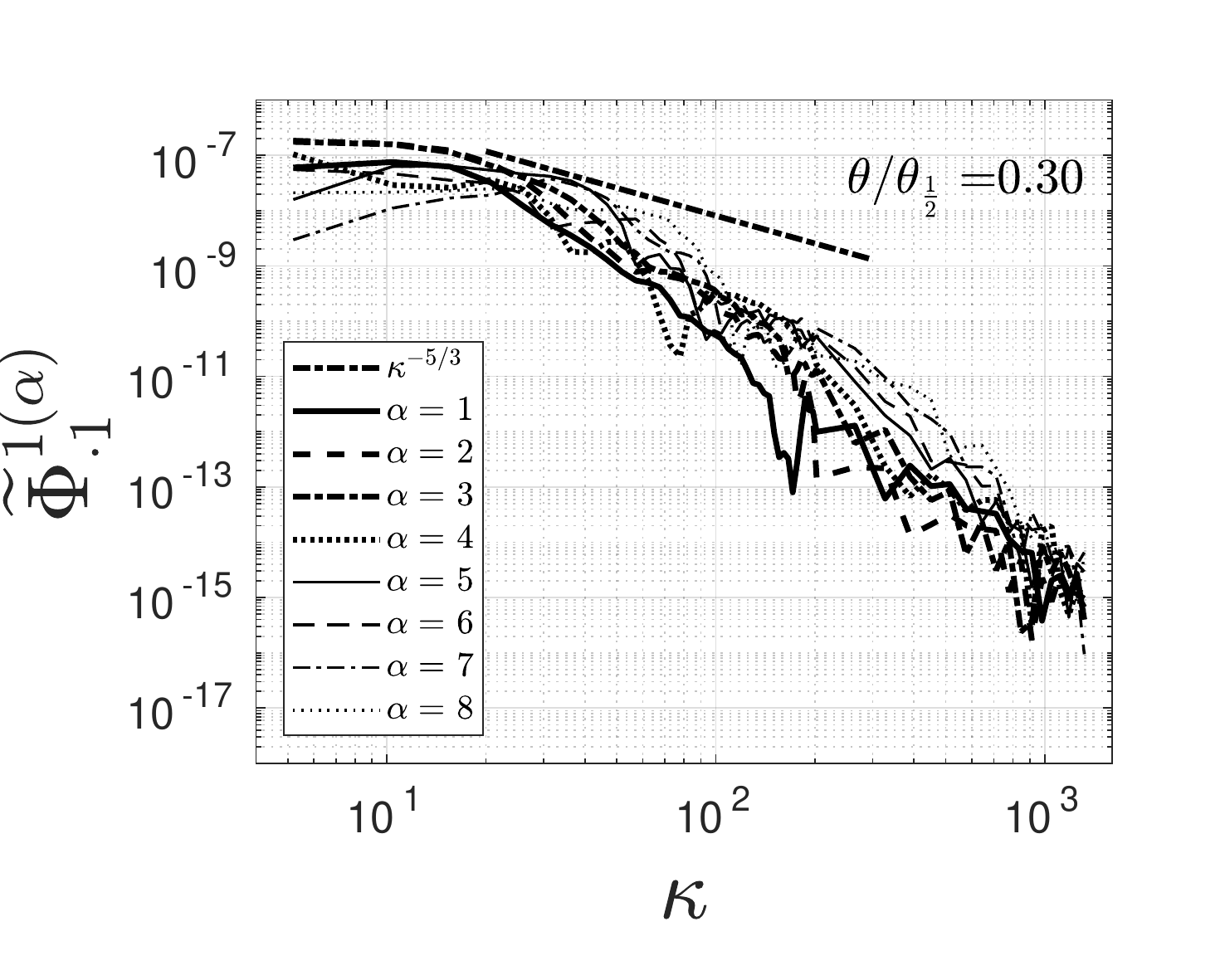}\label{fig:app_single_spectra_uu_30_SSC}}\\
    \subfloat[]{\includegraphics[width=0.40\linewidth]{figs/spectra/SSC/reconstructed/single_spectra_uu_60}\label{fig:app_single_spectra_uu_60_SSC}}
    \subfloat[]{\includegraphics[width=0.40\linewidth]{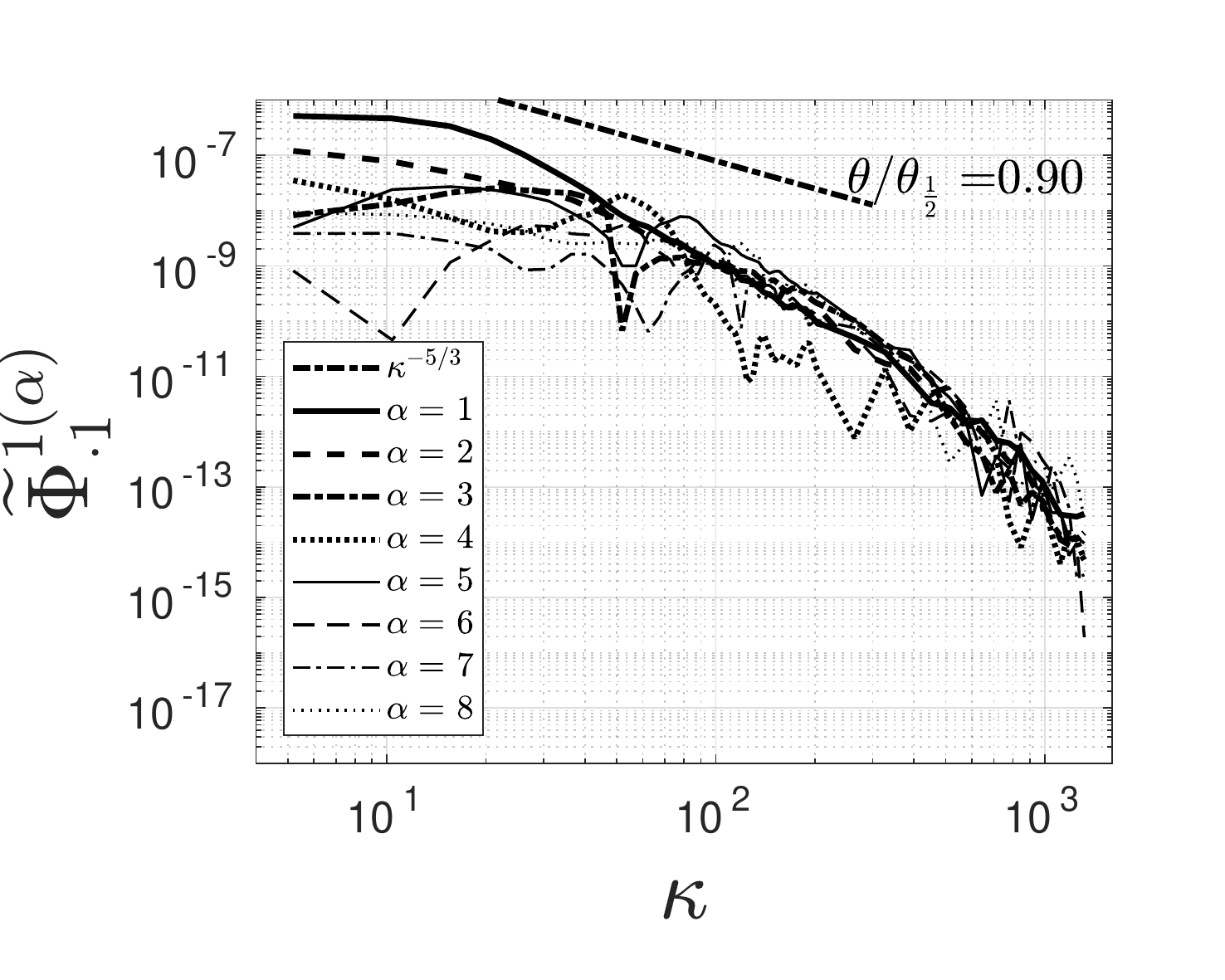}\label{fig:app_single_spectra_uu_90_SSC}}\\
    \subfloat[]{\includegraphics[width=0.40\linewidth]{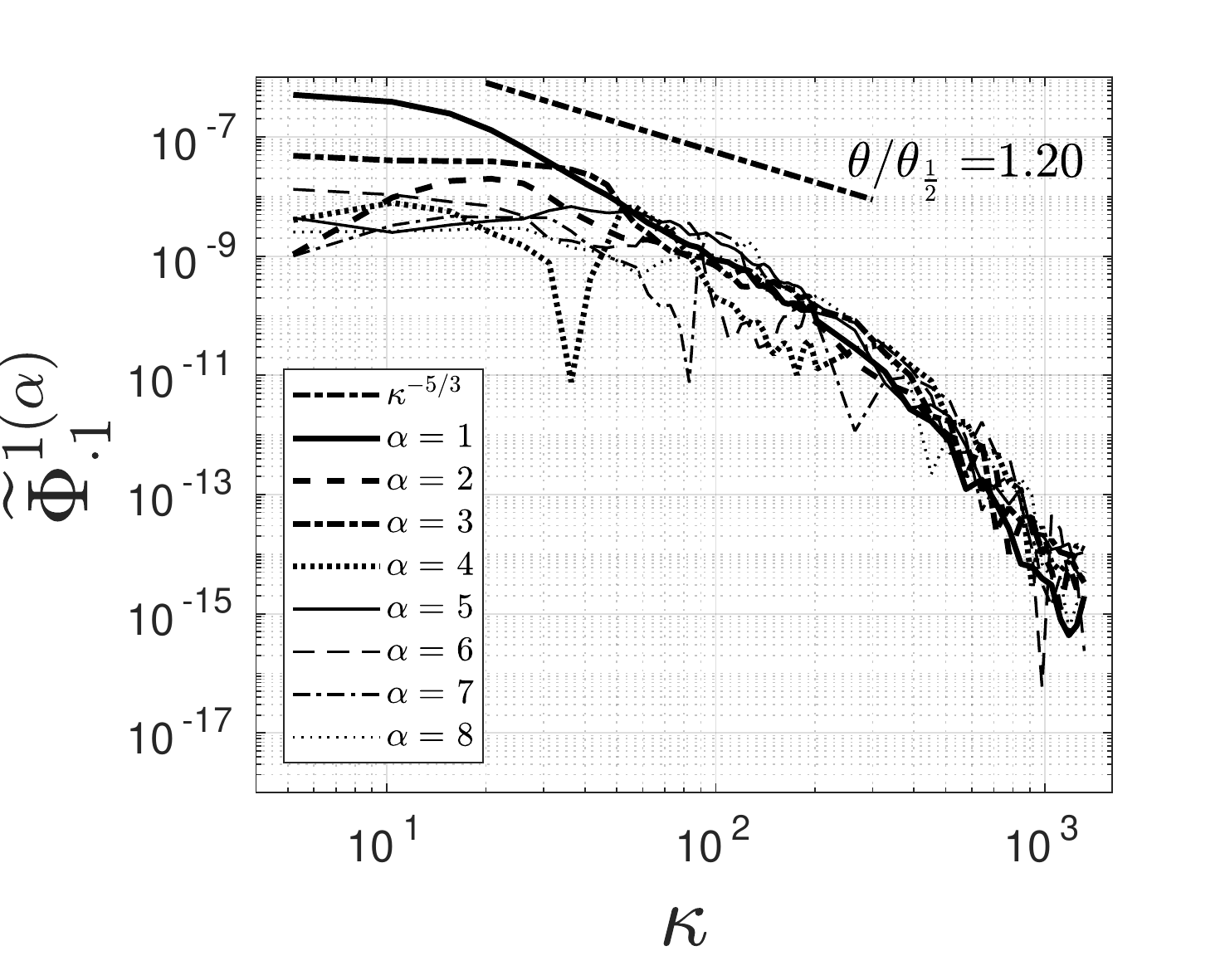}\label{fig:app_single_spectra_uu_120_SSC}}
    \subfloat[]{\includegraphics[width=0.40\linewidth]{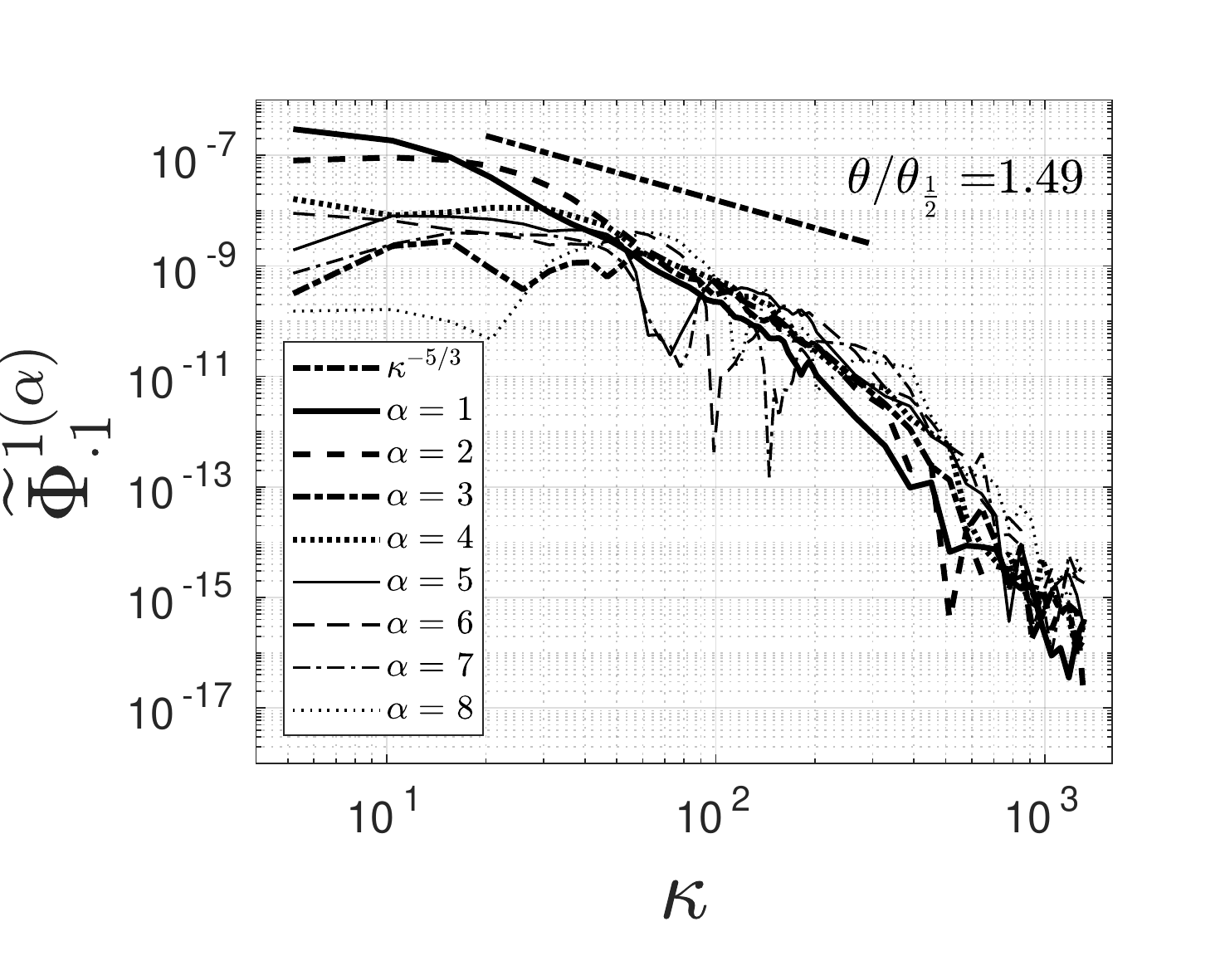}\label{fig:app_single_spectra_uu_149_SSC}}\\
    \subfloat[]{\includegraphics[width=0.40\linewidth]{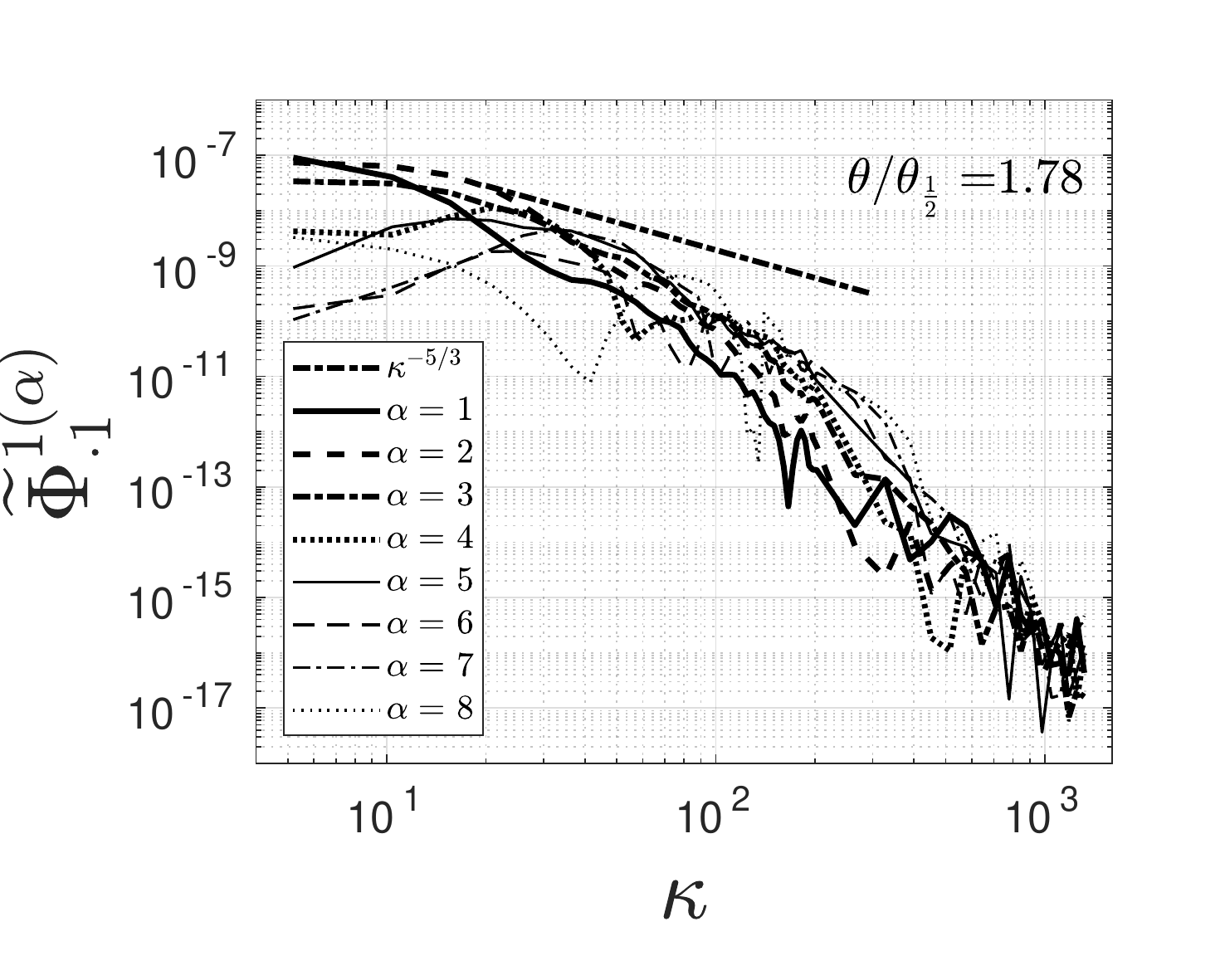}\label{fig:app_single_spectra_uu_178_SSC}}
    \subfloat[]{\includegraphics[width=0.40\linewidth]{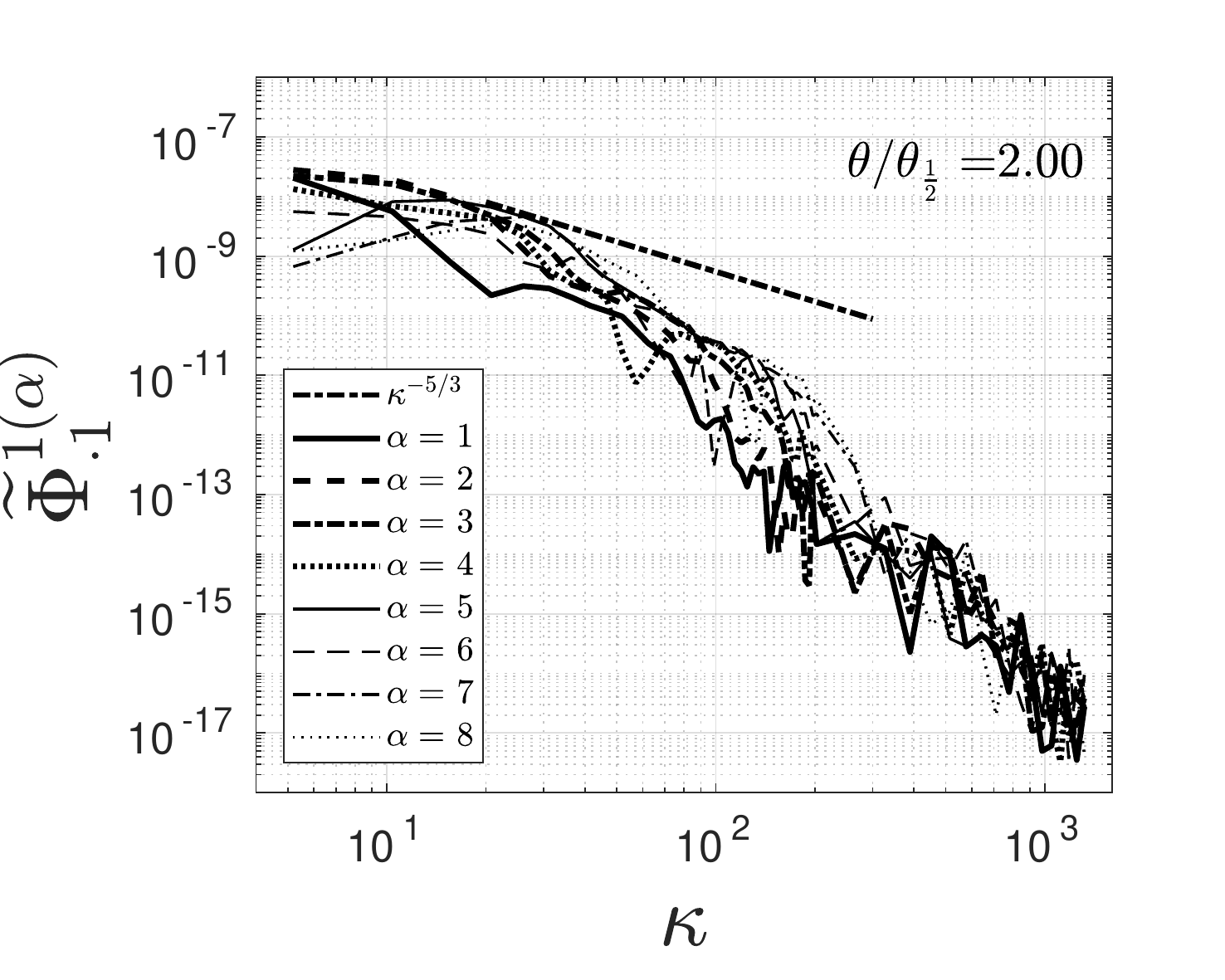}\label{fig:app_single_spectra_uu_200}}
\caption{Modal components of single-point spatial spectra, $\widetilde{\Phi}^{1(\alpha)}_{\cdot,1}$, at various spanwise coordinates, $\theta/\theta_{\frac{1}{2}}$. \label{fig:app_single_spectra_uu_reconstructed_SSC}}
\end{figure}

\begin{figure}[h]
    \centering      
\subfloat[]{\includegraphics[width=0.40\linewidth]{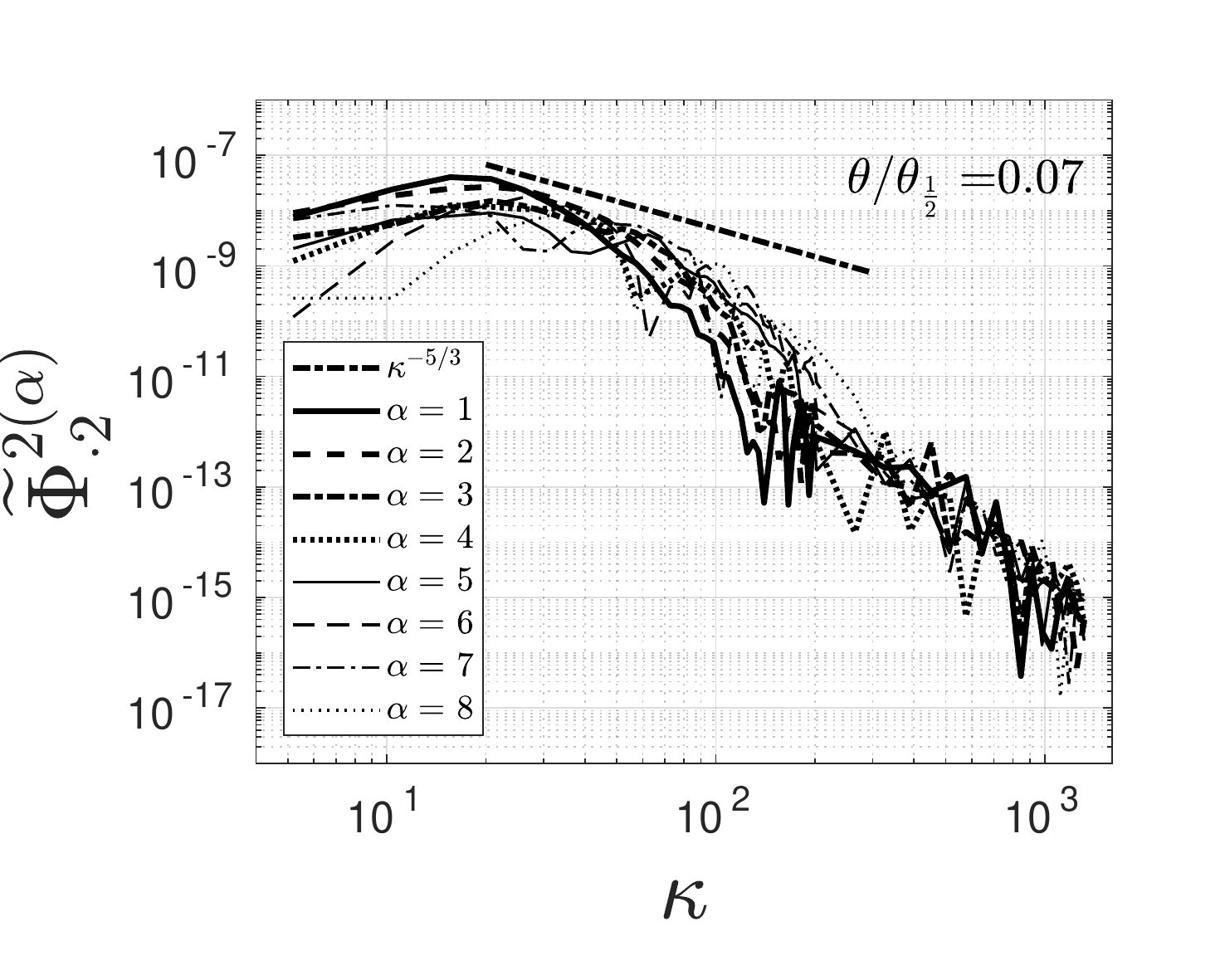}\label{fig:app_single_spectra_vv_7_SSC}}
\subfloat[]{\includegraphics[width=0.40\linewidth]{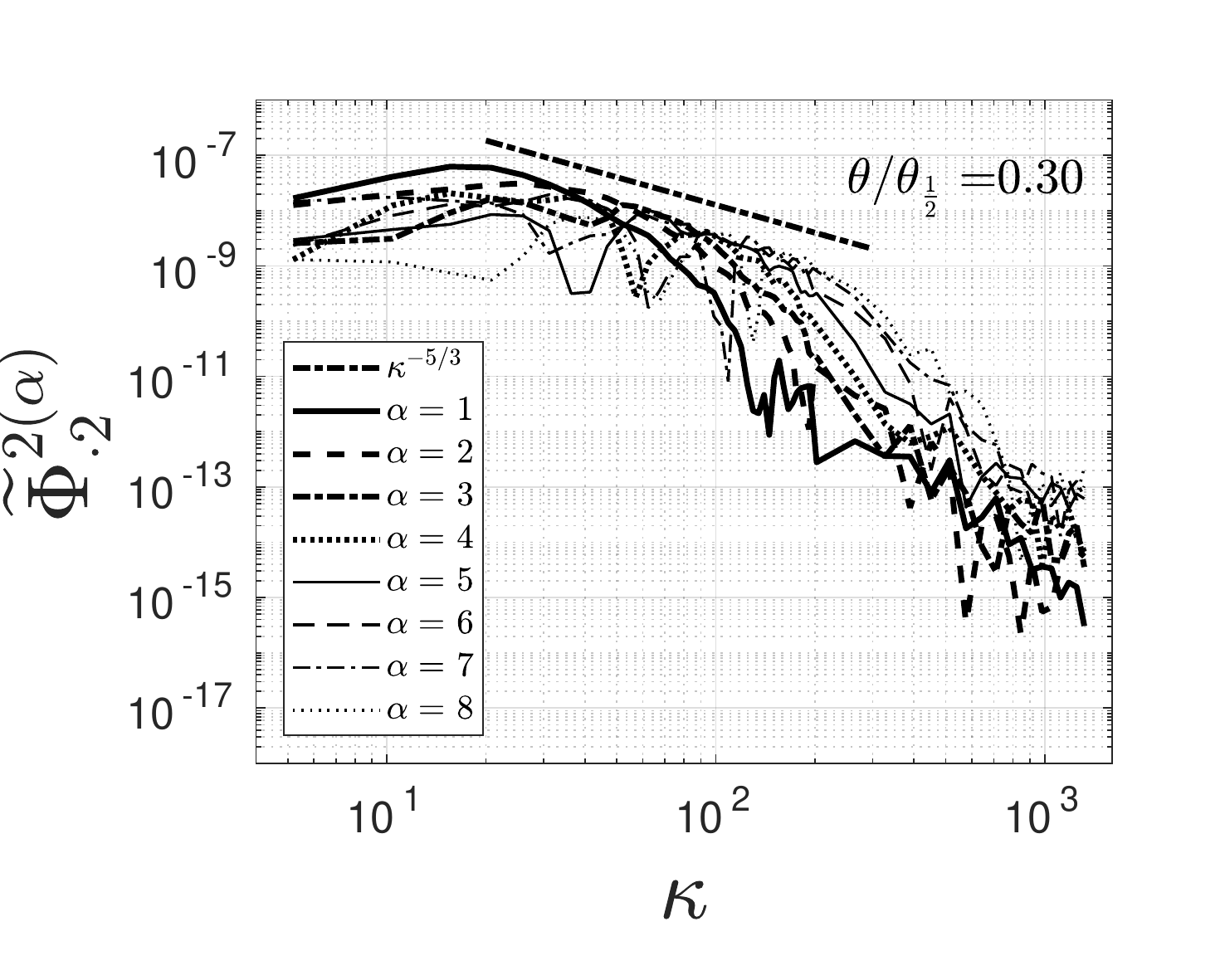}\label{fig:app_single_spectra_vv_30_SSC}}\\
    \subfloat[]{\includegraphics[width=0.40\linewidth]{figs/spectra/SSC/reconstructed/single_spectra_vv_60}\label{fig:app_single_spectra_vv_60_SSC}}
    \subfloat[]{\includegraphics[width=0.40\linewidth]{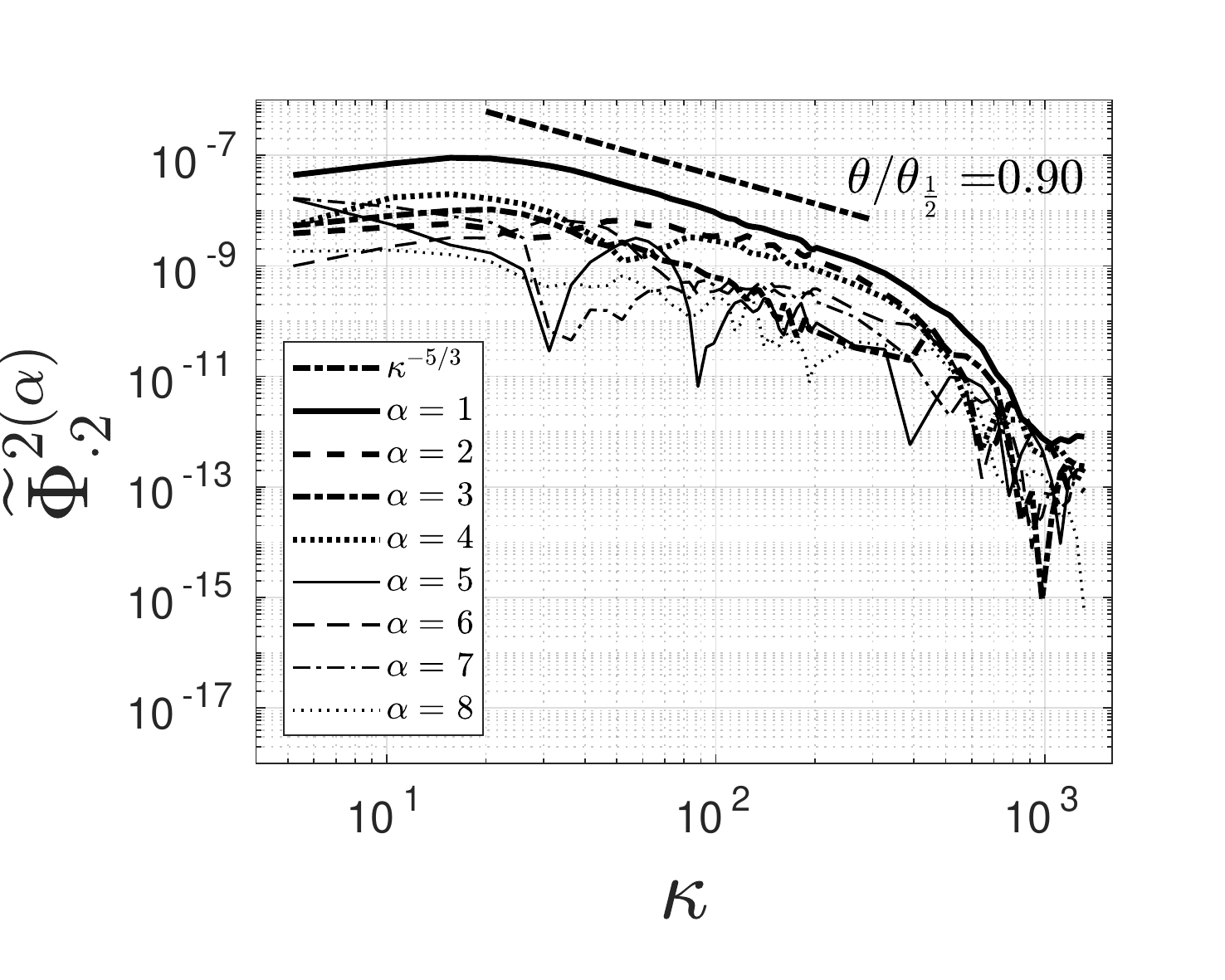}\label{fig:app_single_spectra_vv_90_SSC}}\\
    \subfloat[]{\includegraphics[width=0.40\linewidth]{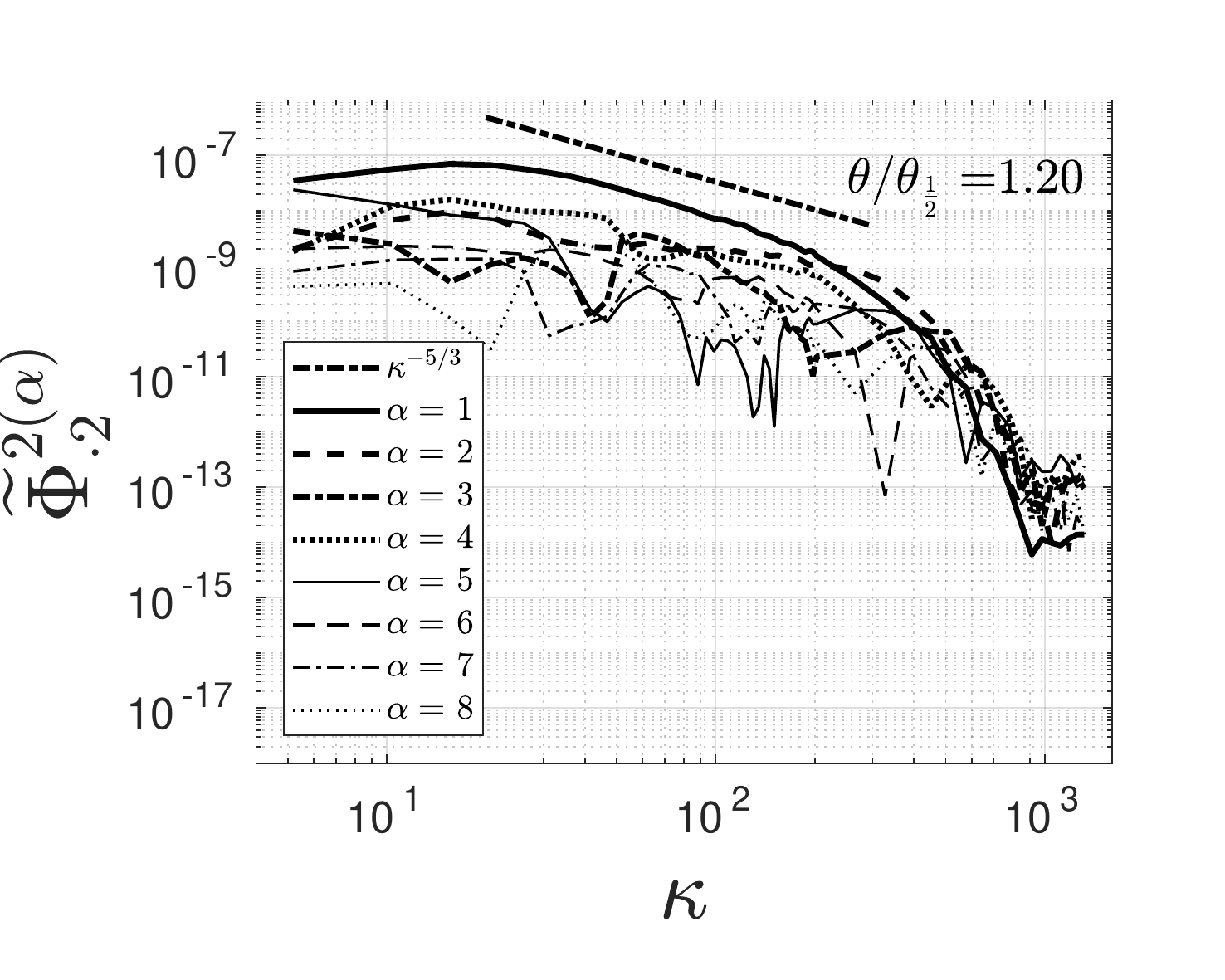}\label{fig:app_single_spectra_vv_120_SSC}}
    \subfloat[]{\includegraphics[width=0.40\linewidth]{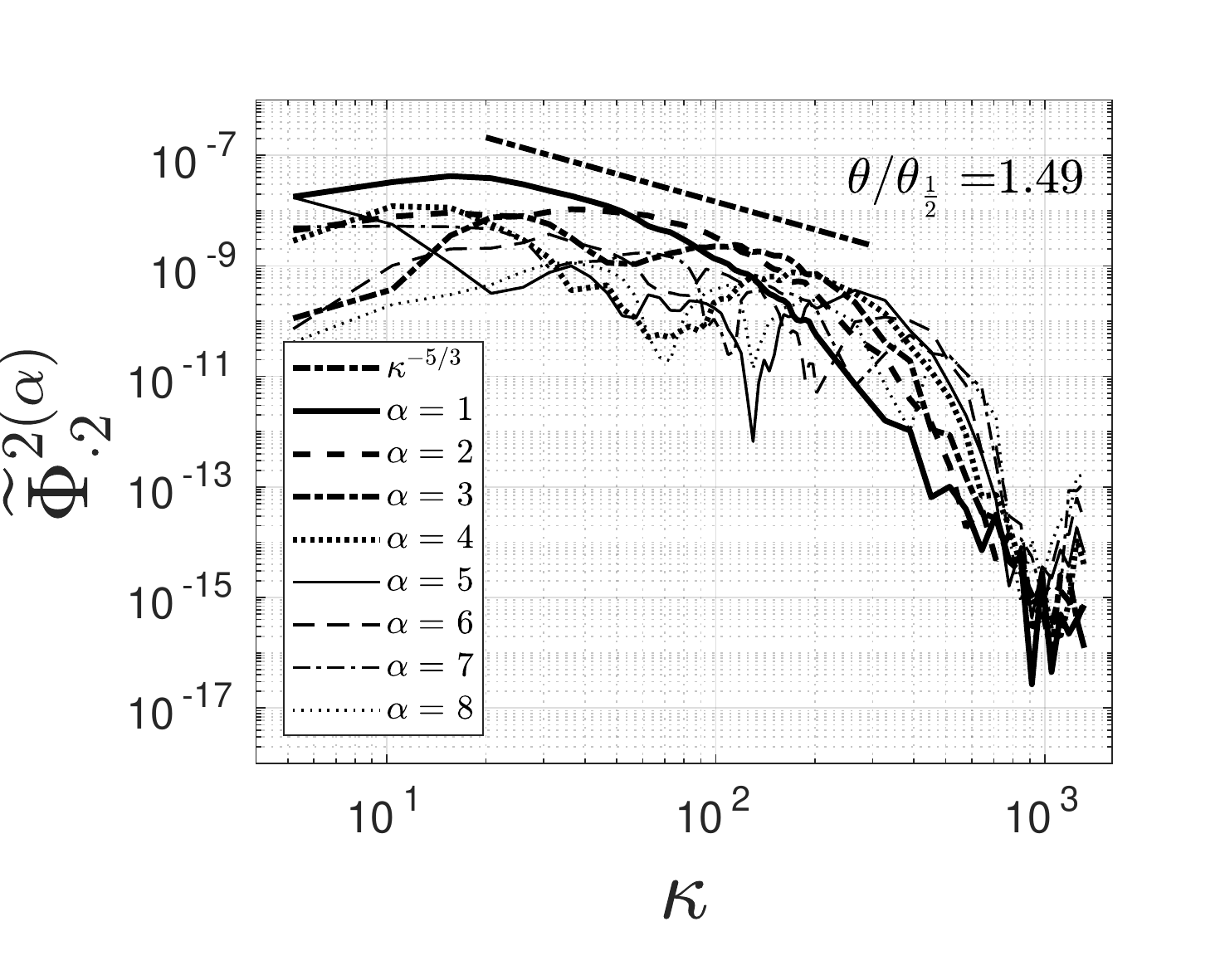}\label{fig:app_single_spectra_vv_149_SSC}}\\
    \subfloat[]{\includegraphics[width=0.40\linewidth]{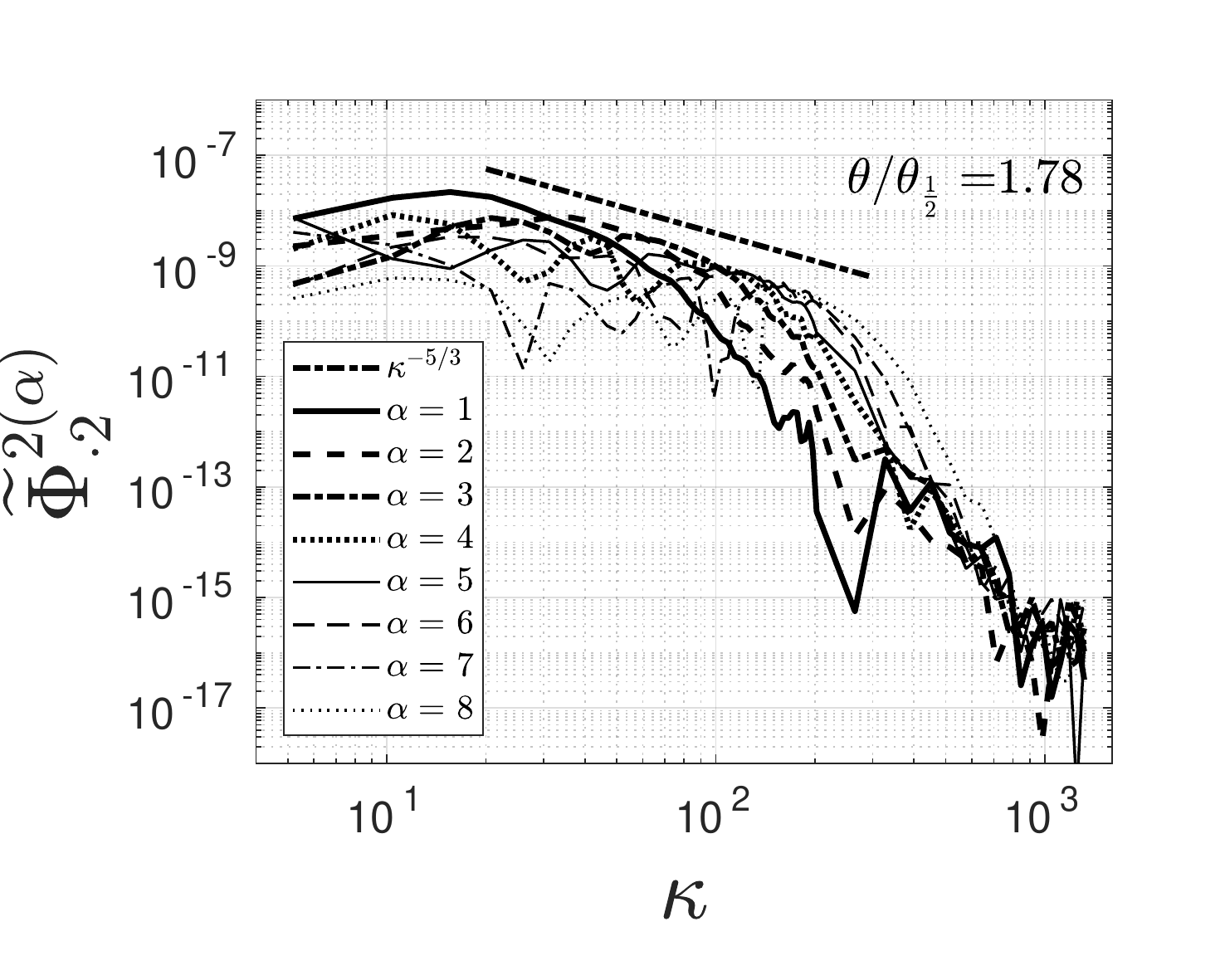}\label{fig:app_single_spectra_vv_178_SSC}}
    \subfloat[]{\includegraphics[width=0.40\linewidth]{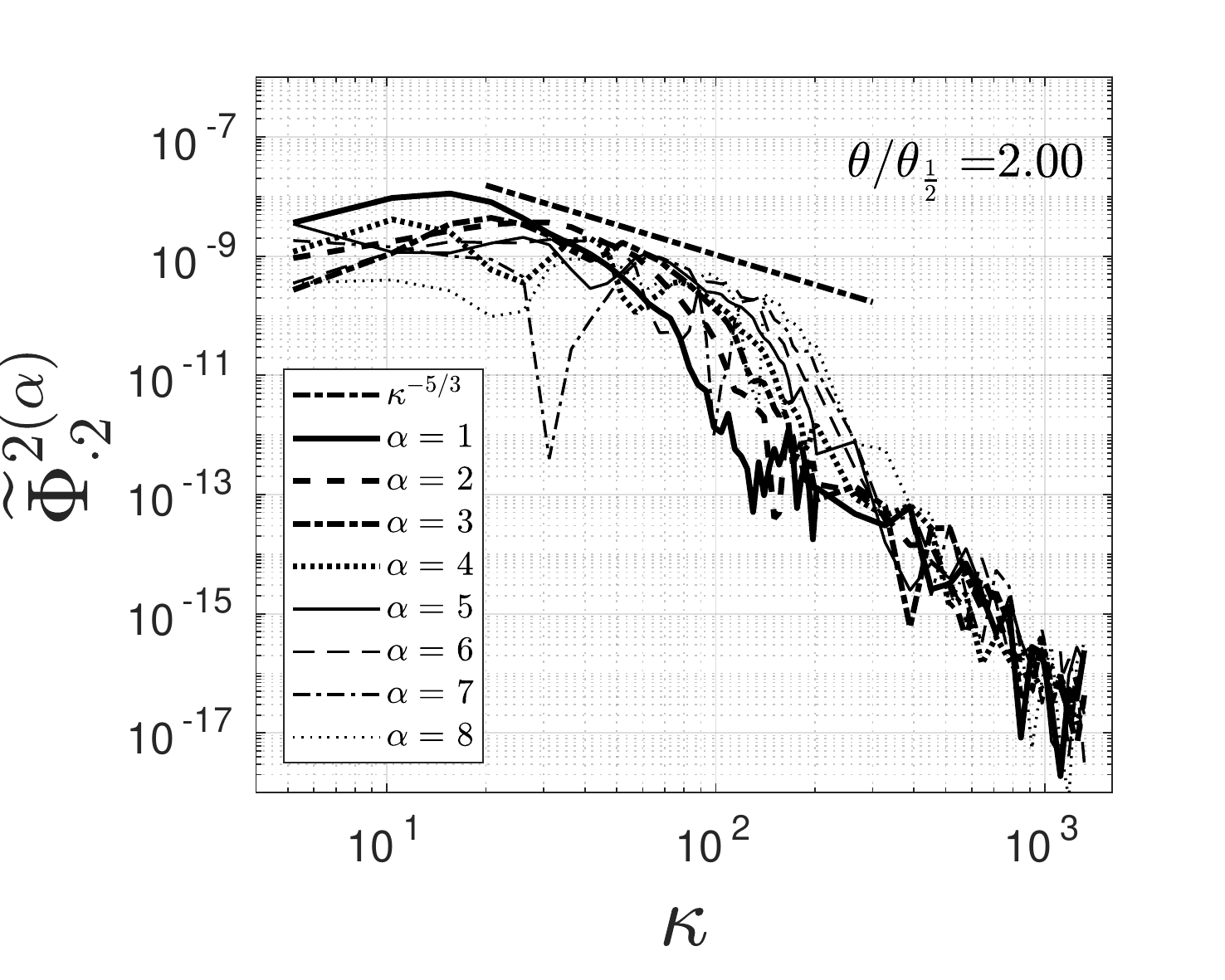}\label{fig:app_single_spectra_vv_200}}
\caption{Modal components of single-point spatial spectra, $\widetilde{\Phi}^{2(\alpha)}_{\cdot,2}$, at various spanwise coordinates, $\theta/\theta_{\frac{1}{2}}$. \label{fig:app_single_spectra_vv_reconstructed_SSC}}
\end{figure}

\begin{figure}[h]
    \centering      
\subfloat[]{\includegraphics[width=0.40\linewidth]{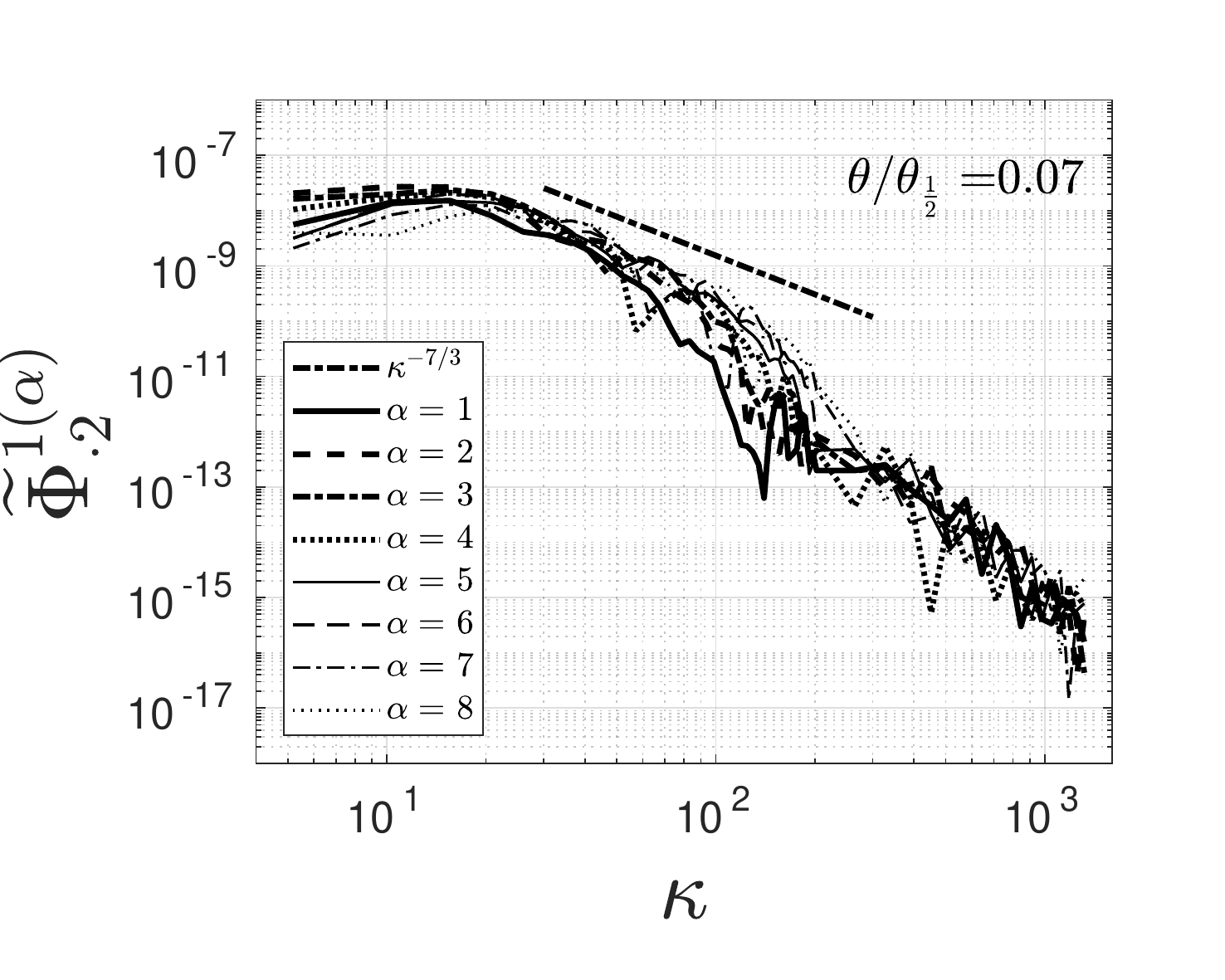}\label{fig:app_single_spectra_uv_7_SSC}}
\subfloat[]{\includegraphics[width=0.40\linewidth]{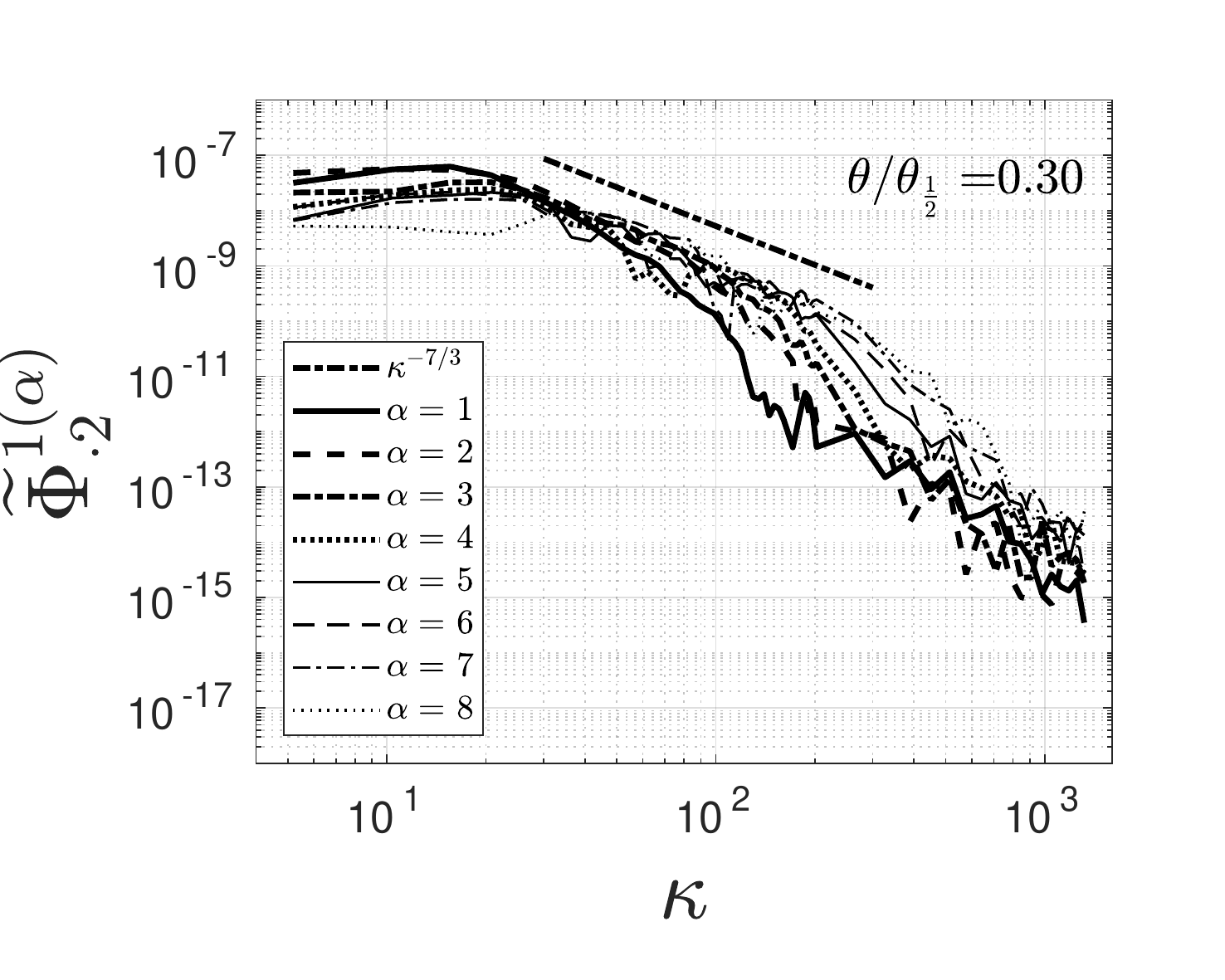}\label{fig:app_single_spectra_uv_30_SSC}}\\
    \subfloat[]{\includegraphics[width=0.40\linewidth]{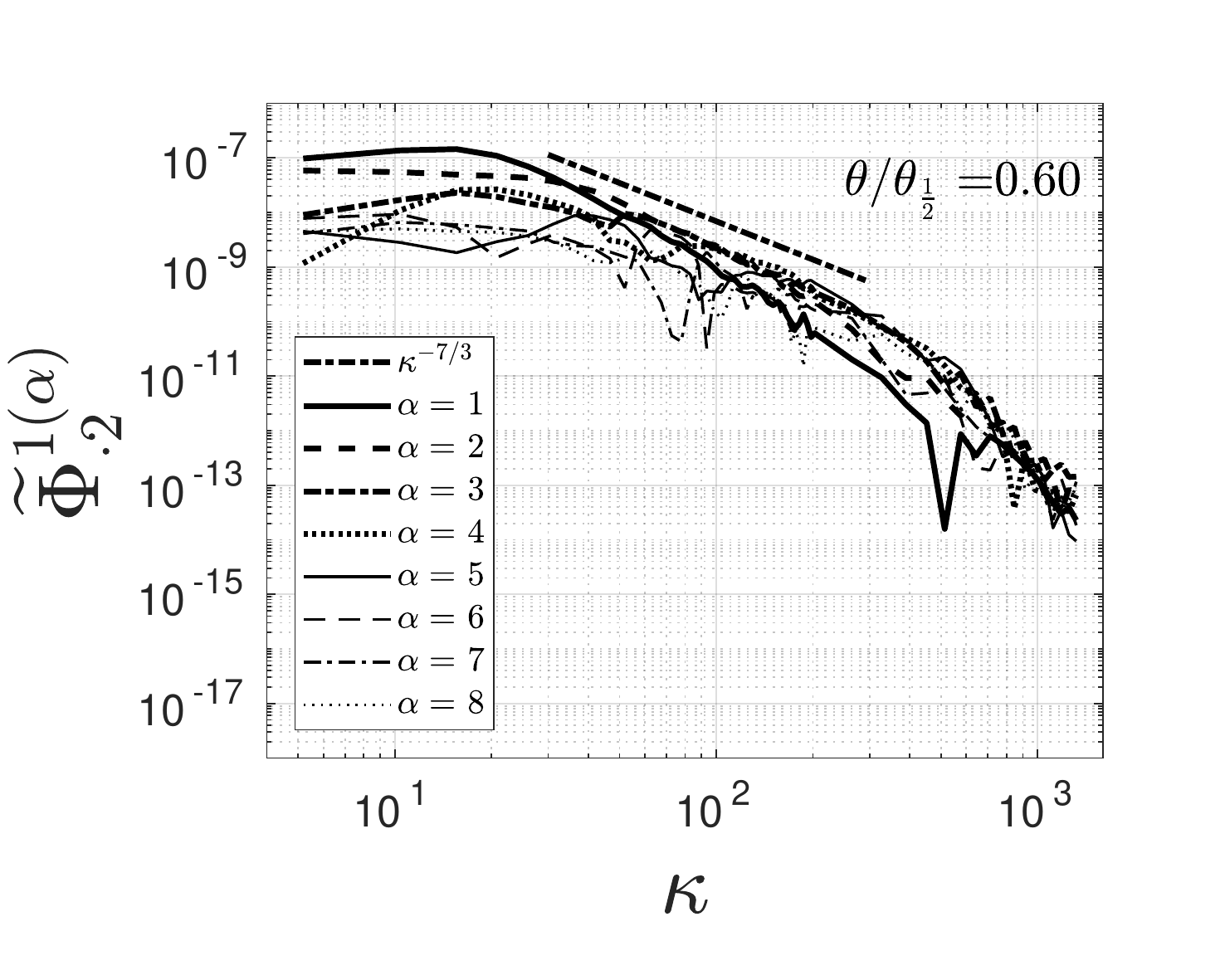}\label{fig:app_single_spectra_uv_60_SSC}}
    \subfloat[]{\includegraphics[width=0.40\linewidth]{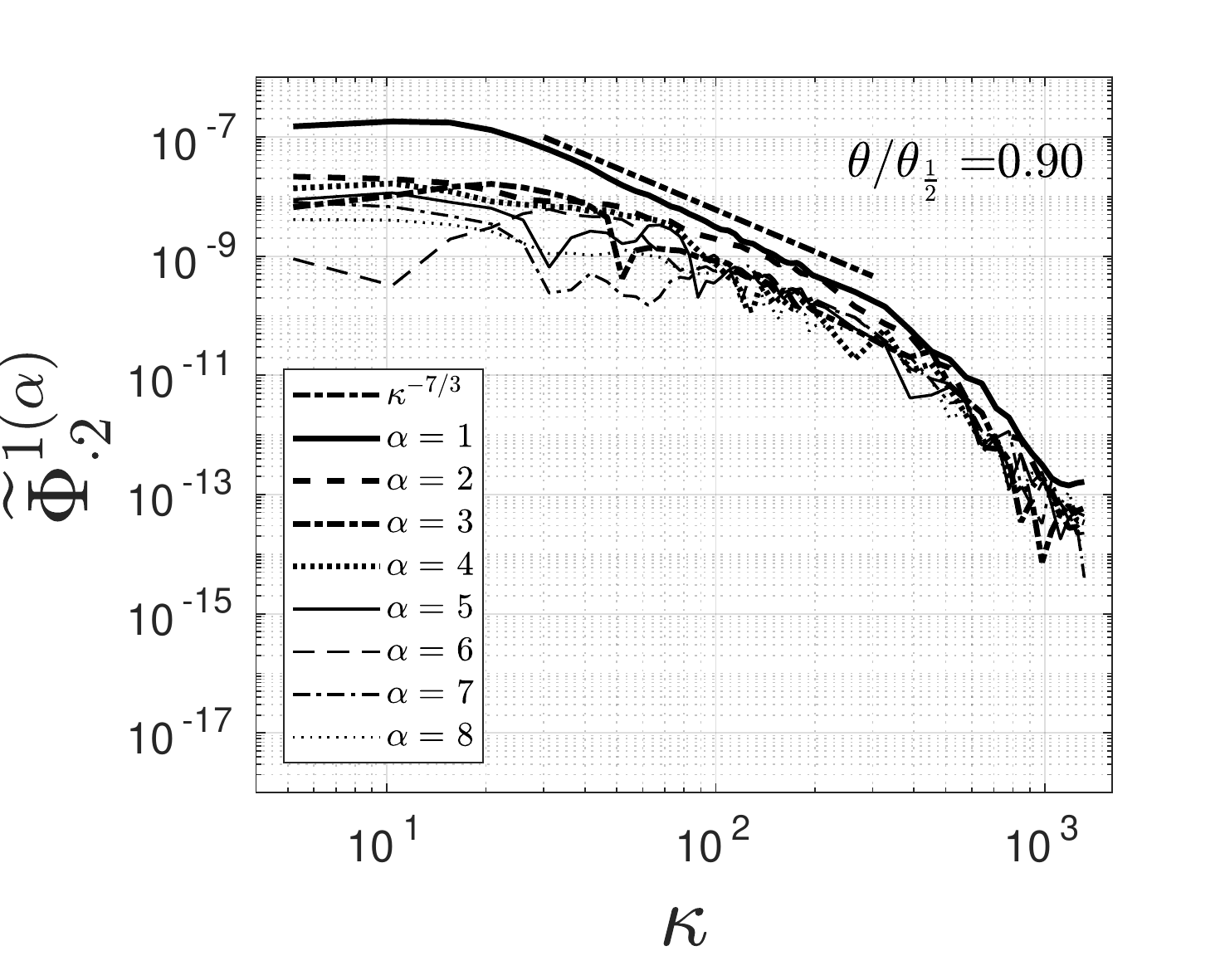}\label{fig:app_single_spectra_uv_90_SSC}}\\
    \subfloat[]{\includegraphics[width=0.40\linewidth]{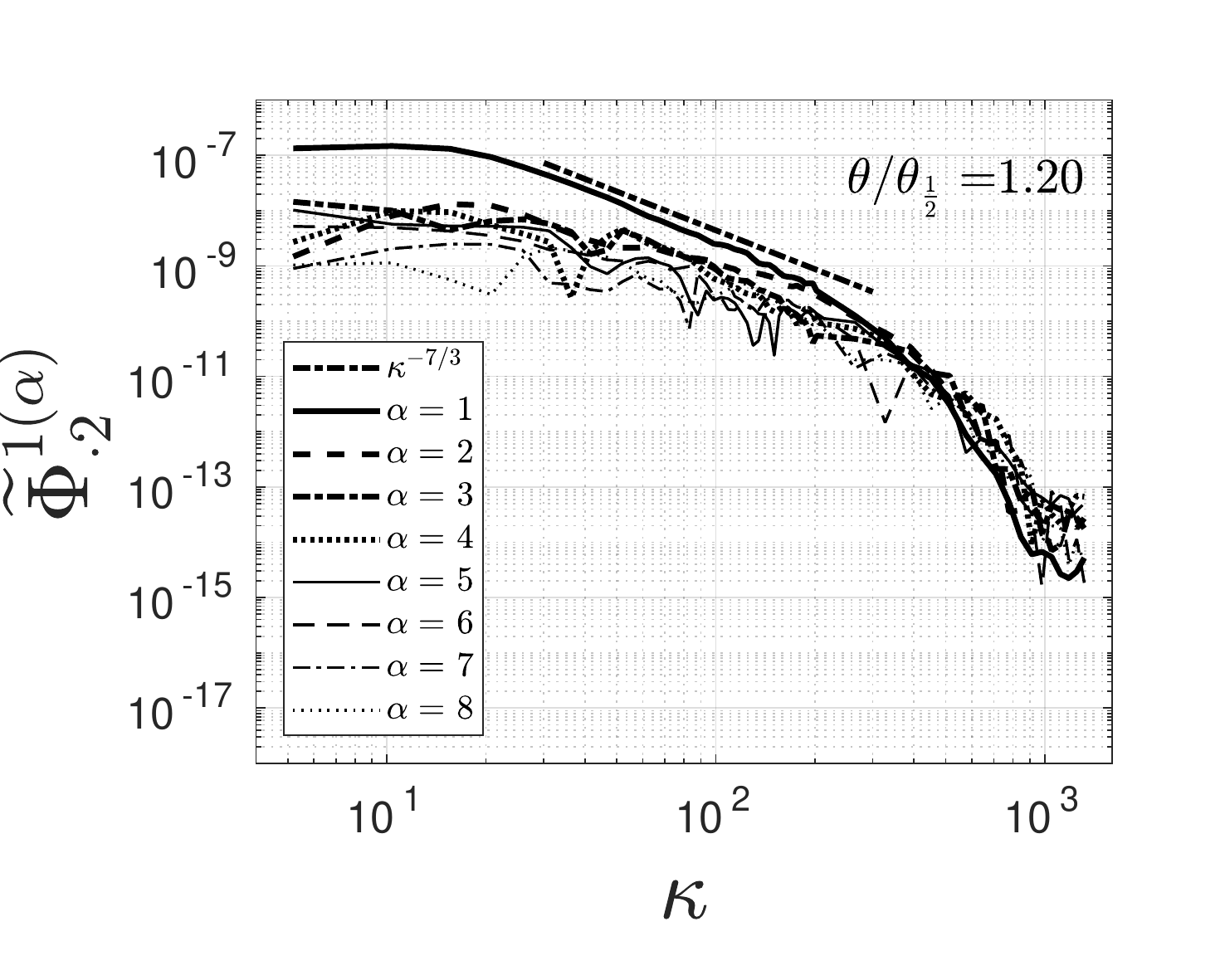}\label{fig:app_single_spectra_uv_120_SSC}}
    \subfloat[]{\includegraphics[width=0.40\linewidth]{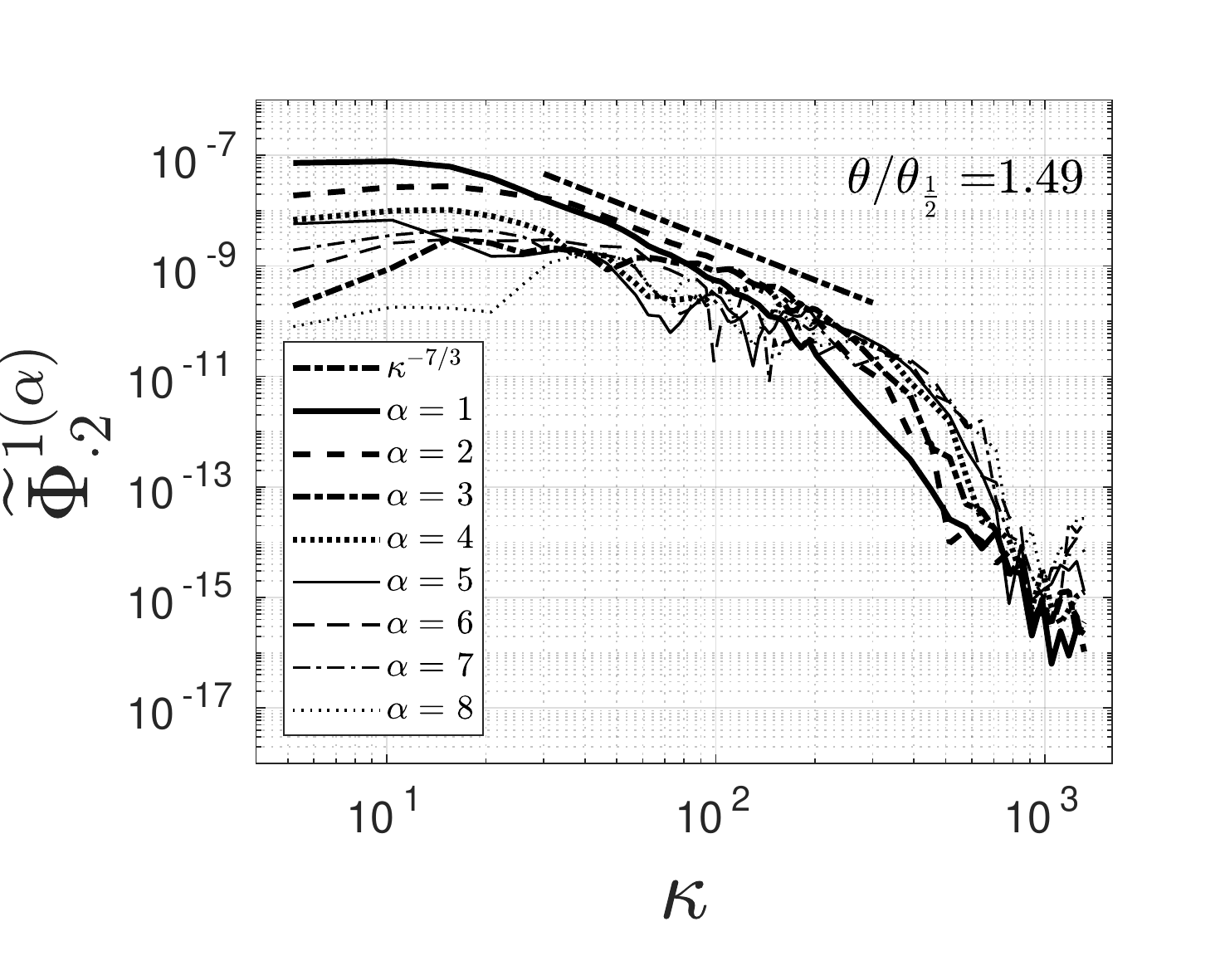}\label{fig:app_single_spectra_uv_149_SSC}}\\
    \subfloat[]{\includegraphics[width=0.40\linewidth]{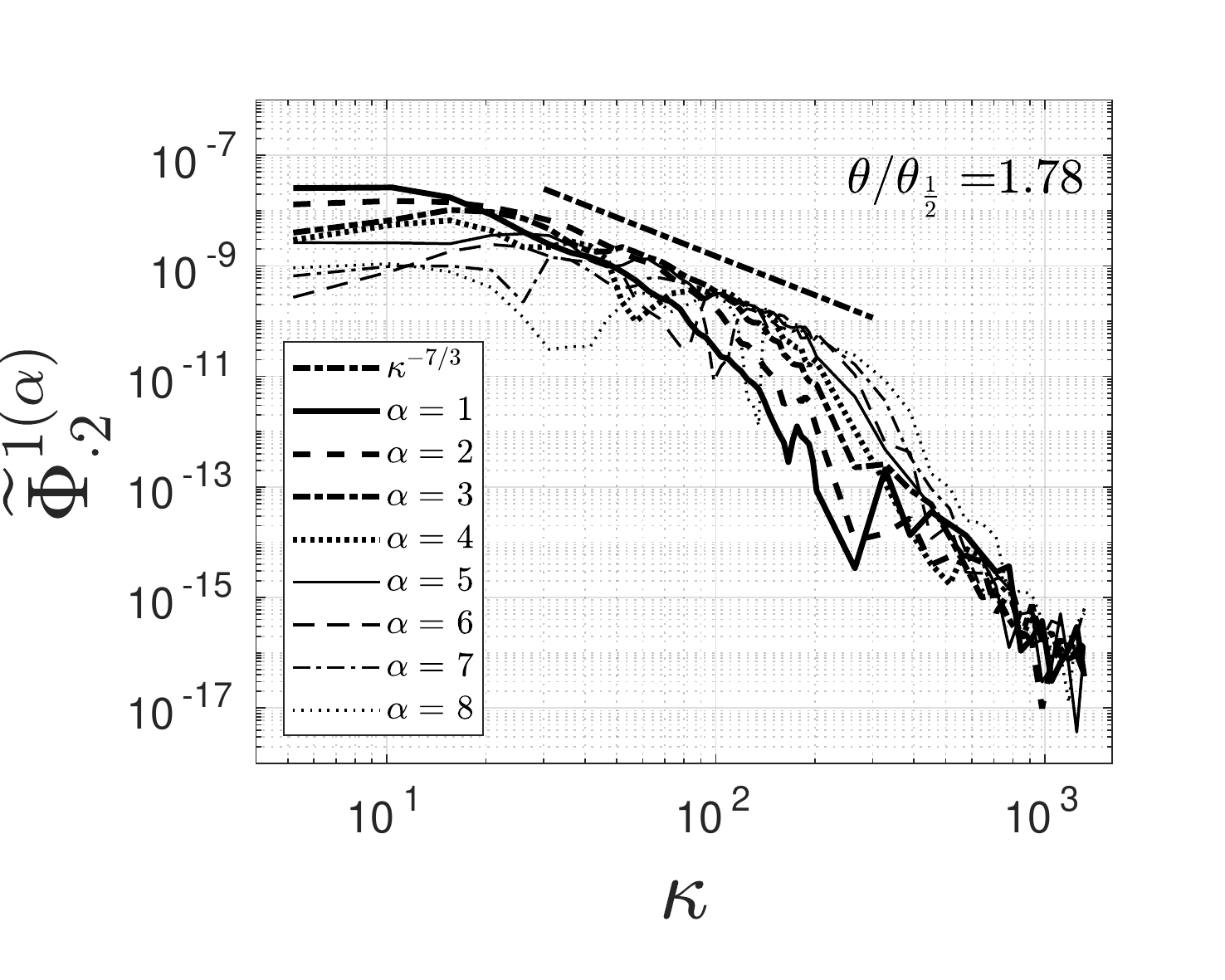}\label{fig:app_single_spectra_uv_178_SSC}}
    \subfloat[]{\includegraphics[width=0.40\linewidth]{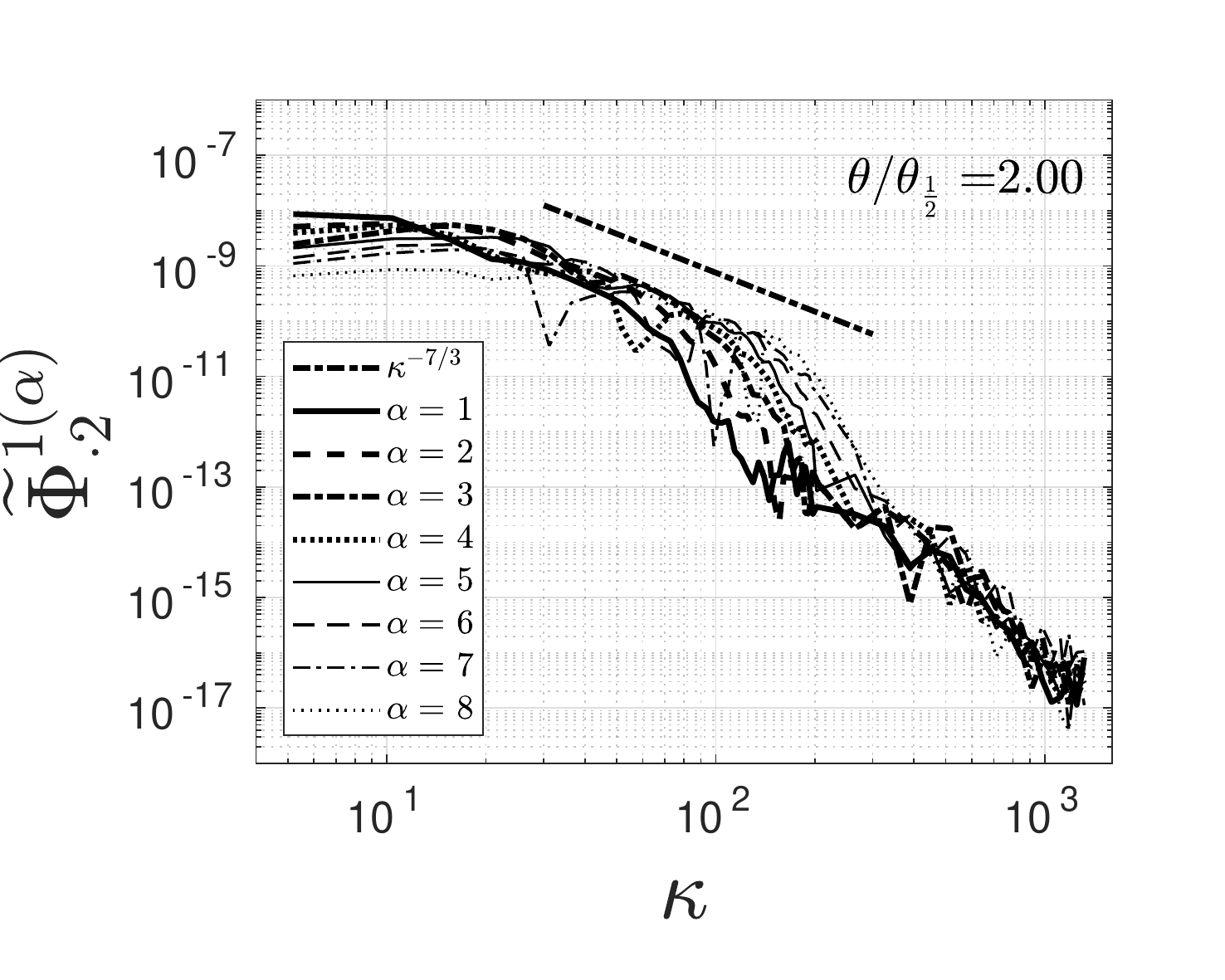}\label{fig:app_single_spectra_uv_200}}
\caption{Modal components of single-point cross-spectra, $\widetilde{\Phi}^{1(\alpha)}_{\cdot,2}$, at various spanwise coordinates, $\theta/\theta_{\frac{1}{2}}$. \label{fig:app_single_spectra_uv_reconstructed_SSC}}
\end{figure}

Modal reconstruction of the SADFM single-line spectra at various spanwise coordinates, $\theta/\theta_{\frac{1}{2}}$ are shown. The reconstruction is of the component energy- and cross spectra is seen in figures \ref{fig:app_single_spectra_uu_reconstructed_SSC}-\ref{fig:app_single_spectra_uv_reconstructed_SSC}.
%
%
\noindent
%
%

%

%
\FloatBarrier

\end{document}